\documentclass{article}

\usepackage{jheppub}

\makeatletter
\providecommand\@fpheader{} 
\providecommand\@preprint{} 
\makeatother

\usepackage{amsmath}
\usepackage{braket}
\usepackage[table]{xcolor}
\usepackage{mathrsfs}
\usepackage[english]{babel}
\usepackage{graphicx,multirow}
\usepackage{caption}
\usepackage{subcaption}
\usepackage{amssymb}
\usepackage{slashed}
\usepackage{makecell}
\usepackage{hhline}
\usepackage{tcolorbox}

\usepackage{upgreek}

\usepackage[makeroom]{cancel}

\def\nn{\nonumber}

\def\bea{\begin{eqnarray}}
\def\eea{\end{eqnarray}}
\def\ba{\begin{eqnarray}}
\def\ea{\end{eqnarray}}
\def\be{\begin{equation}}
\def\ee{\end{equation}}

\def \bal#1\eal  {\begin{align} #1 \end{align}}
\def\({\left(}
\def\){\right)}
\def\[{\left[}
\def\]{\right]}
\def\<{\langle}
\def\>{\rangle}
\def\d{\mathrm{d}}

\def\nn {\nonumber}
\def\d{\mathrm{d}}

\DeclareMathOperator{\Tr}{Tr}
\DeclareMathOperator{\tr}{tr}

\newcommand{\roughly}[1]{\mathrel{\raise.3ex\hbox{$#1$\kern-0.85em
\lower1ex\hbox{$\sim$}}}}

\newcommand{\stu}{St\"uckelberg }
\newcommand{\Sch}{Schr\"odinger }

\def\funcd{\mathrm{d}} 
\def\pathd{\mathcal{D}} 

\numberwithin{equation}{section}
\title{Schwinger-Keldysh Path Integral for Gauge theories}

\author[a,b]{Greg Kaplanek,}
\author[c,d]{Maria Mylova,}
\author[e,f]{and Andrew J. Tolley}

\affiliation[a]{Department of Electrical Engineering and Computer Science, Syracuse University, NY 13210, USA}
\affiliation[b]{Institute for Quantum \& Information Sciences, Syracuse University, NY 13210, USA}
\affiliation[c]{Kavli Institute for the Physics and Mathematics of the Universe (WPI)
the University of Tokyo, Kashiwa, Chiba 277-8583, Japan}
\affiliation[d]{Perimeter Institute for Theoretical Physics, Waterloo, Ontario, N2L 2Y5, Canada}
\affiliation[e]{Abdus Salam Centre for Theoretical Physics, Imperial College, London, SW7 2AZ, UK}
\affiliation[f]{CERCA, Department of Physics, Case Western Reserve University, 10900 Euclid Ave, Cleveland,
OH 44106, USA}

\emailAdd{gkaplane@syr.edu, maria.mylova@ipmu.jp, a.tolley@imperial.ac.uk}

\date{today}

\begin{document}

\sloppy

\abstract{We develop the Schwinger-Keldysh path-integral formalism for open non-Abelian gauge theories that are gauge-fixed via the BRST method in covariant gauges. We focus on generic initial states, pure and mixed, specified at finite times suitable for non-equilibrium processes. We pay particular attention to the handling of the indefinite Hilbert space, the construction of BRST-invariant \Sch picture wavefunctionals, density matrices and inner product, the implementation of the Hata-Kugo prescription, and the role of boundary terms at both the initial and final times. We highlight the advantages of the Nakanishi-Lautrup field representation in dealing with initial/final conditions. The resulting Schwinger-Keldysh path integral is manifestly invariant under a diagonal (retarded) BRST symmetry for arbitrary physical initial states, whether pure or mixed. From this, we obtain the corresponding Ward-Takahashi-Slavnov-Taylor identities, valid perturbatively. Non-perturbatively the Gribov ambiguity is expected to break or modify the BRST symmetry. The naive advanced BRST symmetry is shown to be explicitly violated by the in-in boundary conditions. We show that the Feynman-Vernon influence functional derived by integrating out charged matter and/or hard gluon modes remains (perturbatively) BRST invariant. When the Open EFT action is expanded to second order in advanced fields it exhibits an exact symmetry under a contraction of the original BRST symmetry. This Keldysh BRST symmetry is equivalent to the BRST associated with the retarded gauge transformations together with a linearly realised BRST transformation of the advanced fields.
Together these govern the structure of the leading terms in an Open EFT. We illustrate this with the explicit example of Hard Thermal Loop Effective Theory, and construct the general form of the Open EFT in a Higgs phase when all gauge symmetries are spontaneously broken.}

\maketitle


\section{Introduction}

The Schwinger-Keldysh, or closed time path (CTP) formalism, was introduced by Schwinger as a means to compute real-time expectation values, or in-in correlation functions \cite{Manoharan:1960jca,Schwinger:1960qe}. However, it found immediate application in the description of open quantum systems in non-vacuum states and fully out of equilibrium, specifically for maser/laser systems \cite{Manoharan:1960jca,Schwinger:1960qe,KORENMAN196672}. 
The formalism was soon applied to field theory, in particular QED in \cite{Bakshi:1962dv,Bakshi:1963bn,Mahanthappa:1962ex,PhysRev.154.1233}. 
The central trick of the CTP method is to double the number of fields associated with two branches of a CTP.  The two branches are identified at the final time, which can be taken to be any time in the future of any measurements. The initial state can be specified at any finite initial time and can be chosen to be an arbitrary pure or mixed state. 

The equivalent path integral formalism was developed by Feynman and Vernon \cite{Feynman:1963fq}, who identified the influence functional encoding interactions between two branches of the CTP contour, as the characteristic of open quantum systems.
Keldysh's subsequent work introduced a rotation of the field variables that greatly simplifies calculations \cite{Keldysh:1964ud} and focused on thermal states. This was highly influential in condensed matter physics, as was the earlier work of Kadanoff and Baym, who following Schwinger and Dyson emphasised the evolution of Green's functions \cite{Baym:1961zz,kadanoff1962quantum,fujita1964partial,Danielewicz:1982kk}. Later, the work of Caldeira and Leggett \cite{Caldeira:1982iu,Caldeira:1981rx} revitalised the path-integral treatment of open quantum systems, an approach that continues to be actively developed today.

In Schwinger's original example, he considered an arbitrary time-translation invariant mixed state, of which thermal states are a special case. Thus, the Schwinger-Keldysh formalism can easily describe the evolution of an arbitrary mixed state at finite time, to any later time, for a system that can be closed or open and whose action can be time-dependent.
It is because of this flexibility that the Schwinger-Keldysh formalism is the basis for modern treatments of non-equilibrium systems, including transport theory, hydrodynamic effective actions, and open-system descriptions in cosmology and high-energy physics.

The application of the formalism is straightforward for quantum mechanical systems and field theories without symmetries. Here, the central feature is the doubling of fields, which leads to matrices of propagators whose components reflect the different time ordering boundary conditions. Nevertheless, when evolution is viewed along the CTP, and boundary conditions are absorbed into the action, the path integral SK formalism behaves identically to the more familiar in-out path integrals \cite{DeWitt:2003pm}.
This is in particular true for states that are space-time translation invariant for which the boundary conditions can be encoded in $i \epsilon$ terms \cite{Kaplanek:2025moq}.

The inclusion of global and local symmetries introduces new features related to the doubling of the fields. When the underlying closed system has a global symmetry, the SK formalism naively admits two copies of the underlying global symmetry, which act separately on each branch. However, the late time boundary condition, which identifies the two branches generically breaks the doubled global symmetry to the diagonal subgroup.
This means that on integrating out any part of the system, the resulting open system only exhibits a single diagonal copy of the global symmetry. This manifests itself due to the emergence of interactions between the two branches of the CTP that are characteristic of open quantum systems, i.e.~terms in the Feynman-Vernon influence functional that break the off-diagonal global symmetry.

For theories with local gauge symmetries, the application of the SK formalism is more subtle. As for the global case, the doubling of the number of fields suggests a doubling of the gauge symmetry. To understand to what extent this survives, it is necessary to understand how the two branches of the contour are tied together at the initial and final times. For this, it is helpful to remember that gauge symmetries are better thought of as redundancies, i.e.~they are not real symmetries. Indeed, a consistent way to deal with gauge theories is to fix the gauge and solve the constraints to reduce the system to the physical (or reduced) phase space. For example, in electromagnetism, we can fix to Coulomb gauge, solve the Gauss constraint for the scalar part of the conjugate momenta, and reduce the theory to the physical phase space of transverse electromagnetic potentials and their associated conjugate momenta. With the gauge symmetry removed, we can follow the usual SK recipe for non-gauge theories. Although valid, this procedure is rather unwieldy and obscures the locality of the theory. In addition, in nonlinear and non-Abelian gauge theories, solving the constraints can become impractical (see for example \cite{Reinhardt:2017pyr} for an explicit treatment of QCD in Coulomb gauge). Furthermore, fixing the gauge in a non-Abelian theory will lead to the Gribov ambiguity \cite{Gribov:1977wm}.

The most successful procedure to quantise gauge theories is the Becchi-Rouet-Stora-Tyutin (BRST) formalism \cite{Becchi:1975nq,Tyutin:1975qk} (see \cite{Henneaux:1992ig} for an exhaustive review). This has the virtue of allowing locality to be maintained and, if desired, Lorentz invariance, whilst providing a clear recipe to identify the physical states and observables of the system. The application of BRST methods to gauge theories at finite temperature in both imaginary time and real time formalisms is well known \cite{Ojima:1981ma,Matsumoto:1983gk,Landsman:1986uw,Calzetta:2008iqa,DAttanasio:1996psq}. In thermal equilibrium, the Kubo-Martin-Schwinger (KMS) conditions \cite{Kubo:1957mj,Martin:1959jp} provide a recipe for the treatment of boundary conditions for gauge and ghost fields. There are two main competing prescriptions, the Bernard-Hata-Kugo \cite{Bernard:1974bq,Hata:1980yr} one which gives a thermal spectrum to all fields, including ghosts, or the Landshoff-Rebhan \cite{Landshoff:1992ne} one where only physical modes are thermalised.
Thermal methods rely heavily on a deformation of the Schwinger-Keldysh contour, which extends the usual CTP with an imaginary segment that accounts for the thermal state \cite{mills1969propagators,craig1968perturbation,Niemi:1983nf}. When the imaginary contour segment is included at finite time, its interactions must be included within perturbation theory. However, when the initial time is sent to $- \infty$ its interactions may be dropped \cite{Landsman:1986uw} and it serves only to fix the form of the free theory Green's functions.\footnote{More precisely the contribution from the imaginary part of the contour factorises in this limit. This is because if the time integrals are taken between $\pm \infty$, space-time translation invariance is preserved by the interactions. Then if the free theory satisfies the KMS boundary conditions, the interacting theory will automatically do so, and the only thing is undetermined is the overall normalisation of the path integral. There are, however, special cases where this factorisation may not occur \cite{Landsman:1986uw,Matsumoto:1985af,Evans:1994gb}.} Schwinger-Keldysh methods have been used extensively in thermal and non-equilibrium gauge theory, scalar QED, heavy-ion transport, and real-time photon effective actions \cite{Landsman:1986uw,Calzetta:2008iqa,Weldon:1982aq,Blaizot:1995kg}.

In comparison, a fully general BRST formulation for arbitrary non-equilibrium states has received relatively little discussion. This is despite the widespread use of Schwinger-Keldysh methods in non-equilibrium gauge theory, heavy-ion physics, cosmology, and open-system EFTs. 

Recent years have seen significant discussion of Schwinger-Keldysh effective descriptions of open quantum systems, including scalar and cosmological Open EFTs \cite{Sakagami:1987mp,morikawa1990dissipation,Calzetta:1995ys,Lombardo:1995fg,Casini:2009sr,Franco:2011fg,LopezNacir:2011kk,Balasubramanian:2011wt,sieberer2016keldysh,marino2016quantum,Boyanovsky:2015tba,Boyanovsky:2015jen,Baidya:2017eho,Burrage:2018pyg,Hongo:2018ant,Hongo:2019qhi,Zhang:2019wqo,Nagy:2020bal,Pinol:2020cdp,Chaykov:2022pwd,Tinwala:2024wod,Ota:2024mps,Salcedo:2024smn,Burgess:2024heo,Green:2024cmx,Panda:2025tpu,Akyuz:2025bco,Kading:2025cwg,Wang:2025hlz,Cespedes:2025zqp,Cespedes:2025ple,Lee:2025kgs,Colas:2025ind,Burrage:2025xac,Cespedes:2026fdp,Li:2026lwl}, applications to photons and gravitons \cite{Boyanovsky:1998pg,Boyanovsky:1999jh,Blaizot:2015hya,Bao:2019ghe,DeLisle:2019dyw,He:2021jna,Salcedo:2024nex,Lau:2024mqm,Salcedo:2025ezu,Christodoulidis:2025ymc,Christodoulidis:2025vxz,Yoshimura:2026vil,Salcedo:2026cqb,Bu:2026mxr}, as well as holography and hydrodynamics \cite{Agon:2014uxa,Yeh:2014mfa,Harder:2015nxa,Haehl:2015uoc,Crossley:2015evo,Jensen:2017kzi,Jana:2020vyx,Loganayagam:2020eue,Loganayagam:2022zmq,Akyuz:2023lsm,Pelliconi:2023ojb,Jain:2023obu,Baggioli:2023tlc,Ota:2024yws,Liu:2024tqe,Hongo:2024brb,Abbasi:2024pwz,Vardhan:2024qdi,Huang:2024rml,Bu:2025zad,Kawamoto:2025kfu,Sharma:2025hbk,Firat:2025upx,Jain:2026obh,Hauser:2026sgr}. Related open-system methods have been widely used in cosmology, including studies of decoherence, inflationary fluctuations, reheating, and stochastic semiclassical gravity \cite{Brandenberger:1990bx,Matacz:1992mk,Calzetta:1993qe,Hu:1994ep,Polarski:1995jg,Hu:2002jm,Burgess:2006jn,Prokopec:2006fc,Martineau:2006ki,Sharman:2007gi,Kiefer:2008ku,Hu:2008rga,Kiefer:2010pb,Bachlechner:2012dg,Burgess:2014eoa,Nelson:2016kjm,Shandera:2017qkg,boddy2017decoherence,Hollowood:2017bil,Martin:2018zbe,Martin:2018lin,Gong:2019yyz,Brahma:2020zpk,Brahma:2021mng,Zarei:2021dpb,Banerjee:2021lqu,Martin:2021znx,Burgess:2022nwu,Colas:2022hlq,DaddiHammou:2022itk,Colas:2022kfu,Sou:2022nsd,Brahma:2022yxu,Boutivas:2023mfg,Kading:2023mdk,Ning:2023ybc,Sharifian:2023jem,Colas:2024xjy,Colas:2024ysu,Bowen:2024emo,Burgess:2024eng,Bhattacharyya:2024duw,Lopez:2025arw,Sano:2025ird,deKruijf:2024ufs,Brahma:2024ycc,Sharifian:2025olk,Burgess:2025dwm,Pipa:2025pjp,Li:2025azq,Takeda:2025cye,Cielo:2025ibc,deKruijf:2025jya,Christie:2025knc,Zhou:2025mbq,Dutta:2025ouy,Green:2025hmo,Quispitupa:2025ayu,Micheli:2025yux,Bhattacharyya:2025cxv,Nandi:2026sww,Liu:2026mzz,Amoruso:2026txw}. Also of relevance to this work are applications of open system methods to QCD, particularly for quarkonium  \cite{Borghini:2011ms,Akamatsu:2011se,Akamatsu:2014qsa,Blaizot:2017ypk,Kajimoto:2017rel,Yao:2018nmy,Yao:2021lus,Scheihing-Hitschfeld:2023tuz,Hammou:2024dtj,BR:2025lhx,Brambilla:2025sis}, where dynamics are typically formulated via master equations or transport approaches rather than an explicit influence functional (although see \cite{Akamatsu:2012vt,Hongo:2024brb,Abbasi:2024pwz,Abe:2026zlv}).

In practice, most non-equilibrium treatments of gauge theories invoke the SK formalism only briefly, typically as a tool to derive transport/kinetic/Boltzmann equations for Wigner functions or phase-space distributions \cite{Blaizot:1999xk,Arnold:2002zm}, or instead rely on lattice methods, where gauge symmetry is handled in a rather different manner \cite{Kasper:2014uaa,Tanji:2017xiw,Hoshina:2020gdy}, or on classical-statistical real-time simulations \cite{Jeon:2013zga}, and these approaches usually sidestep a detailed treatment of ghost degrees of freedom. Explicit applications of the BRST framework in genuinely non-equilibrium settings include \cite{Kao:2001vt,Cooper:2002rh,Cooper:2002ff,Cooper:2002td}. As is already evident from the competing Bernard-Hata-Kugo and Landshoff-Rebhan prescriptions in the thermal case, a careful specification of the ghost-sector state is crucial; at the heart of this issue lies a proper analysis of BRST invariance for generic states.

The purpose of this paper will be to fill that gap and specify the complete BRST path integral recipe for the generic initial states defined at finite times. Doing so requires paying particular attention to the boundary conditions at the initial and final times and to the role of the BRST symmetry.
The present work will extend the discussion given in the Abelian case in \cite{Kaplanek:2025moq} to the non-Abelian case, and will clarify several technical points related to the BRST symmetry and the open quantum description of gauge theories.

The key virtue of the BRST formalism is that it replaces the local gauge symmetry with a global (albeit Grassmann odd) symmetry. As such, the SK rules for global symmetries apply. The fields of the SK contour are doubled, including doubled Faddeev-Popov-DeWitt ghosts, and doubled Nakanishi-Lautrup fields. Nevertheless, the in-in boundary conditions guarantee that only the diagonal (retarded) BRST symmetry survives, with the off-diagonal (advanced) BRST symmetry broken even in vacuum. The diagonal BRST symmetry can be used to define a CTP extension of the Zinn-Justin equation which captures the BRST symmetry at the quantum level. This allows us to give a precise definition of the effective SK action (or Feynman-Vernon influence functional) by matching the CTP analogue of the 1PI effective action between the open and closed systems. In this way we guarantee that the open system necessarily exhibits a diagonal BRST symmetry, which is the open system realisation of that present in the closed system. The only caveat in this discussion is the thorny issue of the Gribov ambiguity which arises in non-perturbative discussions \cite{Gribov:1977wm,Vandersickel:2012tz}. Naive considerations suggest that the BRST symmetry is broken, at least in its standard form due to the Gribov ambiguity, and there is a wide literature of possible resolutions to this, for example the Gribov-Zwanziger action \cite{Vandersickel:2012tz}. There are also suggestions that the BRST formalism remains intact for a judicious choice of the gauge fixing formalism \cite{Hirschfeld:1978yq,Scholtz:1997jp,Shabanov:1999mu,Rogers:1999zj}. Although not yet fully clarified, these subtleties are probably irrelevant for the relationship between closed and open systems.

In section \ref{Sec2}, we first provide a detailed discussion of the indefinite (Krein) Hilbert space that underpins BRST quantisation. This entails introducing the Schrödinger picture, specifying an inner product, and clarifying the procedure for identifying physical states. With this in place, in section \ref{Sec3} we present the CTP path integral formalism, paying particular attention to the treatment of boundary terms. This framework then enables a clear derivation of the Ward-Takahashi-Slavnov-Taylor identities. Finally, in section \ref{Sec4} we examine how these structures inform the construction of Open EFTs, in which selected degrees of freedom are traced over or integrated out, discussing the explicit example of Hard Thermal Loop EFTs. Finally, we construct the general form, up to quadratic order in advanced fields, of the Open EFT for a gauge theory for which all gauge symmetries are spontaneously broken.

\section{Indefinite Hilbert Space Quantisation}

\label{Sec2}

The desire to quantise relativistic gauge theories whilst maintaining manifest Lorentz invariance results in the necessity to address the issue of negative norm states. This was first observed in QED, where quantisation in the Lorenz gauge implies that the kinetic (and gradient) term for the timelike polarisation of the photon $A_0$ has the wrong sign. Either we quantise in a way that preserves unitarity but permits negative-energy states, or we insist on positive-energy states and then must accept states with negative norm. Preserving Lorentz invariance requires the second choice. In QED this is not fundamentally problematic since the negative norm states are not part of the physical Hilbert space. As long as we can argue that they do not contribute to physical processes, we can live with the negative norm, and this is dealt with adequately by the Gupta-Bleuer quantisation. 
In a non-Abelian theory such as Yang-Mills, we can only guarantee that interactions do not excite unphysical states if there is a symmetry present to forbid them. Fortunately, the BRST symmetry which arises when the Faddeev-Popov-DeWitt (FPDW) ghosts are included does the job. The ghosts themselves have negative (or zero) norm states, but they serve to cancel unphysical contributions from $A_0$ and the longitudinal polarisations by means of the Kugo-Ojima quartet mechanism \cite{Kugo:1977zq,Kugo:1977yx,Kugo:1977mk,Kugo:1977mm}.  

To set notation, we shall focus on Yang-Mills with coupling constant $g$ coupled to charged/coloured matter (bosonic or fermionic) transforming in some chosen representation under a group ${\cal G}$ with structure constants $f^{abc}$. We first start with the BRST action for Yang-Mills in Lorenz gauge in the off-shell representation that includes the Nakanishi-Lautrup fields 
\be
S = \int_{t_{\rm i}}^{t_{\rm f}} \d^4 x \left[-\frac{1}{4} F^{a}_{\mu\nu}{}^2 + B^a (\partial_{\mu} A_a^{\mu}) + \frac{\xi}{2} B_a^2 + \partial_{\mu}\bar c^a D^{\mu} c^a \right] \, ,
\ee
with gauge fields $F_{\mu\nu}^a = \partial_{\mu}A_{\nu}^a -\partial_{\nu}A_{\mu}^a+g f^{abc} A_{\mu}^b A_{\nu}^c$.
We shall for most part set $\xi=1$ (i.e.~Feynman 't Hooft gauge) since this form of the free theory is the simplest.
Since our concern is to develop the Schwinger-Keldysh/in-in path integral we shall be careful to deal with the action between two finite times and track the boundary terms at the initial and final times. By contrast, since we will focus on fields in $R^3$ we will assume that all spatial boundary contributions can be neglected so that we may freely integrate by parts in space.

The off-shell form of the BRST transformation is $\delta = \eta \hat s$
\begin{equation}
\begin{split}
& \hat s A^a_{\mu} 
= D_{\mu}c^a
= \partial_{\mu} c^a + g f^{abc}A_{\mu}^b c^c \,, \\
&\hat s \bar c^a 
= B^a \,,
\end{split}
\qquad
\begin{split}
&\hat s c^a 
= - \frac{1}{2}g f^{abc} c^b c^c \,, \\
&\hat s B^a 
= 0 \, ,
\end{split}
\end{equation}
with $\eta$ a constant Grassmann parameter and $\hat s $ a Grassmann odd variation.
The Nakanishi-Lautrup field $-B^a$ has the interpretation as the momentum conjugate to $A_0^a$.
The above form of the action is BRST invariant up to a boundary term at the initial and final times
\be
\hat s S =- \Big[ \int \d^3 {\bf x} \, B^a D_0 c^a \Big]_{t_{\rm i}}^{t_{\rm f}} \, .
\ee
In the usual in-out scattering theory we are interested in the limit $t_{i} \rightarrow -\infty$ and $t_{\rm f} \rightarrow +\infty$. Including $i \epsilon$ terms to enforce the adiabatic switching off of interactions the boundary terms can be neglected. In the Schwinger-Keldysh formalism with states specified at finite times, this is no longer the case, except for space-time translation invariant states for which a similar $i \epsilon$ prescription may be defined (see \cite{Kaplanek:2025moq}). 
We can alternatively integrate by parts from the outset and work with
\be \label{SNL_sec2}
S_{\mathrm{NL} } = \int_{t_{\rm i}}^{t_{\rm f}} \d^4 x \left[-\frac{1}{4} F^{a}_{\mu\nu}{}^2 -\partial_{\mu} B^a A_a^{\mu} + \frac{1}{2} B_a^2 + \partial_{\mu}\bar c^a D^{\mu} c^a \right] = S + \Big[ \int \d^3 {\bf x} \; B^a A_0^a \Big]_{t_{\rm i}}^{t_{\rm f}} \, .
\ee
for which $\hat s S_{\mathrm{NL} }=0$. The implication is that BRST invariance will be more naturally treated in a formalism in which $B^a$ is considered the field coordinate and $A_0^a$ as its conjugate momentum. We shall refer to this as the Nakanishi-Lautrup representation \cite{Nakanishi:1990qm} given the central importance of $B^a$.
We shall later take advantage of this  to formulate the action in a manner for which the correct initial and final boundary contributions are included. 

The conserved BRST charge that generates this transformation is
\be
Q = \int \d^3 {\bf x} \[ F_{0i}^a D^i c^a-B^a D_0 c^a+ \partial_t \bar c^a \(  \frac{1}{2}g f^{abc} c^b c^c\)\] \, .
\ee
At the level of free theory $g=0$, Yang-Mills is equivalent to $N$ copies of Maxwell.
Hence, to understand the quantisation of the free theory, we first focus on the quantisation of Maxwell (i.e.~suppress the index $a$)
\be
S_{\rm Maxwell} = \int_{t_{\rm i}}^{t_{\rm f}} \d^4 x \left[-\frac{1}{4} F_{\mu\nu}^2 + B (\partial_{\mu} A^{\mu}) + \frac{1}{2} B^2 + \partial_{\mu} \bar c \partial^{\mu} c \right] \, .
\ee
with now $\delta = \eta \hat s$
\be
 \hat s A_{\mu} = \partial_{\mu} c  \,, \quad \hat s c = 0  \,, \quad  \hat s \bar c = B  \,, \quad \hat s B = 0 \, .
\ee
 The Nakanishi-Lautrup fields will play an important role later in defining the path integral, but for now we choose to integrate it out so that the path integrals sets $B=-\partial^{\mu} A_{\mu}$
\ba
S_{\rm Maxwell} &=& \int_{t_{\rm i}}^{t_{\rm f}} \d^4 x \left[-\frac{1}{4} F_{\mu\nu}^2 -\frac{1}{2}(\partial_{\mu} A^{\mu})^2  + \partial_{\mu} \bar c \partial^{\mu} c \right] \, . \\
&=& \int_{t_{\rm i}}^{t_{\rm f}} \d^4 x \left[-\frac{1}{2} (\partial_\mu A_\nu)^2 +\frac{1}{2}\partial_\mu A_\nu \partial^\nu A^\mu-\frac{1}{2}(\partial_{\mu} A^{\mu})^2  + \partial_{\mu} \bar c \partial^{\mu} c \right] \, .
\ea
Remembering that we can freely integrate by parts spatial derivatives, but need to keep track of contributions at the initial and final times, we have
\be
S_{\rm Maxwell} = \tilde S_{\rm Maxwell} + \Big[ \int \d^3 {\bf x} \, (\partial_i A_i) A_0 \Big]_{t_{\rm i}}^{t_{\rm f}} \, ,
\ee
where $\tilde S$ is the action in which each degree of freedom cleanly decouples
\be
\tilde S_{\rm Maxwell} = \int_{t_{\rm i}}^{t_{\rm f}} \d^4 x \left[-\frac{1}{2} \partial_{\nu} A_{\mu} \partial^{\nu} A^{\mu} + \partial^{\nu}\bar c \partial_{\nu} c \right] \, .
\ee
However, the BRST transformation is now
\be
\hat s \tilde S_{\rm Maxwell} =  \Big[ \int \d^3 {\bf x} \, \partial_{\nu} c \partial^{\nu} A_0 \Big]_{t_{\rm i}}^{t_{\rm f}} \, .
\ee
Thus, the form of the action $\tilde S_{\rm Maxwell}$ that is cleanest to quantise, in the sense that each off-shell degree of freedom decouples both in the equations and boundary conditions, is not the one in which the BRST symmetry is manifest. Implicitly most textbook treatments of the quantisation in Feynman 't Hooft gauge $\xi=1$ use $\tilde S_{\rm Maxwell} $ and its extension to include charged matter as the starting point. In the in-out formalism where the boundary conditions are dealt with via the $i \epsilon$ prescription, this distinction is not important.

Following the standard canonical quantisation, using $\tilde S_{\rm Maxwell}$ as the definition of the action, the conjugate momenta are \be \label{adjghost}
\pi^{\mu} =\frac{\partial \tilde {\cal L}_{\rm Maxwell}}{\partial \dot A_{\mu}}=\partial_t A^{\mu} \, , \quad \pi_c= - \partial_t \bar c \, , \quad \pi_{\bar c} = \partial_t c. 
\ee
In order for the action to be Hermitian, we make the choice that
\begin{equation}
\label{ghostsherm}
c^\dagger = c\, , \quad \text{and} \quad  \bar c^{\dagger} = -\bar c \, .
\end{equation}
These choices are consistent if the ghost conjugate momenta have the opposite Hermiticity assignment and $\eta^{\dagger}= - \eta$. Let us look at canonical quantisation prior to imposing the BRST physical state condition,
this gives the  free field mode expansions
\ba
&&\hat A_{\mu}(x) = \int \d \tilde {\bf k} \left[ \hat a_{\mu}({\bf k}) e^{ik.x}+ \hat a_{\mu}^{\dagger}({\bf k}) e^{-ik.x} \right] \, , \\
&& \hat c(x) = \int \d \tilde {\bf k} \left[ \hat b({\bf k}) e^{ik.x}+ \hat b^{\dagger}({\bf k}) e^{-ik.x} \right] \, , \\
&& \hat{\bar c}(x) = \int \d \tilde {\bf k} \left[\hat d({\bf k}) e^{ik.x}- \hat d^{\dagger}({\bf k}) e^{-ik.x} \right] \, ,
\ea
with frequency  $k^0=\omega_k$ and $\d \tilde {\bf k} \equiv \d^3 {\bf k}/[ (2\pi)^3 2 \omega_k ]$, and
with commutation relations 
\ba
&& [\hat a_{\mu}({\bf k}), \hat a_{\nu}^{\dagger}({\bf k}')] = \eta_{\mu\nu} 2 \omega_k (2\pi)^3 \delta^3({\bf k}-{\bf k}') \, ,\\
&& \{ \hat b({\bf k}), \hat d^{\dagger}({\bf k}') \} = 2 \omega_k (2\pi)^3 \delta^3({\bf k}-{\bf k}') \, , \\
&& \{ \hat d({\bf k}), \hat b^{\dagger}({\bf k}') \} =  2 \omega_k (2\pi)^3 \delta^3({\bf k}-{\bf k}') \, .
\ea
Defining the vacuum in the standard way $\hat a_{\mu}({\bf k})|0 \rangle = \hat b({\bf k}) |0 \rangle =\hat d({\bf k}) |0 \rangle=0 $, the single particle photon states  $\hat a^{\dagger}_{\mu}({\bf k}) | 0 \rangle $ are normalised as
\be \label{aanorm}
\langle 0 | \hat a_{\mu}({\bf k}) \hat a^{\dagger}_{\nu}({\bf k}') | 0 \rangle = \eta_{\mu\nu}2\omega_k (2\pi)^3 \delta^3({\bf k}-{\bf k}') \,, 
\ee
so that in particular $a^{\dagger}_{0}({\bf k}) | 0 \rangle$, or more precisely a wavepacket built out of it, has negative norm. The single-particle ghost states have zero norm 
\be
\langle 0 | \hat b({\bf k}) \hat b^{\dagger}({\bf k}') | 0 \rangle=0 \, , \quad \langle 0 | \hat d({\bf k}) \hat d^{\dagger}({\bf k}') | 0 \rangle=0 \, .
\ee
We can alternatively redefine the ghost operators as 
\be
\hat f_{\pm}({\bf k})= \frac{1}{\sqrt{2}}\( \hat b({\bf k}) \pm \hat d({\bf k}) \) \, ,
\ee
such that 
\be
\{ {\hat f}_I({\bf k}), \hat f_J^{\dagger}({\bf k}') \}=2 \kappa_{IJ} \omega_k (2\pi)^3 \delta^3({\bf k}-{\bf k}') \, , 
\ee
with $\kappa_{++}=-\kappa_{--}=1$ and $\kappa_{+-}=0$. In this representation single-particle ghost states have positive and negative norms. However, doing so is undesirable as $\hat f_{\pm}$ mix under ghost number transformations, and the latter play an important role later. 

The operator expression for the hermitian BRST charge is
\ba
\hat Q &=& \int \d^3 {\bf x} \( \hat F_{0i} \partial_i \hat c + (\partial_{\mu} \hat A^{\mu}) \partial_t \hat c\) \, , \\
&=&\int \d^3 {\bf x} \( \hat \pi_i \partial_i \hat c+\nabla^2 \hat A_0 \hat c + \hat \pi^0 \hat \pi_{\bar c} + (\partial_i \hat A_i) \hat \pi_{\bar c} \) \, \\
&=& \int \d \tilde {\bf k} \(   b^{\dagger}({\bf k}) k_{\mu} \hat a^{\mu}({\bf k}) + k_{\mu} \hat a^{\mu}{}^\dagger({\bf k})\hat b({\bf k})  \) \, ,
\ea
which generates the on-shell $\hat s A_{\mu} =\partial_{\mu} c$, $\hat s c=0$, $\hat s \bar c=-(\partial_{\mu}A^{\mu})$ form of the BRST transformation via
\be
\delta \hat O = \eta \hat s \hat O = i [\eta \hat Q ,\hat O  ] \, .  
\ee
The Hermitian ghost charge operator $\hat Q_G$ is
\ba
\hat Q_G &=&   -\int \d^3 {\bf x} \; ( \pi_c c + \bar{c} \pi_{\bar{c}}) \\
&=&-i  \int \d \tilde {\bf k} \(  \hat d^{\dagger}({\bf k}) \hat b({\bf k})- \hat b^{\dagger}({\bf k}) \hat d({\bf k}) \)\, .
\ea
It generates transformations of the ghost as
\ba
&& \delta \hat c = i [\hat c ,\theta \hat Q_G ] = \theta \hat c \,, \\
&& \delta \hat{\bar c} = i [\hat{\bar c}, \theta \hat Q_G ] = -\theta \hat{\bar c} \, .
\ea
where $\theta$ is the constant (Grassmann even) parameter of the ghost number transformation. Nonlinearly we have
\be
e^{-i \theta \hat Q_G} c e^{i \theta \hat Q_G} = e^{\theta} c \, , \quad e^{-i \theta \hat Q_G} \bar c e^{i \theta \hat Q_G} = e^{-\theta} \bar c \, .
\ee
Despite being a Hermitian operator, the eigenvalues of the ghost charge are pure imaginary. For example, for the single particle ghost states
\be
\hat Q_G \hat b^{\dagger}|0 \rangle = i \hat b^{\dagger}|0 \rangle \, , \quad \hat Q_G \hat{ d}^{\dagger}|0 \rangle =- i \hat{d}^{\dagger}|0 \rangle \, .
\ee
This is not a contradiction precisely because we are dealing with a vector space with negative norm states. Hence, we may identify $\hat Q_G = i \hat N_G$ with $\hat N_G$ the ghost number operator whose eigenvalues are integers despite being anti-Hermitian.
The commutation relation of the charges is
\be \label{QGQ_comm}
[\hat Q_G, \hat Q] = i \hat Q \, , \quad  \hat Q^2 =0 \, .
\ee
The ghost number operator acts as
\be
[ \hat N_G, \hat Q] = \hat Q \, , \quad [ \hat N_G, \hat c] = \hat c
 \,, \quad [ \hat N_G, \hat {\bar c}] = -\hat {\bar c} \, ,
\ee
so that $\hat Q$ and $\hat c$ are ghost number $+1$ and $\bar c$ is $-1$.

\subsection{\Sch Representation for the photon}

Path integrals are most naturally realised in either the \Sch representation or a coherent state representation. For the spatial polarisations of the photon it is easiest to work in a \Sch representation with eigenstates
\be
\hat A_i({\bf x}) | A_i \rangle = A_i({\bf x}) | A_i \rangle \, .
\ee
The annihilation and creation operators have explicit representations
\ba
\hat a_i({\bf k})
& \equiv \int \mathrm{d}^3 {\bf x} \, 
e^{-i \mathbf{k}\cdot\mathbf{x}}
\left(
\sqrt{-\nabla^2}\, A_i({\bf x})
+ \frac{\delta}{\delta A_i({\bf x})}
\right), \\ \label{adagger2}
\hat a_i^{\dagger}({\bf k})
&\equiv \int \mathrm{d}^3 {\bf x} \, 
e^{i \mathbf{k}\cdot\mathbf{x}}
\left(
\sqrt{-\nabla^2}\, A_i({\bf x})
- \frac{\delta}{\delta A_i({\bf x})}
\right).
\ea
acting on wavefunctionals, so that the vacuum is defined by
\be
\hat a_i({\bf k}) |0 \rangle =0 \quad \Rightarrow \quad \left(
\sqrt{-\nabla^2}\, A_i({\bf x})
+ \frac{\delta}{\delta A_i({\bf x})}
\right) \langle A_i | 0 \rangle=0\, .
\ee
The vacuum wavefunctional is real, Gaussian and normalisable with respect to the naive inner product
\be
 \langle A_i | 0 \rangle = (\det(-\nabla^2))^{3/8} e^{-\frac{1}{2} \int \d^3 {\bf x}\;  A_i({\bf x}) \sqrt{-\nabla^2} A_i({\bf x})} \, , 
\ee
where we normalised the 3 dimensional functional integral measure 
\be
\int  \funcd \phi[{\bf x}] \, e^{-\int \d^3{\bf x}\phi({\bf x})^2}=1 \, .
\ee
With this choice
\be
\Big( \prod_{i=1}^3 \int \funcd A_i \Big) \langle 0| A_i \rangle \langle A_i | 0 \rangle =1 \, . 
\ee
Multi-photon states can then be obtained by applying the representation of $\hat a_i^{\dagger}({\bf k})$ in \eqref{adagger2}, and are similarly normalisable with positive norm. In this respect, each spatial component of the photon behaves like an independent scalar field.

The problem arises as soon as we consider the timelike component of the photon. Manifest Lorentz invariance would force us to follow the same procedure and define \Sch eigenstates as 
\be
\hat A_0({\bf x}) | A_0 \rangle = A_0({\bf x}) | A_0 \rangle \, ,
\ee
where now 
\ba \label{a0defs}
\hat a_0({\bf k})
& \equiv \int \mathrm{d}^3 \mathbf{x} \, 
e^{-i \mathbf{k}\cdot\mathbf{x}}
\left(
\sqrt{-\nabla^2}\, A_0({\bf x})
- \frac{\delta}{\delta A_0({\bf x})}
\right), \\
\hat a_0^{\dagger}({\bf k})
&\equiv \int \mathrm{d}^3 \mathbf{x} \, 
e^{i \mathbf{k}\cdot\mathbf{x}}
\left(
\sqrt{-\nabla^2}\, A_0({\bf x})
+ \frac{\delta}{\delta A_0({\bf x})}
\right).
\ea
The difference in sign in front of the functional derivative is directly related to the fact that for $A_0$ its momentum conjugate is $-\partial_t A_0$ which is the origin of the negative norm that follows from \eqref{aanorm}, and the reason why these states have a negative norm. The immediate consequence is that the vacuum wavefunctional now satisfies 
\be
\left(
\sqrt{-\nabla^2}\, A_0({\bf x})
- \frac{\delta}{\delta A_0({\bf x})}
\right) \langle A_0 | 0 \rangle=0 \, ,
\ee
whose solution is 
\be
 \langle A_0 | 0 \rangle = (\det(-\nabla^2))^{1/8} e^{\frac{1}{2} \int \d^3 {\bf x} \; A_0({\bf x}) \sqrt{-\nabla^2} A_0({\bf x})} \, ,
\ee
which is not normalisable. This was inevitable since all states in the \Sch picture have a positive norm, but we know that states with an odd number of timelike polarisations must have a negative norm.
Thus, the \Sch picture simply fails, as evidenced by the non-normalizable wavefunction. 

At first glance then it appears that we cannot give a meaningful path integral representation for a photon in Lorenz gauge. Fortunately, there is a simple way out: since we are dealing with an indefinite Hilbert space, it is no longer necessary that the eigenvalues of a Hermitian operator are real. We have already seen this with the ghost-number, the operator $\hat Q_G$ is Hermitian, but its eigenvalues are pure imaginary. As noted by Pauli \cite{Pauli:1943sqd}, there exists an alternative Schr\"odinger-like representation in which the eigenvalues are pure imaginary
\be
\hat A_0 | A_4 \rangle = i A_4 | A_4 \rangle \, .
\ee
The notation is suggestive of what happens under a Wick rotation $x^0 \rightarrow -i x^4$, $A_0 \rightarrow i A_4$, however, it is important to stress that we are not Wick rotating the operator, we are simply defining a different representation for the same original Hermitian operator. Stated differently, if we redefine $\hat A_0 = i \hat A_4$ then $\hat A_4$ is anti-Hermitian, but it is nevertheless self-adjoint with respect to a redefined inner product. Specifically, the resolution of unity associated with this redefined inner product is 
\be
\hat 1 = \int \funcd A_4\;  |-A_4 \rangle \langle A_4 | \, ,
\ee
and the overlap of two \Sch eigenstates is given by a functional delta function
\be
\langle A_4 | A_4 ' \rangle = \delta(A_4+A_4') \, .
\ee
The sign flip is what is necessary to ensure that $\hat A_4$ is self-adjoint despite being anti-Hermitian, specifically
\be
\langle A_4'| \hat A_0 | A_4 \rangle = i A_4 \langle A_4'| A_4 \rangle = i A_4 \delta(A_4+A_4') = -i A_4' \delta(A_4+A_4') \, .
\ee
so that  
\be
\langle A_4'| \hat A_0 | A_4 \rangle^*=\langle A_4| \hat A_0 | A_4' \rangle \, ,  \langle A_4'| \hat A_4 | A_4 \rangle^*=-\langle A_4| \hat A_4 | A_4' \rangle \, ,
\ee
or more abstractly 
\be
\langle A_4| \hat A_0 =\langle A_4|( - i A_4) \, , \quad \quad \langle A_4| \hat A_4 =\langle A_4|( -  A_4)\, ,
\ee
as required by the Hermiticity of $\hat A_0$ and anti-Hermiticity of $\hat A_4$.
The associated definition of the trace of an operator is
\be
\Tr[\hat O] = \int \funcd A_4\;  \langle -A_4 | \hat O | A_4 \rangle \, , 
\ee
so that
\be
\Tr[|\psi \rangle \langle \psi| \hat O] = \langle \psi | \hat O | \psi \rangle \, , 
\ee
consistent with the rules for computing expectation values of pure state density matrices.
In the language of wavefunctionals we compute the inner product as
\be
\langle \psi_1| \psi_2 \rangle = \int \funcd A_4 \, \psi_1^*(-A_4) \psi_2(A_4) =\langle \psi_2| \psi_1 \rangle^*\, .
\ee
It is this sign flip that allows states in the \Sch picture to still have negative norm. Specifically, since 
\be
\langle \psi| \psi \rangle = \int \funcd A_4 \, \psi^*(-A_4) \psi(A_4) \, ,
\ee
we see that even wavefunctionals have a positive norm and odd wavefunctionals have a negative norm. 

The creation and annihilation operators are now defined by 
\ba
\hat a_0({\bf k})
& \equiv i \int \mathrm{d}^3 \mathbf{x} \, 
e^{-i \mathbf{k}\cdot\mathbf{x}}
\left(
\sqrt{-\nabla^2}\, A_4({\bf x})
+ \frac{\delta}{\delta A_4({\bf x})}
\right), \\
\hat a_0^{\dagger}({\bf k})
&\equiv i \int \mathrm{d}^3 \mathbf{x} \, 
e^{i \mathbf{k}\cdot\mathbf{x}}
\left(
\sqrt{-\nabla^2}\, A_4({\bf x})
- \frac{\delta}{\delta A_4({\bf x})}
\right).
\ea
and the vacuum wavefunctional now satisfies
\be
\left(
\sqrt{-\nabla^2}\, A_4({\bf x})
+ \frac{\delta}{\delta A_4({\bf x})}
\right) \langle A_4 | 0 \rangle=0 \, ,
\ee
with solution 
\be
\langle A_4| 0 \rangle =(\det(-\nabla^2))^{1/8} e^{-\frac{1}{2} \int \d^3 {\bf x}\; A_4({\bf x}) \sqrt{-\nabla^2} A_4({\bf x})} \, . 
\ee
This wavefunctional is now clearly normalisable and since it is an even function of $A_4$ has a positive norm as required $\langle 0 | 0 \rangle=1$. It is straightforward to see that any state that contains $n$ timelike photons, i.e.~
\be
| \mathbf{k}_1 , \dots \mathbf{k}_n \rangle = \hat a_0^{\dagger}(\mathbf{k}_n) \dots \hat a_0^{\dagger}(\mathbf{k}_1) | 0 \rangle \, ,
\ee
will have a wavefunctional which satisfies
\be
\langle -A_4| \mathbf{k}_1 , \dots \mathbf{k}_n \rangle  = (-1)^n \langle A_4| \mathbf{k}_1 , \dots \mathbf{k}_n \rangle \, ,
\ee
implying that states with odd numbers of $A_0$ polarisations have negative norm
\be
\langle  \mathbf{k}_1 , \dots \mathbf{k}_n | \mathbf{k}_1 , \dots \mathbf{k}_n \rangle < 0 \,, \quad \text{for $n$ odd} \, ,
\ee
exactly as we expect from the commutation relations. This confirms that the pure imaginary representation provides a consistent representation of the operator algebra for which generic states are normalisable. If we use particle eigenstates to compute the trace of an operator, then schematically, we have
\be
\Tr[\hat O] = \sum_{n} (-1)^n \langle n| \hat O | n \rangle = \tr \[ e^{i \pi \int \d \tilde {\bf k} \; \hat a_0^{\dagger}({\bf k}) \hat a_0({\bf k})}\hat O\]\, ,
\ee
with $\tr$ the conventional `matrix' trace with only additive terms.
\be
\tr[\hat O]=\sum_{n}  \langle n| \hat O | n \rangle \, .
\ee

\subsection{Relation with path integral contour rotation}

\label{sec:contourrotation}
An alternative way to understand Pauli's representation is to return to the original \Sch representation $\psi(A_0)$ with its naively non-normalizable wavefunctions. We then define the inner product of two such wavefunctionals via
\be
\langle \psi_1 | \psi_2 \rangle = \int \funcd A_4\, \left[ \psi_1(A_0)^* \psi_2(A_0) \right]\big|_{A_0 \rightarrow i A_4} \, .
\ee
The sign flip comes from the fact that
\be
\psi_1(A_0)^*|_{A_0 \rightarrow i A_4} = \( \psi_1(-iA_4)\)^* \, ,
\ee
since the complex conjugate on the LHS is taken assuming that $A_0$ is real, while on the RHS it assumes that $A_4$ is real. This leads to a very pragmatic way to understand quantisation. We can work in the naive \Sch representation with $A_0$ real, for example, to compute the dynamical evolution of the wavefunctional, but at the final step when we need to compute an expectation value, we do so by first constructing the naive probability density and then rotating the field space contour of integration to ensure it is convergent. In the language of the Schwinger-Keldysh formalism, for which we naturally have two fields on each branch, the implication is that unitarity, {\it i.e}. conservation of probability, is only maintained if both branches were rotated in the same manner
\be \label{rotationchoice}
A_0^{\pm} \rightarrow i A_4^{\pm}  \, .
\ee
The opposite choice 
\be
A_0^{\pm} \rightarrow \pm i A_4^{\pm}  \, ,
\ee
would correspond to working with the inner product 
\be
\langle \psi_1| \psi_2 \rangle = \int \funcd A_4 \, \psi_1^*(A_4) \psi_2(A_4) \, ,
\ee
which, despite being positive for $\psi_1=\psi_2$, would give results inconsistent with the canonical commutation relations since no states would have a negative norm. In fact, we shall see that unitarity and causality can only be maintained with the former choice \eqref{rotationchoice}. For a related recent discussion see \cite{Witten:2025ayw}.

\subsection{Ghost \Sch Representation}

Physical fermions do not admit an analogue of the \Sch representation, even allowing for Grassmann valued arguments. The basic problem is that fermions need to come in pairs and a single Grassmann parameter for each mode is insufficient to describe the physical Hilbert space. The usual resolution is to work with fermionic coherent states, which gives the desired pairing. 
However, for ghosts in the Lorenz gauge, the situation is different. Since ghosts satisfy second order equations, and since their momentum conjugates are truly independent, the pairing of $c$ and $\bar c$ already provides what is necessary to construct an inner product without the need to introduce coherent states.\footnote{For related discussions, see \cite{Kashiwa:1983tr,Jackiw:1995be,Grensing:2001dc}.}

\subsubsection{\Sch states for a single pair of ghosts}

To see why, let us consider the simple example of a single mode of a pair of quantum mechanics ghosts with unit angular frequency and second order action 
\be
S_{\rm ghosts} = \int_{t_{\rm i}}^{t_{\rm f}} \d t \(- \partial_t  \bar c \partial_t  c + \bar c c \)\, .
\ee
The non-trivial commutation relations are $ \{ \hat c, \hat \pi_c \} = i $, $ \{ \hat{\bar c}, \hat \pi_{\bar c} \} = i$ with $  \hat \pi_c=-\partial_t \hat{\bar c}$, $  \hat \pi_{\bar c}=\partial_t \hat c$.
Given the equation of motion $\partial_t^2 c = - c$, and $\partial_t^2 \bar c = - \bar c$ and recalling from (\ref{ghostsherm}) that $c$ is Hermitian and $\bar c$ is anti-Hermitian, we can split the ghost operators into positive and negative frequency solutions
\be
\hat c(t) = \frac{1}{\sqrt{2}}(\hat b e^{-it } + \hat b^{\dagger} e^{it}) \, , \quad  \hat { \bar c }(t) = \frac{1}{\sqrt{2}}(\hat d e^{-it } - \hat d^{\dagger} e^{it}) \, ,
\ee
so that at $t=0$:
\begin{equation}
  \begin{split}
    \hat c &= \frac{1}{\sqrt{2}}(\hat b + \hat b^{\dagger})\, , \\
    \hat \pi_c & = \frac{1}{\sqrt{2}} i (\hat d + \hat d^{\dagger})\, ,
  \end{split} \qquad \qquad
  \begin{split}
    \hat{\bar c} & = \frac{1}{\sqrt{2}}(\hat d - \hat d^{\dagger})\, , \\
    \hat \pi_{\bar c} & = -\frac{1}{\sqrt{2}}i (\hat b - \hat b^{\dagger} )\, ,
  \end{split}
\end{equation}
from which it follows
\begin{equation}
  \begin{split}
  \label{LadderOpGhosts}
    \hat b & = \frac{1}{\sqrt{2}} (\hat c + i \hat \pi_{\bar c} )\, , \\
    \hat d & = \frac{1}{\sqrt{2}} (\hat{\bar c} - i \hat \pi_{c} )\, ,
  \end{split} \qquad \qquad
  \begin{split}
    \hat b^{\dagger} & = \frac{1}{\sqrt{2}} (\hat c - i \hat \pi_{\bar c} ) \, ,\\
    \hat d^{\dagger} & = \frac{1}{\sqrt{2}} (-\hat{\bar c} - i \hat \pi_{c} ) \, ,
  \end{split} 
\end{equation}
and finally the commutation relations for the raising and lowering operators are
\ba
&& \{ \hat b , \hat d^{\dagger} \} = 1 \, , \quad  \{ \hat d , \hat b^{\dagger} \} =1\,, \\
&& \{ \hat b , \hat b^{\dagger} \} =  \{ \hat b , \hat b \}=\{ \hat d , \hat d^{\dagger} \} =  \{ \hat d , \hat d \}=0 \,.
\ea
Due to the Fermi statistics, there are only four states which we denote
\be
|0 \rangle \, , \quad |1 \rangle = \hat b^{\dagger} |0 \rangle \, , \quad |2 \rangle = \hat d^{\dagger} |0 \rangle \, , \quad |3 \rangle = \hat b^{\dagger} \hat d^{\dagger} |0 \rangle \, .
\ee
These have a non-standard inner product 
\be
\label{metricghost}
\eta_{\alpha\beta}= \langle \alpha | \beta \rangle
=
\begin{pmatrix}
1 & 0 & 0 & 0 \\
0 & 0 & 1 & 0 \\
0 & 1 & 0 & 0 \\
0 & 0 & 0 & -1
\end{pmatrix},
\qquad
\alpha,\beta = 0,1,2,3.
\ee
which has two positive eigenvalues and two negative. This is again a reminder that we are dealing with an indefinite Hilbert space. We can define the resolution of unity as
\be
\hat 1 = \sum_{\alpha =0}^3\sum_{\beta =0}^3 | \alpha \rangle \eta^{\alpha\beta} \langle \beta | \, , 
\ee
with $\eta^{\alpha\beta}$ the inverse metric (which in this case is identical). 
Similarly, the trace of an operator is defined as
\be
\Tr[\hat O] = \sum_{\alpha =0}^3\sum_{\beta =0}^3 \eta^{\alpha\beta} \langle \alpha | \hat O | \beta \rangle \, ,
\ee
which differs from the conventional matrix trace.
This is the appropriate definition of the trace for an indefinite Hilbert space, which together with the inner product preserves the cyclicity
\ba
\Tr[\hat A \hat B ] &=&  \sum_{\alpha =0}^3\sum_{\beta =0}^3 \eta^{\alpha\beta} \langle \alpha | \hat A \hat B | \beta \rangle \\
 &=&  \sum_{\alpha =0}^3\sum_{\beta =0}^3  \sum_{\alpha' =0}^3\sum_{\beta' =0}^3\eta^{\alpha\beta} \langle \alpha | \hat A | \beta' \rangle \eta_{ \beta' \alpha'} \langle \alpha' |\hat B | \beta \rangle 
\\
 &=&  \sum_{\alpha =0}^3\sum_{\beta =0}^3  \sum_{\alpha' =0}^3\sum_{\beta' =0}^3\eta^{\alpha'\beta'} \langle \alpha' | \hat A | \beta \rangle \eta^{ \beta \alpha} \langle \alpha |\hat B | \beta' \rangle \\
 &=&\Tr[\hat B \hat A ] \, , 
\ea
given $\eta^{\alpha \beta}=\eta^{\beta \alpha}$.
Our goal then is to construct fermionic wavefunctions that are consistent with the above inner product.
To do this we define simultaneous eigenstates $|c,\bar c \rangle$ with Grassmann odd eigenvalues\footnote{If two nilpotent operators anti-commute, they can be simultaneously diagonalised with eigenstates that have Grassmann-odd eigenvalues.}
\be
 \hat c |c,\bar c \rangle = c |c,\bar c \rangle \, , \quad  \hat{\bar c} |c,\bar c \rangle = \bar c |c,\bar c \rangle \, .\ee
We are allowed to do this since $\hat c$ and $\hat{\bar c}$ anticommute, as a consequence of the second order equation of motion, and hence they can be simultaneously realised as Grassmann odd parameters. The eigenvalues $c$ and $\bar c$ inherit the hermiticity properties of the operators from Eq.~(\ref{ghostsherm}), so that with the notational convention $|c,\bar c \rangle^{\dagger}=\langle c,\bar c |$, we have
\be
 \langle c,\bar c |\hat c  =\langle c,\bar c | c  \, , \quad  \langle c,\bar c |\hat{\bar c}  = \langle c,\bar c | \bar c  \, ,\ee
which is to say both the left and right eigenstates have the same Grassmann odd eigenvalue.
We define the inner product 
\be
\langle \psi_1 | \psi_2 \rangle = \int \d c \int \d \bar c \, \langle \psi_1 |c,\bar c\rangle \langle c, \bar c | \psi_2 \rangle \, ,
\ee
and 
\be
\langle c,\bar c| c', \bar c' \rangle = \delta(\bar c-\bar c')\delta(c-c') \, ,
\ee
where $\delta(c-c')=c-c'$ is the Grassmann delta function
and consistent with our convention on $\langle c , \bar c|$. Note this condition is not entirely trivial since $\langle c,\bar c | \psi \rangle$ is generically a Grassmann number with nontrivial Hermiticity properties. We are making a very specific choice in defining the inner product, consistent with the notation
\be
(\langle c,\bar c | \psi \rangle )^{\dagger}= \langle \psi |c,\bar c\rangle\, .
\ee
Our goal is to show that these definitions are consistent with the commutation relations. 

In \Sch representation $\pi_c = i \partial_c$, $\pi_{\bar c} = i \partial_{\bar c}$  and so the creation and annihilation operators in (\ref{LadderOpGhosts}) become
\ba
&& \hat b \equiv \frac{1}{\sqrt{2}} ( c - \partial_{\bar c} ) \, , \quad \hat b^{\dagger} \equiv \frac{1}{\sqrt{2}} (c + \partial_{\bar c} ) \, , \\
&& \hat d \equiv \frac{1}{\sqrt{2}} ({\bar c} +\partial_{c} ) \, , \quad \hat d^{\dagger} \equiv \frac{1}{\sqrt{2}} (-{\bar c} + \partial_{c} ) \, .
\ea
The vacuum wavefunctional then satisfies
\be
(c - \partial_{\bar c} ) \langle c, \bar c | 0 \rangle=({\bar c} + \partial_{c} ) \langle c, \bar c | 0 \rangle =0 \, , 
\ee
which has the unique and consistent solution 
\be
\langle c, \bar c | 0 \rangle = \frac{1}{\sqrt{2}}e^{ \bar c c} \, .
\ee
The hermiticity properties are designed so that $(\bar c c)^{\dagger}=\bar c c$ and so
\be
\langle 0 | 0 \rangle = \int \d c \int \d \bar c  \,  \langle 0 | c, \bar c \rangle \langle c, \bar c | 0 \rangle = \frac{1}{2} \int \d c \int \d \bar c e^{ 2\bar c c} =1 \, .
\ee
It is now straightforward to compute the wavefunction for the remaining states by applying  $\hat b^{\dagger}$ and $\hat d^{\dagger}$:
\be
\langle c, \bar c | 1 \rangle= c \, e^{\bar c c}  \qquad \qquad \langle c, \bar c | 2 \rangle= - \bar c\, e^{ \bar c c} \qquad \qquad \langle c, \bar c | 3 \rangle=\frac{1}{\sqrt{2}} (-1+2\bar c c) \, e^{ \bar c c} 
\ee
To verify that the representation is consistent, we must check the inner product
\be
\eta_{\alpha \beta}=  \int \d c \int \d \bar c  \,  \langle \alpha | c, \bar c \rangle \langle c, \bar c | \beta \rangle \, .
\ee
A short calculation shows this works out, for example,
\be
\langle 3|3 \rangle = \int \d c \int \d \bar c\, \frac{1}{2} (1-2\bar c c)^2 e^{2\bar c c}=\int \d c \int \d \bar c \, \frac{1}{2}(1-4 \bar c c+ 2\bar c c)=-1 \, ,
\ee
and
\be
\langle 1|2 \rangle = \int \d c \int \d \bar c \, - c \bar c e^{2\bar c c}=\int \d c \int \d \bar c \,  \bar c c=1 \, .
\ee
Using that the matrix (\ref{metricghost}) is its own inverse $\eta^{-1} =\eta$, the inner product of the \Sch eigenstates is
\begin{align}
\langle c, \bar c \mid c' , \bar c' \rangle
&=
\langle c, \bar c \mid 0 \rangle \langle 0 \mid c' , \bar c' \rangle
+ \langle c, \bar c \mid 2 \rangle \langle 1 \mid c' , \bar c' \rangle
+ \langle c, \bar c \mid 1 \rangle \langle 2 \mid c' , \bar c' \rangle
- \langle c, \bar c \mid 3 \rangle \langle 3 \mid c' , \bar c' \rangle \nn \\
&=
\frac{1}{2} e^{\bar c c + \bar c' c'}
- \bar c\,c' \, e^{\bar c c + \bar c' c'}
+ c\,\bar c' \, e^{\bar c c + \bar c' c'} \quad
- \frac{1}{2} (1-2\bar c c)(1-2\bar c' c') e^{\bar c c + \bar c' c'} \nn \\
&=
\left[-\bar c c'+c \bar c'+\bar c c+\bar c' c' -2 \bar c c \bar c' c'
\right] e^{\bar c c + \bar c' c'} 
=(\bar c-\bar c')(c-c') \nn \\
&=
\delta(\bar c-\bar c') \, \delta(c-c') \, ,
\end{align}
which confirms the consistency of the definition of the inner product.

Finally, we should give an expression for the trace in terms of \Sch eigenstates. This is achieved by inserting a complete set of eigenstates 
\ba
\Tr[\hat O] &=& \int \d c \int \d \bar c \, \Tr\[|c,\bar c \rangle \langle c,\bar c |\hat O\] \\ &=&\sum_{\alpha=0}^3 \sum_{\beta=0}^3 \eta^{\alpha\beta} \int \d c \int \d \bar c \langle \alpha |c,\bar c \rangle \langle c,\bar c| \hat O | \beta \rangle \, .
\ea
It is straightforward to check from the explicit form of the wavefunctionals that 
\be
\langle \alpha |c,\bar c \rangle \langle c,\bar c| \beta \rangle=\langle -c,-\bar c| \beta \rangle \langle \alpha |c,\bar c \rangle \,, 
\ee
as a consequence of the Grassmann statistics. Similarly 
\be
\langle \alpha |c,\bar c \rangle \langle c,\bar c| \hat O |\beta \rangle=\langle -c,-\bar c| \hat O|\beta \rangle \langle \alpha |c,\bar c \rangle \,, 
\ee
and so repackaging the trace we have
\ba
\Tr[\hat O] &=&  \int \d c \int \d \bar c \langle -c, -\bar c|\hat O | c,\bar c \rangle \, .
\ea
This is the equivalent of the usual relation for physical fermions which is applied at the level of coherent states (see e.g. \cite{Jackiw:1995be,Srednicki:2007qs}).

\subsubsection{\Sch states for ghost fields}

Returning now to the field theory context, we define simultaneous eigenstates
\be
\hat c({\bf x}) | c, \bar c \rangle = c({\bf x}) | c, \bar c \rangle\, , \quad \hat{\bar c}({\bf x}) | c, \bar c \rangle =\bar c({\bf x}) | c, \bar c \rangle\, .
\ee
Again this is allowed since $c({\bf x})$ and $\bar c({\bf y})$ anticommute at equal times as a consequence of the second order equations of motion for the ghosts in Lorenz gauge. The Hermiticity properties of the operators are inherited by the Grassmann odd eigenvalues
\be \label{hermiticity}
c^{\dagger}({\bf x})=c({\bf x})\, , \quad \bar c^{\dagger}({\bf x}) =-\bar c({\bf x}) \, .
\ee
We define the inner product in the naive Grassmann functional integral way
\be
\hat 1= \int \funcd  c \int \funcd \bar c \, | c, \bar c \rangle \langle  c, \bar c | \, .
\ee
where now 
\be
\int \funcd  c = \prod_{\bf x} \int \d c({\bf x}) \, , \quad \int \funcd  \bar c = \prod_{\bf x} \int \d \bar c({\bf x}) \, , 
\ee 
and the inner product is
\be
\langle  c', \bar c' | c, \bar c \rangle = \delta(\bar c-\bar c') \delta(c-c')=\prod_{\bf x} \delta(\bar c({\bf x})-\bar c'({\bf x}))\delta(c({\bf x})-c'({\bf x}))
\ee
where the RHS are now functional versions of the Grassmann delta function. The trace contains the minus sign characteristic of fermions
\be
\Tr[\hat O] = \int \funcd  c \int \funcd \bar c \, \langle -c, -\bar c|\hat O | c, \bar c \rangle \, ,
\ee
and is cyclic $\Tr[\hat A \hat B]=\Tr[\hat B \hat A]$ even when $\hat A$ and $\hat B$ are fermionic.
To construct the wavefunctionals corresponding to the Fock vacuum and its first excitations, it is convenient to work in momentum space. We define the Fourier transforms at fixed time
\be
c({\bf x})=\int \frac{\d^3 k}{(2\pi)^3} \, c({\bf k}) e^{i{\bf k}\cdot{\bf x}} \, ,\qquad 
\bar c({\bf x})=\int \frac{\d^3 k}{(2\pi)^3} \, \bar c({\bf k}) e^{i{\bf k}\cdot{\bf x}} \, ,
\ee
and similarly for the operators $\hat c({\bf x})$, $\hat{\bar c}({\bf x})$. The hermiticity properties \eqref{hermiticity}
imply the momentum space reality conditions
\be
c^{\dagger}({\bf k})=c(-{\bf k})\, , \qquad \bar c^{\dagger}({\bf k})=-\bar c(-{\bf k}) \, .
\ee
In the \Sch representation (with left functional derivatives) we take
\be
\hat \pi_c({\bf x}) = i \frac{\delta}{\delta c({\bf x})}\, ,\qquad
\hat \pi_{\bar c}({\bf x}) = i \frac{\delta}{\delta \bar c({\bf x})}\, ,
\ee
so that $\{ \hat c({\bf x}), \hat \pi_c({\bf y})\}= i \delta^3({\bf x}-{\bf y})$ and $\{ \hat{\bar c}({\bf x}), \hat \pi_{\bar c}({\bf y})\}= i \delta^3({\bf x}-{\bf y})$. Using the same linear combinations as in the single-mode example, we define the annihilation and creation operators mode-by-mode as
\ba
&& \hat b({\bf k}) \equiv \left( \omega_{k} c({\bf k}) - \frac{\delta}{\delta \bar c(-{\bf k})}\right)\, ,\qquad
\hat b^{\dagger}({\bf k}) \equiv \left(\omega_{k} c(-{\bf k}) + \frac{\delta}{\delta \bar c({\bf k})}\right)\, , \\
&& \hat d({\bf k}) \equiv \left(\omega_{k}\bar c({\bf k}) + \frac{\delta}{\delta c(-{\bf k})}\right)\, ,\qquad
\hat d^{\dagger}({\bf k}) \equiv \left(-\omega_{k}\bar c(-{\bf k}) + \frac{\delta}{\delta c({\bf k})}\right)\, .
\ea
With these conventions, one reproduces the momentum-space anticommutators consistent with
\be
\{ \hat b({\bf k}), \hat d^{\dagger}({\bf k}')\}=2\omega_k(2\pi)^3\delta^3({\bf k}-{\bf k}')\, ,\qquad
\{ \hat d({\bf k}), \hat b^{\dagger}({\bf k}')\}=2\omega_k(2\pi)^3\delta^3({\bf k}-{\bf k}')\, ,
\ee
given the choice of measure $\d\tilde {\bf k}$ in the mode expansions.
The vacuum $|0\rangle$ is defined by
\be
\hat b({\bf k})|0\rangle=0\, ,\qquad \hat d({\bf k})|0\rangle=0\, ,\qquad \forall\,{\bf k}\, ,
\ee
which, in the \Sch basis, becomes a pair of functional differential equations for the vacuum wavefunctional
\be
\Psi_0[c,\bar c]\equiv \langle c,\bar c|0\rangle\, :
\qquad
\left(\omega_{k}c({\bf k})-\frac{\delta}{\delta \bar c(-{\bf k})}\right)\Psi_0[c,\bar c]=0\, ,\qquad
\left(\omega_{k}\bar c({\bf k})+\frac{\delta}{\delta c(-{\bf k})}\right)\Psi_0[c,\bar c]=0\, .
\ee
These are solved by a Gaussian functional, 
\be
\Psi_0[c,\bar c]= \frac{1}{(\det[-\nabla^2])^{1/4}} \exp\!\left[\int \d^3{\bf x}\, \bar c({\bf x})\sqrt{-\nabla^2}c({\bf x})\right]
=\frac{1}{(\det[-\nabla^2])^{1/4}}   \exp\!\left[\int \frac{\d^3 \mathbf{k}}{(2\pi)^3} \, \bar c(-{\bf k})\omega_{k}c({\bf k})\right] \, ,
\ee
where the normalisation is fixed (up to powers of $2$, which can be absorbed in the measure) by
\be
\langle 0|0\rangle
=\int \funcd c \int \funcd \bar c \, \( \Psi_0[c,\bar c]\)^{\dagger}\Psi_0[c,\bar c] =1 \, .
\ee
The first excited one-particle ghost states are obtained by acting with $\hat b^{\dagger}$ and $\hat d^{\dagger}$ on the vacuum,
\be
|1;{\bf p}\rangle \equiv \hat b^{\dagger}({\bf p})|0\rangle\, ,\qquad
|2;{\bf p}\rangle \equiv \hat d^{\dagger}({\bf p})|0\rangle\, ,
\ee
and their \Sch wavefunctionals follow directly from the differential operator realisations:
\ba
\langle c,\bar c|1;{\bf p}\rangle
&=&
\hat b^{\dagger}({\bf p})\,\Psi_0[c,\bar c]
=\left(\omega_{p}c(-{\bf p})+\frac{\delta}{\delta \bar c({\bf p})}\right)\Psi_0[c,\bar c]
= 2\omega_{p}c(-{\bf p})\,\Psi_0[c,\bar c] \, ,\\
\langle c,\bar c|2;{\bf p}\rangle
&=&
\hat d^{\dagger}({\bf p})\,\Psi_0[c,\bar c]
=\left(-\omega_p \bar c(-{\bf p})+\frac{\delta}{\delta c({\bf p})}\right)\Psi_0[c,\bar c]
= -2\omega_{p}\bar c(-{\bf p})\,\Psi_0[c,\bar c] \, .
\ea
Similarly, a two-particle excitation at fixed momenta is
\be
|3;{\bf p},{\bf q}\rangle \equiv \hat b^{\dagger}({\bf p})\hat d^{\dagger}({\bf q})|0\rangle \, ,
\ee
with wavefunctional
\be
\langle c,\bar c|3;{\bf p},{\bf q}\rangle
=
\hat b^{\dagger}({\bf p})\hat d^{\dagger}({\bf q})\,\Psi_0[c,\bar c] \, ,
\ee
which reduces to a polynomial in $c,\bar c$ multiplying $\Psi_0[c,\bar c]$ by repeated use of the above differential representations and the nilpotency of Grassmann variables giving
\be\langle c,\bar c|3;{\bf p},{\bf q}\rangle
=
\hat b^{\dagger}({\bf p})\hat d^{\dagger}({\bf q})\,\Psi_0[c,\bar c] = \left[ 4 \omega_q \omega_p \bar{c}(-\mathbf{q}) c(-\mathbf{p})  - 2 \omega_q (2\pi)^3 \delta^{(3)}(\mathbf{q}+\mathbf{p}) \right] \Psi_0[c,\bar{c}] \, .
\ee 

\subsection{BRST Invariance of Physical Wavefunctionals}

A physical state is BRST closed 
\be
\hat Q|\psi_{\rm phys} \rangle =0 \, ,
\ee
and is defined up to a BRST exact form $|\psi_{\rm phys} \rangle \sim |\psi_{\rm phys} \rangle+ \hat Q | \chi \rangle$.
In addition a physical state should have zero ghost number 
\be
\hat N_G | \psi_{\rm phys} \rangle =0 \, .
\ee
Temporarily neglecting matter, our state space is described by \Sch eigenstates $|A_i, A_4 , c, \bar c \rangle$ with resolution of unity 
\be
\hat 1 = \int \funcd[A_i ,A_4 , c ,\bar c ]\,  |A_i, A_4 , c, \bar c \rangle \langle A_i, -A_4 , c, \bar c   |\, .
\ee
To simplify notation, we will label the states with $A_0$, i.e.~work in the non-normalisable representation, so that we can write
\be
\psi_{\rm phys} [A_{\mu},c,\bar c] = \langle A_{\mu},c,\bar c | \psi_{\rm phys}  \rangle \, ,
\ee
with the understanding that we rotate $A_0 \rightarrow i A_4$ before computing any inner product.
It is easy to satisfy the condition of vanishing ghost number by demanding that the wavefunctional depends on the ghosts with $c$ and $\bar c$ always paired together. The non-trivial constraint is the vanishing of the BRST charge.
In this representation, the BRST charge is given by
\ba
\hat Q &=& \int \d^3 {\bf x} \left[F_{0i} \partial_i c - B \partial_t c \right] \\
&=& \int \d^3 \mathbf{x} \left[ (\partial_t A_i - \partial_i A_0) \partial_i c + ( -\partial_t A_0 + \partial_i A_i)  \partial_t c \right] \\
&=& \int \d^3 {\bf x} \left[(\pi_i-\partial_i A_0) \partial_i c +( \pi^0+\partial_i A_i) \pi_{\bar c} \right] \, .
\ea
So, remembering that $\pi^{\mu}= - i \frac{\delta}{\delta A_{\mu}}$, $\pi_c= i \frac{\delta}{\delta c}$, $\pi_{\bar c}= i \frac{\delta}{\delta \bar{c}}$, 
we have in the \Sch picture
\ba
&& i \langle A_{\mu},c,\bar c | \hat Q| \psi_{\rm phys}  \rangle  \\
&& =\int \d^3 {\bf x}  \left[ \partial_i c({\bf x}) \frac{\delta }{\delta A_i({\bf x})}-\(\partial_i A^i \)({\bf x}) \frac{\delta }{\delta \bar c({\bf x})}
-i \partial_i A_0 \partial_i c+i \frac{\delta}{\delta A_0({\bf x})} \frac{\delta}{\delta \bar c({\bf x})}
\right]  \psi_{\rm phys} [A_{\mu},c,\bar c] = 0 \, . \nn
\ea
We have already constructed the vacuum wavefunctional 
\be
\psi_0 [A_{\mu},c,\bar c] = (\det[-\nabla^2])^{1/4} e^{-\frac{1}{2} \int \d^3 {\bf x} A_{\mu} \sqrt{-\nabla^2} A^{\mu}+\int \d^3 {\bf x} \bar c \sqrt{-\nabla^2} c } \, ,
\ee
for which 
\be
\frac{\delta }{\delta \bar c({\bf x})}\psi_0 [A_{\mu},c,\bar c]= \sqrt{-\nabla^2} c({\bf x}) \psi_0 [A_{\mu},c,\bar c] \, ,
\ee
and
\be
\frac{\delta }{\delta A_{\mu}({\bf x})}\psi_0 [A_{\mu},c,\bar c]=-\sqrt{-\nabla^2} A^{\mu}({\bf x}) \psi_0 [A_{\mu},c,\bar c] \, ,
\ee
so that 
\be
i \hat Q \psi_0 [A_{\mu},c,\bar c] = - \int \d^3 {\bf x}  \bigg[ \partial_i \( c({\bf x}) \sqrt{-\nabla^2} A_i({\bf x})\)+i \partial_i\(A_0 \partial_i c)\) \bigg] \psi_0 [A_{\mu},c,\bar c] =0 \, .
\ee
vanishes via the divergence theorem, neglecting spatial boundary terms. 
 This is a useful consistency check on the signs in the ghost wavefunctional. Because we are still considering an Abelian theory it is easy to write down a generic wavefunctional which is BRST invariant in the form 
\be
\psi_{\rm phys} [A_{\mu},c,\bar c] = e^{-\frac{1}{2} \int \d^3 {\bf x} A_{\mu} \sqrt{-\nabla^2} A^{\mu}+\int \d^3 {\bf x} \bar c \sqrt{-\nabla^2} c } \Psi[A_i] \, , 
\ee
where $\Psi[A_i]$ is independent of ghosts and is invariant under (small) spatial gauge transformations, i.e.~
\be
\partial_i \( \frac{\delta \Psi[A_j]}{\delta A_i({\bf x})}\)=0 \, .
\ee
This is equivalent at the operator level to constructing a generic state by applying creation operators only for the two transverse polarisation modes 
\be
| \Psi_{\rm  phys} \rangle =\int \funcd \alpha({\bf k}) \,  \tilde \Psi[\alpha_i({\bf k})] e^{\int \d \tilde {\bf k} \sum_{i} \alpha_i({\bf k}) \hat a_i^{\dagger}({\bf k})} | 0 \rangle \, ,
\ee
with $ {\bf k} \cdot \alpha({\bf k})=0$ to enforce the transverse condition. Using 
\be
\langle A_i | e^{\int \d \tilde {\bf k} \sum_{i} \alpha_i({\bf k}) \hat a_i^{\dagger}({\bf k})} | 0 \rangle = e^{-\frac{1}{2}\int \d \tilde {\bf k} \; \alpha({\bf k}) \cdot \alpha({-\bf k})} e^{\int \d \tilde {\bf k} \; 2 \omega_k \alpha(-{\bf k})\cdot A({\bf k})}\langle A_i  | 0 \rangle \, ,
\ee
with $A_i({\bf x})=\int \frac{\d^3 {\bf k}}{(2\pi)^3} \, e^{i{\bf k}\cdot {\bf x}} A_i({\bf k})$,
we have
\be
\Psi[A_i]=\int \funcd \alpha({\bf k}) \, \tilde \Psi[\alpha_i({\bf k})]  e^{-\frac{1}{2}\int \d \tilde {\bf k}\; \alpha({\bf k}) \cdot \alpha({-\bf k})} e^{\int \d \tilde {\bf k} \; 2 \omega_k \alpha(-{\bf k})\cdot A({\bf k})} \, .
\ee
Gauge invariance may be made more manifest by writing $\alpha_i({\bf k}) = i\epsilon_{ijk} k_j \beta_k(\bf k) $, which is guaranteed transverse, so that
\be
\Psi[A_i]=\int \funcd \beta({\bf k}) \, \check \Psi[\beta_i({\bf k})]  e^{-\frac{1}{2}\int \d \tilde {\bf k}\; ({\bf k} \times \beta(-{\bf k})).({\bf k} \times \beta({\bf k})) } e^{+\int \d \tilde {\bf k} \; 2 \omega_k \beta_i(-{\bf k})B_i({\bf k})} \, .
\ee
and we have defined the magnetic field $B_i = \frac{1}{2}\epsilon_{ijk}F_{jk}$. This is equivalent to a functional Fourier/Laplace transform and effectively merely states that the wave functional can be an arbitrary normalisable function of the gauge invariant magnetic field $B_i({\bf x})$. 

\subsection{Wavefunctionals for Charged States}

It is straightforward to extend the previous discussion to include charged bosonic matter (fermions require coherent states and so must be treated at the level of the density operator). To illustrate, let us introduce a single charged complex scalar $\Phi$ with charge $q$ so that the BRST transformation is $\hat s \Phi = i q c \Phi$.
We now work with wavefunctionals of the form
\be
\psi_{\rm phys} [A_{\mu},c,\bar c,\Phi,\Phi^\dagger] = \langle A_{\mu},c,\bar c ,\Phi,\Phi^\dagger| \psi_{\rm phys}   \rangle \, ,
\ee
and the condition for BRST invariance is now
\ba
&& i \langle A_{\mu},c,\bar c, \Phi,\Phi^\dagger| \hat Q| \psi_{\rm phys}  \rangle  \\
&& =\int \d^3 {\bf x}  \left[ \partial_i c({\bf x}) \frac{\delta }{\delta A_i({\bf x})}-\partial_i A^i ({\bf x}) \frac{\delta }{\delta \bar c({\bf x})} -i \partial_i A_0 \partial_i c+i \frac{\delta}{\delta A_0({\bf x})} \frac{\delta}{\delta \bar c({\bf x})}\right. \nn\\
 &&  \left. +  i q c({\bf x})  \Phi({\bf x}) \frac{\delta }{\delta \Phi({\bf x})}- i q c({\bf x})\Phi^\dagger({\bf x}) \frac{\delta }{\delta \Phi^\dagger({\bf x})}\right]  \psi_{\rm phys} [A_{\mu},c,\bar c,\Phi,\Phi^\dagger ] = 0 \, . \nn
\ea
Making the ansatz
\be \label{wavansatz}
\psi_{\rm phys} [A_{\mu},c,\bar c, \Phi,\Phi^\dagger] = e^{-\frac{1}{2} \int \d^3 {\bf x} A_{\mu} \sqrt{-\nabla^2} A^{\mu}+\int \d^3 {\bf x} \bar c \sqrt{-\nabla^2} c } \Psi[A_i,\Phi,\Phi^\dagger] \, ,
\ee
we see this is a consistent solution provided
\be
\partial_i \( \frac{\delta \Psi[A_i,\Phi,\Phi^\dagger]}{\delta A_i({\bf x})}\)= i q  \Phi({\bf x}) \frac{\delta \Psi[A_i,\Phi,\Phi^\dagger]}{\delta \Phi({\bf x})}- i q \Phi^\dagger({\bf x}) \frac{\delta \Psi[A_i,\Phi,\Phi^\dagger]}{\delta \Phi^\dagger({\bf x})}\, ,
\ee
which is just the condition that $\Psi[A_i,\Phi,\Phi^\dagger]$ is gauge invariant under (small) spatial gauge transformations. Since the form of a wavefunctional is typically non-local in space, the dependence on $\Phi$ and $\Phi^\dagger$ will typically be ensured  gauge invariant by Wilson lines that lie in the constant time surface.
Even in vacuum, the naive wavefunctional for a charged scalar $\sim e^{- \int \d^3 {\bf x} \Phi^\dagger({\bf x}) \sqrt{-\nabla^2} \Phi({\bf x})}$ is (at leading order) replaced by the following gauge invariant form
\be
(\det[-{\bf D}[A]^2])^{1/4}e^{- \int \d^3 {\bf x} \Phi^\dagger({\bf x}) \sqrt{-{\bf D}[A]^2} \Phi({\bf x})} \, , 
\ee
where 
\be
{\bf D}[A]^2 = D_i[A]D^i[A] = (\partial_i - i q A_i)^2 \, ,
\ee
is the covariant version of the Laplacian under spatial gauge transformations.
This has the form 
\be
 \sqrt{-{\bf D}[A]^2} \delta^3({\bf x}-{\bf y}) = e^{i q \int_C A_i(z) \d z^i} S({\bf x},{\bf y};F) \, ,
\ee
with $S({\bf x},{\bf y};F)$ spatially gauge invariant, i.e.~a (generically non-local) functional of $F_{ij}$. Here, $C$ is the straight Wilson line that extends from ${\bf y}$ to ${\bf x}$. Putting this together, a BRST invariant physical state that (approximately) corresponds to the vacuum in the presence of a charge complex scalar is
\ba
&& \psi_{\rm phys} [A_{\mu},c,\bar c, \Phi,\Phi^\dagger]= \\
&&(\det[-\nabla^2] )^{1/4} \sqrt{\det[S[F]]} e^{-\frac{1}{2} \int \d^3 {\bf x} A_{\mu} \sqrt{-\nabla^2} A^{\mu}+\int \d^3 {\bf x} \bar c \sqrt{-\nabla^2} c }e^{- \int \d^3 {\bf x} \int \d^3 {\bf y}e^{i q \int_C A_i(z) \d z^i} \Phi^\dagger({\bf x}) S({\bf x},{\bf y};F) \Phi({\bf y})} \, . \nn
\ea
This is only an approximation since the theory is interacting and so it is not the true ground state of the interacting Hamiltonian, but it illustrates the Wilson lines that will emerge in any description of the matter within the wavefunctional.
In the decoupling limit $q \rightarrow 0$, the Wilson line disappears and this state reduces to the true free theory vacuum
\be
\lim_{q \rightarrow 0}\psi_{\rm phys} [A_{\mu},c,\bar c, \Phi,\Phi^\dagger]=\sqrt{\det[-\nabla^2]} \, e^{-\frac{1}{2} \int \d^3 {\bf x} A_{\mu} \sqrt{-\nabla^2} A^{\mu}+\int \d^3 {\bf x} \bar c \sqrt{-\nabla^2} c }e^{- \int \d^3 {\bf x}  \Phi^\dagger({\bf x})\sqrt{-\nabla^2} \Phi({\bf x})} \, .
\ee
To describe states of finite charge $N_q q$,  we need the wavefunctional to scale as $N_q$ powers of $\Phi$ or more generally include powers such as $\Phi^{N_q+r} (\Phi^{\dagger})^r$. However, $\Phi$ itself is not BRST invariant, and so the naive charged states are not physical. The solution is to dress the field $\Phi$ appropriately. One way to do that is to extend a Wilson line within the fixed time surface to spatial infinity, but a more physical way is to include the Coulomb or relativistic dressing associated with inserting a charge $q$ at ${\bf x}$.
Defining the gauge invariant dressed field
\be
\tilde \Phi({\bf x}) = \Phi({\bf x}) e^{i q \int \d^3 y \, \tfrac{1}{4 \pi |\bf x-\bf y|} (\partial_i A_i({\bf y}))} \, ,
\ee
then assuming gauge transformations and, by extension, $c({\bf x})$ vanish at spacelike infinity, this is BRST invariant. Thus, an example of a physical BRST invariant state with charge $N_q q$, is 
\ba
&& \psi_{\rm phys}^{N_q} [A_{\mu},c,\bar c, \Phi,\Phi^\dagger]= \nn \\ \nn
&& {\cal N} \int \d^{3N_q} \mathbf{x} \, H({{\bf x}_1 , \dots ,{\bf x}_{N_q}}) \( \prod_{r=1}^{N_q} \tilde \Phi({\bf x}_r) \)  e^{-\frac{1}{2} \int \d^3 {\bf x} A_{\mu} \sqrt{-\nabla^2} A^{\mu}+\int \d^3 {\bf x} \bar c \sqrt{-\nabla^2} c - \int \d^3 {\bf x}  \Phi^\dagger({\bf x})\sqrt{-D[A]^2} \Phi({\bf x})} =  \\ 
&& {\cal N} \int \d^{3N_q} \mathbf{x} \,  H({{\bf x}_1 , \dots ,{\bf x}_{N_q}}) \( \prod_{r=1}^{N_q}  \Phi({\bf x}_r) \)  e^{i  \sum_{r=1}^{N_q}q \int \d^3 y \frac{\partial_i A_i({\bf y})}{4 \pi |{\bf x}_r-{\bf y}|} } \nn \\
&& \quad \quad \times e^{-\frac{1}{2} \int \d^3 {\bf x} A_{\mu} \sqrt{-\nabla^2} A^{\mu}+\int \d^3 {\bf x} \bar c \sqrt{-\nabla^2} c - \int \d^3 {\bf x}  \Phi^\dagger({\bf x})\sqrt{-D[A]^2} \Phi({\bf x})} \, . 
\ea
with $H({{\bf x}_1 , \dots ,{\bf x}_{N_q}}) $ arbitrary functions of all $N_q$ spatial positions.
This type of construction does not generalise to the non-Abelian case in general due to confinement, there are no physical states of non-zero colour charge. Formally it is possible to define BRST invariant states by inserting Wilson lines which extend to infinity, but in a confining gauge theory these states have infinite energy and so are not included as part of the physical Hilbert space.
However, whenever the gauge symmetry is spontaneously broken, we can easily construct gauge invariant states by utilising the field in unitary gauge, even for a non-Abelian theory. The mass associated with symmetry breaking cuts off the IR growth of the potential energy, rendering the states finite in energy and hence physical.

\subsection{Inner Product of Physical States (Abelian Case)}

In the Abelian case we do not need to worry about Gribov ambiguities
that can potentially spoil the  non-perturbative BRST symmetry and definition of physical states. It is straightforward to construct the inner product of two physical states and show that the norm of any physical state is positive. To be concrete, let us continue with the example of Maxwell coupled to charged bosonic matter. Using the form \eqref{wavansatz} the norm of a physical state is
\ba
\langle \psi_{\rm phys} | \psi_{\rm phys} \rangle &=& \int \funcd [A_4,A_i,c, \bar c, \Phi , \Phi^{\dagger}]  \, |\psi_{\rm phys} [A_{\mu},c,\bar c, \Phi,\Phi^\dagger] |^2 \Big|_{A_0 \rightarrow i A_4} \\
&=& \int \funcd [A_i,c, \bar c, \Phi , \Phi^{\dagger}] \int \funcd A_4  \Big|e^{-\frac{1}{2}\int \d^3 {\bf x}  A_{\mu} \sqrt{-\nabla^2} A^{\mu}+\int \d^3 {\bf x} \bar c \sqrt{-\nabla^2} c }\Big|^2 \Big|_{A_0 \rightarrow i A_4} |\Psi [A_{i},\Phi,\Phi^\dagger] |^2 \nn \, , 
\ea
with the modulus taken assuming $A_0$ real. Integrating out $A_4$ first we have
\be
\langle \psi_{\rm phys} | \psi_{\rm phys} \rangle = 
\int \funcd [A_i,c, \bar c, \Phi , \Phi^{\dagger}]  \frac{1}{({\det[-\nabla^2]})^{1/4}} e^{-\int \d^3 {\bf x} A_{i} \sqrt{-\nabla^2} A^{i}+2 \int \d^3 {\bf x} \bar c \sqrt{-\nabla^2} c }|\Psi [A_{i},\Phi,\Phi^\dagger] |^2 \,, 
\ee
and further integrating out the ghosts we obtain
\be
\langle \psi_{\rm phys} | \psi_{\rm phys} \rangle = 
\int \funcd [A_i, \Phi , \Phi^{\dagger}]  ({\det[-\nabla^2]})^{1/4} e^{- \int \d^3 {\bf x} A_{i} \sqrt{-\nabla^2} A^{i}}|\Psi [A_{i},\Phi,\Phi^\dagger] |^2 \ge 0 \, , 
\ee
which is manifestly positive definite\footnote{To be precise, this assumes that regularisation and renormalisation do not spoil these properties, nevertheless it is clear that in a cutoff scheme the expression is positive definite.}. This is consistent with the usual operator arguments of \cite{Kugo:1979gm}. 
To clarify the form of this answer, it is helpful to reintroduce the NL field $B$ to rewrite this as
\ba
&& \langle \psi_{\rm phys} | \psi_{\rm phys} \rangle \\ 
&& \quad =
\int \funcd[A_i,c,\bar c,\Phi,\Phi^{\dagger},B]\,
e^{
-\frac{1}{2} \int \d^3 {\bf x} F_{ij}\,\frac{1}{\sqrt{-\nabla^2}}\,F^{ij}
+ 2\,\int \d^3 {\bf x} \bar c\,\sqrt{-\nabla^2}\,c
+\int \d^3 {\bf x}\;  B \frac{1}{\sqrt{-\nabla^2}} ( B + 2\partial_i A_i ) }
\,\bigl|\Psi[A_i,\Phi,\Phi^\dagger]\bigr|^2 \nn
\\
&& \quad =
\int \funcd[A_i,c,\bar c,\Phi,\Phi^{\dagger},B]\,
e^{
-\frac{1}{2}\int \d^3 {\bf x}\;   F_{ij}\,\frac{1}{\sqrt{-\nabla^2}}\,F^{ij}
+ \int \d^3 {\bf x} \; \hat s \big(
\,\bar c\,\frac{1}{\sqrt{-\nabla^2}}\,B
+ 2\,\bar c\,\frac{1}{\sqrt{-\nabla^2}}\,(\partial_i A_i)
\big)
}
\,\bigl|\Psi[A_i,\Phi,\Phi^\dagger]\bigr|^2 . \nn
\ea
where the term in the exponent $\hat s (\cdots)$ is BRST-exact. This acts as a gauge-fixing fermion for the fixed time functional integral. In other words, this is nothing other than a BRST/Faddeev-Popov gauge fixed version of the 3 dimensional functional integral
using the 3 dimensional version of the BRST symmetry
\be
\hat s A_i =\partial_i c \, , \quad \hat s \bar c = B \, , \quad \hat s c =0 \, , \quad \hat s B =0 \, .
\ee
Thus in gauge invariant language we are evaluating
\be
\langle \psi_{\rm phys} | \psi_{\rm phys} \rangle =
\int \frac{\funcd [A_i, \Phi , \Phi^{\dagger}] }{{\cal{V}}ol({\cal G})} e^{-\frac{1}{2}\int \d^3 {\bf x} \;  F_{ij} \frac{1}{\sqrt{-\nabla^2}} F^{ij}}
 |\Psi [A_{i},\Phi,\Phi^\dagger] |^2 \, ,
\ee
with ${\cal{V}}ol({\cal G})$ the volume of the 3 dimensional gauge orbit.
We can interpret this as the statement that the true gauge invariant wavefunctional is 
\be
e^{-\frac{1}{4} \int \d^3 {\bf x} \;  F_{ij} \frac{1}{\sqrt{-\nabla^2}} F^{ij}}\Psi [A_{i},\Phi,\Phi^\dagger] \, ,
\ee
and the correct inner product is the naive Born rule integrating over the gauge fields and charged states on a 3 dimensional slice with the volume of the gauge orbit factored out. This perspective is helpful since it is easily generalised to the non-Abelian case. In addition, it allows us more easily to compare the wavefunction in different gauges, including non-covariant gauges.

More generally for two distinct physical states, their inner product is 
\ba
&& \langle \psi_1| \psi_2 \rangle \\
&& \quad = \int \funcd[A_i,c,\bar c,\Phi,\Phi^{\dagger},B]\,
e^{
-\frac{1}{2} \int \d^3 {\bf x} F_{ij}\,\frac{1}{\sqrt{-\nabla^2}}\,F^{ij}
+ \int \d^3 {\bf x} \;  \hat s \big(
\,\bar c\,\frac{1}{\sqrt{-\nabla^2}}\, ( B + 2 \partial_i A_i )
\big)} \Psi_1 [A_{i},\Phi,\Phi^\dagger] ^* \Psi_2 [A_{i},\Phi,\Phi^\dagger] \nn  \\
&&  \quad = \int \frac{\funcd [A_i, \Phi , \Phi^{\dagger}] }{{\cal{V}}ol({\cal G})}  e^{-\frac{1}{2} \int \d^3 {\bf x} F_{ij} \frac{1}{\sqrt{-\nabla^2}} F^{ij}}
 \Psi_1 [A_{i},\Phi,\Phi^\dagger] ^* \Psi_2 [A_{i},\Phi,\Phi^\dagger]  \, . \nn
\ea

\subsection{Nakanishi-Lautrup Representation}

Within the operator formalism, it is natural to eliminate the Nakanishi-Lautrup field $B$, since it plays the role of an auxiliary field. However, doing so has the consequence that the BRST algebra closes only on-shell; in particular, one finds already in Maxwell theory $\hat s^2 \bar c = -\Box c$. This complication is highly problematic in the non-Abelian case, where one cannot guarantee that loop corrections preserve BRST invariance off-shell. In anticipation of the path-integral formulation, and to establish BRST identities that hold off-shell, it is therefore advantageous to restore the field $B$ to the description.

There is, however, a second reason to reintroduce $B$, The BRST symmetry transformations of the wavefunctionals and the indefinite Hilbert space inner product are much more transparent in the representation in which $A_0$ is replaced by $B$. The reason for this is that the transformation $\hat s A_0 = \partial_t c$ is not a simple field redefinition of the \Sch variables  at a fixed time (generating conjugate momenta), whereas the transformation $\hat s B=0$ is.

To make this clear, let us return to the original off-shell form of the BRST action for Maxwell
\be
S = \int_{t_{\rm i}}^{t_{\rm f}} \d^4 x \left[\frac{1}{2} F_{0i}^2-\frac{1}{4} F_{ij}^2+ B (-\partial_t A_{0}+\partial_i A_i) + \frac{1}{2} B^2 + \partial_{\mu}\bar c \partial^{\mu} c \right] \, .
\ee
Written in this way it is clear that $-B$ is the momentum conjugate to $A_0$. To decouple the degrees of freedom it is better to define $B=\tilde B-(\partial_i A_i)$, 
\be
S = \int_{t_{\rm i}}^{t_{\rm f}} \d^4 x \left[\frac{1}{2} F_{0i}^2-\frac{1}{4} F_{ij}^2+(\partial_i A_i) \dot A_0-\frac{1}{2} (\partial_i A^i)^2+\tilde B (-\partial_t A_{0}) + \frac{1}{2} \tilde B^2 + \partial_{\mu}\bar c \partial^{\mu} c \right] \, .
\ee
To fully decouple the modes we define
\be
\tilde S = S -\int_{t_{\rm i}}^{t_{\rm f}} \d^4 x \,  \partial_t (\partial_i A_i  A_0) \, ,
\ee
so that on freely integrating spatial derivatives by parts 
\be
\tilde S = \int_{t_{\rm i}}^{t_{\rm f}} \d^4 x \left[\frac{1}{2} (\partial_t A_i)^2+\frac{1}{2} A_i \nabla^2 A_i-\frac{1}{2} A_0 \nabla^2 A_0+ \tilde B (-\partial_t A_{0}) + \frac{1}{2} \tilde B^2 + \partial_{\mu}\bar c \partial^{\mu} c \right] \, .
\ee
In this form $-\tilde B$ is the conjugate momentum to $A_0$. We now denote 
\be \label{boundary}
\tilde S=S_{\mathrm{NL} }-\int_{t_{\rm i}}^{t_{\rm f}} \d^4 x \,  \partial_t (\tilde B A_0) \, , \quad  S_{\mathrm{NL} }=S+\int_{t_{\rm i}}^{t_{\rm f}} \d^4 x \,  \partial_t (B A_0)\, , 
\ee
for which 
\be \label{NLaction}
S_{\mathrm{NL} } = \int_{t_{\rm i}}^{t_{\rm f}} \d^4 x \left[\frac{1}{2} (\partial_t A_i)^2+\frac{1}{2} A_i \nabla^2 A_i-\frac{1}{2} A_0 \nabla^2 A_0+ \partial_t \tilde B A_{0} + \frac{1}{2} \tilde B^2 + \partial_{\mu}\bar c \partial^{\mu} c \right] \, ,
\ee
or writing in terms of $B$
\be \label{NLaction2}
S_{\mathrm{NL} } = \int_{t_{\rm i}}^{t_{\rm f}} \d^4 x \left[-\frac{1}{4} F_{\mu\nu}F^{\mu\nu}- \partial_{\mu} B A^{\mu} + \frac{1}{2} B^2 + \partial_{\mu}\bar c \partial^{\mu} c \right] \, .
\ee
This trades $A_0$ and $\tilde B$ as conjugate variables. We can further integrate out $A_0$ in \eqref{NLaction} 
\be
S_{\mathrm{NL} } = \int_{t_{\rm i}}^{t_{\rm f}} \d^4 x \left[\frac{1}{2} (\partial_t A_i)^2+\frac{1}{2} A_i \nabla^2 A_i+ \frac{1}{2}\partial_t \tilde B \nabla^{-2} \partial_t \tilde B + \frac{1}{2} \tilde B^2 + \partial_{\mu}\bar c \partial^{\mu} c \right] \, ,
\ee
so that the propagating degrees of freedom are then $A_i,B,c,\bar c$, each having second order equations of motion. Thus, it is natural to define the Nakanishi-Lautrup representation states via the functional Fourier transform (which implements the boundary term in \eqref{boundary})
\ba
\langle A_i, B, c, \bar c| \psi \rangle &=&  \int \funcd A_0 \, e^{-i \int \d^3 {\bf x} A_0 \tilde B} \langle A_{\mu}, c, \bar c| \psi \rangle \, , \\
&=&  \int \funcd A_0 \, e^{-i \int \d^3 {\bf x} A_0 (B+\partial_i A_i)} \langle A_{\mu}, c, \bar c| \psi \rangle \,, 
\ea
or more precisely defining $B=-iB_4$
\be
\langle A_i, B_4, c, \bar c| \psi \rangle =  \int \funcd A_4 \, e^{-i \int \d^3 {\bf x} A_4 (B_4+i \partial_i A_i)} \langle A_i,A_4, c, \bar c| \psi \rangle \, .
\ee
This is the conversion from the natural representation defined by $\tilde S$ (where all modes of $A_{\mu}$ are decoupled) and $S_{\mathrm{NL} }$ where BRST invariance is manifest.
The resolution of unity for these states is\footnote{Note that the functional integral is taken over $B_4$, not $B_0$.}
\be
\int \funcd[A_i , B_4 ,c ,\bar c]  \, | A_i, -B_4, c, \bar c \rangle \langle A_i, B_4, c, \bar c|  = \int \funcd[A_i , B_4 ,c ,\bar c]  \, | A_i,B,  c, \bar c \rangle \langle A_i,B,  c, \bar c| \Big|_{B \rightarrow -i B_4} = \hat 1\, .
\ee
To check, we note that
\ba
&& \int \funcd[A_i , B_4 ,c ,\bar c] \langle \psi_1| A_i, -B_4, c, \bar c \rangle    \langle A_i, B_4, c, \bar c| \psi_2 \rangle \\
&=&\int \funcd B_4 \int \funcd A_4' \, e^{i \int \d^3 {\bf x} A'_4 (-B_4-i \partial_i A_i)} \int \funcd A_4 \, e^{-i \int \d^3 {\bf x} A_4 (B_4+i \partial_i A_i)} \int \funcd A_i \int \funcd c \int \funcd \bar c \langle \psi_1| A_i, A_4', c, \bar c \rangle   \langle A_i, A_4, c, \bar c| \psi_2 \rangle \nn \\
&=& \int \funcd A_4' \int \funcd A_4\, \delta(A_4+A_4') e^{i \int \d^3 {\bf x} (A'_4+A_4) (-i \partial_i A_i)}  \int \funcd A_i \int \funcd c \int \funcd \bar c \langle \psi_1| A_i, A_4', c, \bar c \rangle   \langle A_i, A_4, c, \bar c| \psi_2 \rangle  \\
&=&  \int \funcd[A_4  , A_i ,c ,\bar c] \langle \psi_1| A_i, -A_4, c, \bar c \rangle   \langle A_i, A_4, c, \bar c| \psi_2 \rangle =\langle \psi_1|\psi_2 \rangle \,.
\ea
Similarly, the trace is 
\ba
\Tr[\hat O]&=& \int \funcd[A_i , B_4 ,c ,\bar c]  \, \langle A_i, -B_4, -c, -\bar c|  \hat O | A_i, B_4, c, \bar c \rangle \\
&=& \int \funcd[A_i , B_4 ,c ,\bar c]  \, \langle A_i, B, -c, -\bar c|  \hat O | A_i, B, c, \bar c \rangle\Big|_{B \rightarrow -i B_4} \, .
\ea
The main benefit of the Nakanishi-Lautrup representation defined by $S_{\rm NL}$ is that the BRST operator acting on wavefunctionals is now simply 
\ba
&& i \langle A_i, B, c, \bar c|\hat Q | \psi_{\rm phys} \rangle =\\
&&\int \d^3 {\bf x} \left[ \hat s A_i({\bf x}) \frac{\delta }{\delta A_i({\bf x})}+ \hat s B({\bf x}) \frac{\delta }{\delta B({\bf x})}+\hat s c({\bf x}) \frac{\delta }{\delta c({\bf x})}+\hat s \bar c ({\bf x}) \frac{\delta }{\delta \bar c({\bf x})}\right] \psi_{\rm phys}[A_i, B,c,\bar c]=0 \, , \nn
\ea
with $\hat s A_i = \partial_i c$, $\hat s c=0$, $\hat s \bar c=B$, $\hat s B=0$, which is the statement 
\be
\hat s \psi_{\rm phys}[A_i, B,c,\bar c]=0 \, ,
\ee
i.e.~the wavefunctionals are BRST invariant under the BRST transformation which can now be realised on a fixed time surface.
For example, for the vacuum wavefunctional constructed earlier, up to a normalisation we have
\be
\psi_0 [A_{\mu},c,\bar c] = (\det[-\nabla^2])^{1/4} e^{-\frac{1}{2} \int \d^3 {\bf x} \; A_{\mu} \sqrt{-\nabla^2} A^{\mu}+\int \d^3 {\bf x}\; \bar c \sqrt{-\nabla^2} c } \, ,
\ee
then on performing the functional Fourier transform we have
\be
\psi_0 [A_{i},B,c,\bar c] = e^{\frac{1}{2}\int \d^3 {\bf x}\;  (B+\partial_i A_i)\frac{1}{\sqrt{-\nabla^2}}(B+\partial_i A_i) }e^{-\frac{1}{2}\int \d^3 {\bf x}\;  A_{i} \sqrt{-\nabla^2} A^{i}+\int \d^3 {\bf x}\; \bar c \sqrt{-\nabla^2} c } \, ,
\ee
which can be rearranged as
\be
\psi_0 [A_{i},B,c,\bar c] =  e^{\frac{1}{2}\int \d^3 {\bf x}\; B\frac{1}{\sqrt{-\nabla^2}}B+\int \d^3 {\bf x}\; B\frac{1}{\sqrt{-\nabla^2}} \partial_i A_i+\int \d^3 {\bf x}\; \bar c \sqrt{-\nabla^2} c  }e^{-\frac{1}{4} \int \d^3 {\bf x}\; F_{ij} \frac{1}{\sqrt{-\nabla^2}}  F^{ij} } \, .
\ee
The exponent naturally splits into an BRST exact term and a gauge invariant term
\be
\psi_0 [A_{i},B,c,\bar c] = (\det[-\nabla^2])^{1/4} e^{\hat s G} e^{-\frac{1}{4} \int \d^3 {\bf x} \; F_{ij} \frac{1}{\sqrt{-\nabla^2}}  F^{ij} } \, , 
\ee
with 
\be
G =\int \d^3 {\bf x} \left[ \frac{1}{2} \bar c \frac{1}{\sqrt{-\nabla^2}} B+ \bar c \frac{1}{\sqrt{-\nabla^2}} \partial_i A_i \right] \, .
\ee
being a gauge-fixing fermion for the state.
From this it is clear that $\hat s \psi_0 [A_{i},B,c,\bar c]=0$.
Similarly a generic physical state at the initial time can be described by 
\be
\psi_{\rm phys} [A_{i},B,c,\bar c] = e^{\hat s G} \Psi[A_i] \, ,
\ee
with $\Psi[A_i]$ gauge invariant under spatial gauge transformations.
It is clear from the usual arguments that the precise form of $ G$ does not matter since the shift $G \rightarrow  G+ \Delta G$ will generate
\begin{eqnarray}
\psi_{\rm phys}[A_i,B,c,\bar c] 
&\rightarrow& 
\psi_{\rm phys}[A_i,B,c,\bar c]
+ (\hat s \Delta G)\,
\psi_{\rm phys}[A_i,B,c,\bar c] \, , \\
&=& 
\psi_{\rm phys}[A_i,B,c,\bar c]
+ \hat s \bigl(
\Delta G \,
\psi_{\rm phys}[A_i,B,c,\bar c]
\bigr) \, .
\end{eqnarray}
which differing by an exact form necessarily corresponds to the same physical state. However, we cannot set $G$ to zero, as it is needed to ensure that the inner product is finite, and it must pick out a unique representative of the gauge orbit (modulo Gribov ambiguities) 
The measure in the inner product is clearly invariant 
\be
\hat s \int \funcd A_i \int \funcd B \int \funcd c \int \funcd \bar c =0 \, , 
\ee
and so the inner product between two BRST invariant states is automatically invariant.

Finally, we note that this representation is also distinguished in that if we construct the phase-space action between two finite times
\be
S_{\mathrm{NL} } = \int_{t_{\rm i}}^{t_{\rm f}} \d^4 x \Big[ \Pi_i \partial_t A_i+\Pi_B \partial_t B + \Pi_c \partial_t c + \Pi_{\bar c} \partial_t \bar c - {\cal H}\[A_i,\Pi_i, B,\Pi_B,c,\Pi_c,\bar c, \Pi_{\bar c}\] \Big] \,, 
\ee
then $\hat s S_{\mathrm{NL} } =0$ without needing to impose conditions on the fields at the initial and final times, i.e.~there are no boundary contributions on the initial and final time surface. 
To demonstrate this explicitly, we note from \eqref{NLaction} that the conjugate momenta (which differ from those implied by $\tilde S$) are now
\be
\Pi_i = \partial_t A_i -\partial_i A_0  \, ,\quad \Pi_B=A_0 \, , \quad \Pi_c = -\partial_t \bar c \,, \quad \Pi_{\bar c} = \partial_t c \,.
\ee
The full set of BRST transformations on the phase space variables can now be taken to be 
\begin{equation}
  \begin{split}
    & \hat s A_i = \partial_i c \ , \\
    & \hat s \Pi_i = 0 \ ,
  \end{split}
\qquad
  \begin{split}
    & \hat s c =0 \ , \\
    & \hat s \Pi_c  =\partial_i \Pi_i \ , 
  \end{split}
  \qquad
  \begin{split}
    &\hat s \bar c = B \ ,  \\
    & \hat s \Pi_{\bar c}=0 \ ,
  \end{split}
  \qquad
  \begin{split}
    & \hat s B =0 \ ,  \\
    & \hat s \Pi_B = \Pi_{\bar c}\ ,
  \end{split}
\end{equation}
which are nilpotent off-shell $\hat s^2=0$ and are generated by the BRST charge
\be
\hat Q = \int \d^3 {\bf x} \[ \Pi_i \partial_i c - \Pi_{\bar c } B \] \, .
\ee
The Hamiltonian is
\be
H\[A_i,\Pi_i, B,\Pi_B,c,\Pi_c,\bar c, \Pi_{\bar c}\] = \int \d^3 {\bf x} \left[ \frac{1}{2}\Pi_i^2+\frac{1}{4} F_{ij}^2+ \hat s \( -(\partial_i A^i)\ \bar c - \frac{1}{2}B \bar c-\Pi_B \Pi_c ) \) \right] \, ,
\ee
which is easily seen to be BRST invariant
\be
\hat s H\[A_i,\Pi_i, B,\Pi_B,c,\Pi_c,\bar c, \Pi_{\bar c}\]=0 \, .
\ee
Crucially, however,
\ba
 \hat s \int \d^3 {\bf x}  \[ \Pi_i \partial_t A_i+\Pi_B \partial_t B + \Pi_c \partial_t c + \Pi_{\bar c} \partial_t \bar c \] 
&=& \int \d^3 {\bf x} \[ \Pi_i \partial_t \partial_i c+\Pi_{\bar c} \partial_t B+(\partial_i \Pi_i) \partial_t c - \Pi_{\bar c} \partial_t B \] \notag
\\
&=& \int \d^3 {\bf x} \[ \partial_i \( \Pi_i \partial_t c \) \]=0\, ,
\ea
which vanishes without needing to integrate by parts in time. Hence $\hat s S_{\mathrm{NL} }=0$ without imposing conditions at the initial and final times. This makes the BRST invariance of the Schwinger-Keldysh path integral formulated with $S_{\mathrm{NL} }$ transparent.

\subsection{BRST Invariant Yang-Mills Wavefunctionals}

Let us now return to the more general case of a non-Abelian theory\footnote{For earlier work see \cite{Lee:1991ukr,Igarashi:1991ds,Nirov:1994mj}.}. Based on the previous discussion the BRST properties of the wavefunctionals are most easily understood in the Nakanishi-Lautrup representation. Thus we begin with 
\be
\label{SNLnonAbel}
S_{\mathrm{NL} } = \int_{t_{\rm i}}^{t_{\rm f}} \d^4 x \left[-\frac{1}{4} F^{a}_{\mu\nu}{}^2 -\partial_{\mu} B^a A_a^{\mu} + \frac{1}{2} B_a^2 + \partial_{\mu}\bar c^a D^{\mu} c^a \right] \, ,
\ee
which in covariant form is BRST invariant without needing to integrate by parts. 
We can rewrite this in phase space
\begin{align}
S_{\mathrm{NL} }
=
\int^{t_{\rm f}}_{t_{\rm i}}  \d^4x \,
\left[
\Pi_i^a \dot{A}_i^a
+\Pi_B^a \dot{B}^a
+\Pi_c^a \dot{c}^a
+\Pi_{\bar c}^a \dot{\bar c}^{\,a}
-H
\right] \, ,
\end{align}
The canonical momenta are 
\be
\label{conjugatemomenta}
\Pi_i^a = F_{0i}^a  \, ,\quad \Pi_B^a=A_0^a \, , \quad \Pi_c^a = -\partial_t \bar c^a \,, \quad \Pi_{\bar c}^a = D_t[A] c^a \,.
\ee
The full set of BRST transformations on the phase space variables can now be taken to be
\begin{equation}
  \begin{split}
    &\hat s A_i^a  = D_i[A] c^a \ , \\
    & \hat s \Pi_i^a = -g f^{abc} c^b \Pi^c_i \ ,
  \end{split}
\qquad
  \begin{split}
    & \hat s c^a =-\frac{g}{2} f^{abc} c^b c^c\  \ , \\
    & \hat s \Pi_c^a =D_i[A] \Pi_i^a- g f^{abc} \Pi_c^b  c^c \ ,
  \end{split}
\qquad
  \begin{split}
    & \hat s \bar c^a =B^a\ , \\
    &\hat s \Pi_{\bar c}^a=0 \ ,
  \end{split}
\qquad
  \begin{split}
    &\hat s B^a=0\ , \\
    &\hat s \Pi_B^a=\Pi^a_{\bar c} \ ,
  \end{split}
\end{equation}
which are nilpotent off-shell $\hat s^2=0$ and are generated by the BRST charge 
\be
\hat Q = \int \d^3 {\bf x} \[ \Pi_i^a D[A]_i c^a - \Pi_{\bar c }^a B^a +\frac{g}{2}\Pi_c^a f^{abc} c^b c^c\] \, .
\ee
The Hamiltonian is 
\begin{align}
H
&= \int \d^3 \mathbf{x} \, \bigg[
\frac{1}{2}\,(\Pi_i^a)^2
+\frac{1}{4}\,(F_{ij}^a)^2
+\Pi_i^a D_i[A]\Pi_B^a
+A_i^a\,\partial_i B^a \nonumber \\
& -\frac{1}{2}\,(B^a)^2 +\Pi_c^a \Pi_{\bar c}^a
-g\,f^{abc}\,\Pi_c^a \Pi_B^b c^c
-(\partial_i \bar c^{\,a})\,D_i c^a \bigg] \,.
\end{align}
which can be written in a more transparent form as
\be \label{YMHamiltonian}
H\[A_i,\Pi_i, B,\Pi_B,c,\Pi_c,\bar c, \Pi_{\bar c}\] = \int \d^3 {\bf x} \left[ \frac{1}{2}{\Pi^a_i}^2+\frac{1}{4} {F^a_{ij}}^2+ \hat s \( -\partial_i A_i^a\ \bar c^a -  \frac{1}{2}B^a \bar c^a-\Pi_B^a \Pi_c^a  \) \right] \, ,
\ee
which being a sum of a gauge invariant term and a BRST exact term is manifestly BRST invariant.
Since in this representation the BRST charge is linear in momenta, the constraint on physical wavefunctions is a linear differential equation
\ba
&& i \langle A_i, B, c, \bar c|\hat Q | \psi_{\rm phys} \rangle = \\
&& \int \d^3 {\bf x} \left[ \hat s A_i^a({\bf x}) \frac{\delta }{\delta A^a_i({\bf x})}+ \hat s B^a({\bf x}) \frac{\delta }{\delta B^a({\bf x})}+\hat s c^a({\bf x}) \frac{\delta }{\delta c^a({\bf x})}+\hat s \bar c^a ({\bf x}) \frac{\delta }{\delta \bar c^a({\bf x})}\right] \psi_{\rm phys}[A_i^a, B^a,c^a,\bar c^a]=0 \, , \nn
\ea
which is simply the statement that
\be
\hat s \,  \psi_{\rm phys}[A_i^a, B^a,c^a,\bar c^a]=0 \, .
\ee
It is useful to have a general form for the class of physical wavefunctionals, but this is inevitably ambiguous due to the ability to shift by an exact form. It is straightforward to provide an ansatz which matches every physical state in the perturbative Fock space. At fixed time, a sufficiently general form of a physical state is
\be \label{NAphys}
\psi_{\rm phys} [A_i^a, B^a,c^a,\bar c^a] = e^{\hat s G[A_i^a, B^a,c^a,\bar c^a]} \Psi[A_i^a] \, , 
\ee
where $G[A_i^a, B^a,c^a,\bar c^a]$ is an arbitrary gauge fixing fermion, and $\Psi[A_i^a]$ is that part of the wavefunctional satisfying the Gauss-law constraint
\be \label{Gausslaw}
D[A]_i \( \frac{\delta \Psi }{\delta A_i^a({\bf x})}\) =0 \, .
\ee 
Since
\be
\hat s \( e^{\hat s G[A_i^a, B^a,c^a,\bar c^a]} \) =0 \, , \quad \text{and } \quad \hat s \Psi[A_i^a] =0 \, ,
\ee
this is guaranteed to be closed, and different choices of $ G$ correspond to shifting by an exact form. As we have seen, all Fock states constructed from the free Hamiltonian which are constructed from the vacuum wavefunctional by exciting only transverse polarisation modes take this form. However, we can argue that non-perturbatively these states reproduce all of the allowed states according to the Dirac procedure in temporal gauge, which is a common gauge choice in variational and wavefunctional analyses of QCD.

The condition \eqref{Gausslaw} guarantees that $\Psi[A_i^a]$ is gauge invariant under `small' gauge transformations. It may however, transform via a phase under large gauge transformations with non-trivial winding without contradicting the Gauss-law constraint \eqref{Gausslaw}. Alternatively we may from the outset add to the Yang-Mills action a $\theta$ term which, being a total derivative, effectively amounts to redefining the wavefunctionals via a Chern-Simons term 
\be
\Psi[A_i^a] \rightarrow e^{i \theta W_{\rm CS}[A_i^a]}\Psi[A_i^a] \, .
\ee
The Chern-Simons term accounts for phase transformation under large gauge transformations so that the redefined wavefunctional is gauge invariant. We will assume this has been done so that $\Psi[A_i]$ is gauge invariant under both small and large gauge transformations. In other words we assume that a $\theta$ term has been included in the action.

The precise form of \eqref{NAphys} that matches the free theory calculation is
\be \label{GNA}
G = \int \d^3 {\bf x}  \left[ \frac{1}{2} \bar c^a \frac{1}{\sqrt{-\nabla^2}} B^a+ \bar c^a \frac{1}{\sqrt{-\nabla^2}} \partial_i A^a_i \right] \, ,
\ee
however, we are free to choose an alternative. More importantly, since in a non-Abelian theory, ghosts have dynamics, the gauge fixing fermion $G$ in the wavefunction \eqref{NAphys} will necessarily evolve dynamically and so \eqref{GNA} is too restrictive a choice except at the initial time. However, its precise form is irrelevant, provided that it can be viewed as selecting a unique (at least perturbatively) representative from each gauge orbit.

A suitable ansatz for the gauge invariant part of the wavefunction that captures the perturbative vacuum is
\be
\Psi[A^a] = {\cal N} e^{-\frac{1}{4} \int \d^3 {\bf x}\, F_{ij}^a \tfrac{1}{\sqrt{-D[A]^2}}F^{ij}_a }\, .
\ee
This is covariant and reduces in the weak field limit to the true vacuum, however it will not be a good approximation in the IR where the true vacuum state should be confinement. There is a wealth of literature \cite{Greensite:1979yn,Greensite:2009iv,Greensite:2011pj,Zarembo:1997ms,Leigh:2006vg,Olejnik:2015eaa,Greensite:2013rva,Karabali:2007mr} on improved ansatze that attempt to capture both the confining and perturbative region. For example, a common choice is to include mass scale in the manner of Greensite 
\be
\Psi[A^a] = {\cal N} e^{-\frac{1}{4}\int \d^3 {\bf x}\,  F_{ij}^a \tfrac{1}{\sqrt{-D[A]^2+m^2}}F^{ij}_a }\, ,
\ee
which `dimensionally reduces' to the form of the wavefunction in the IR (rendering it spatially local) which can account for confining properties \cite{Kawamura:1996us}. A suitable gauge invariant ansatz which describes, in the perturbative region, a coherent state of gluons but still exhibits confinement in the IR is
\be
\Psi[A^a] = e^{- \frac{1}{4}\int \d^3 {\bf x}\,  F_{ij}^a \tfrac{1}{\sqrt{-D[A]^2+m^2}}F^{ij}_a } \int \funcd[\beta_{i}^a] \, \tilde \Psi[\beta_{i}^a] e^{i \int \d^3 {\bf x} \; \epsilon^{ijk} \beta_i^a({\bf x}) F_{jk}^a({\bf x})}\, ,
\ee
provided
\be
\tilde \Psi[(e^{i T^c \theta^c }){}^a{}_b\beta_{i}^b]=\tilde \Psi[\beta_{i}^a] \, ,
\ee 
for all $\theta^b(\bf x)$ to ensure the coherent state is colour neutral/gauge-invariant.

\subsection{Inner Product of Physical States: Non-Abelian Case}

In the NL representation, the inner product of two states in the non-Abelian case is a straightforward generalisation of the Abelian.
Focussing on the gauge fields alone we have
\ba
 \langle \psi_1 | \psi_2 \rangle&=&\int \funcd[A_i^a , B_4^a ,c^a ,\bar c^a]  \, \langle \psi_1 | A_i^a, -B_4^a, c^a, \bar c^a \rangle \langle A_i^a, B_4^a, c^a, \bar c^a| \psi_2 \rangle \\
&=& \int \funcd[A_i^a , B_4^a ,c^a ,\bar c^a]  \, \langle \psi_1| A_i^a,B^a,  c^a, \bar c^a \rangle \langle A_i^a,B^a,  c^a, \bar c^a| \psi_2 \rangle\Big|_{B^a \rightarrow -i B_4^a}   \, .
\ea
As discussed, physical BRST invariant wavefunctionals take the form
\ba
&& \psi_1 [A_i^a, B^a,c^a,\bar c^a] = e^{\hat s G_1} \Psi_1[A_i^a] \, , \\
&& \psi_2 [A_i^a, B^a,c^a,\bar c^a] = e^{\hat s G_2} \Psi_2[A_i^a] \, , 
\ea
with $\Psi_1[A_i^a]$ and $\Psi_2[A_i^a]$ gauge invariant, and $G_1$ and $G_2$ are Grassmann odd with ghost number $-1$. In general $G_1 \neq G_2$. In particular, in the non-Abelian case, because the ghosts have interactions, $G$ will dynamically evolve. Due to the hermiticity assignment of $\bar c$, we have $(\hat s G)^{\dagger} = - \hat s G^{\dagger}$. The inner product is then
\ba
 \langle \psi_1 | \psi_2 \rangle&=&\int \funcd[A_i^a] \int \funcd[ B_4^a ,c^a ,\bar c^a]  \, e^{\hat s (G_2-G_1^{\dagger})}\Big|_{B^a \rightarrow -i B_4^a}  \Psi_1[A_i^a]^* \Psi_2[A_i^a]\, .
\ea
We recognise this as the Faddeev-Popov gauge fixed version of a functional integral over 3 dimensional gauge fields with gauge fixing fermion $G_2- G_1^{\dagger}$. Thus, we identify
\be \label{BRSTVol}
\int \funcd[A_i^a]\int \funcd[ B_4^a ,c^a ,\bar c^a]  \, e^{\hat s (G_2-G_1^{\dagger})}\Big|_{B^a \rightarrow -i B_4^a} = \int \frac{\funcd[A_i^a]}{{\cal{V}}ol({\cal G})}\, ,
\ee
regardless of the precise choice of $G_2-G_1^{\dagger}$, provided that it picks out (at least perturbatively) a unique representative of the gauge group. Taking $G_1=G_2$ to be of the form implied by the vacuum wavefunctionals
\be
G =\int \d^3 {\bf x} \left[ \frac{1}{2} \bar c^a \frac{1}{\sqrt{-\nabla^2}} B^a+ \bar c^a \frac{1}{\sqrt{-\nabla^2}} \partial_i A_i^a \right] \, ,
\ee
this imposes the 't Hooft averaged version of the spatial gauge 
\be
\frac{1}{\sqrt{-\nabla^2}} \partial_i A_i^a =0 \, \quad \rightarrow \quad  \partial_i A_i^a =0 \, .
\ee
In summary, this is the reverse of the usual Faddeev-Popov argument now applied to the 3 dimensional functional integral rather than the path integral, and so the inner product may be stated as
\ba \label{NAinner}
 \langle \psi_1 | \psi_2 \rangle&=&\int \frac{\funcd[A_i^a]}{{\cal{V}}ol({\cal G})} \Psi_1[A_i^a]^* \Psi_2[A_i^a]\, .
\ea
This final form is the naive gauge invariant definition of the inner product. This confirms the interpretation of $\Psi_{1,2}[A_i^a]$ as the gauge invariant notion of wavefunctionals. Physical states automatically have positive norms
\ba
 \langle \psi | \psi \rangle&=&\int \frac{\funcd [A_i^a]}{{\cal{V}}ol({\cal G})} |\Psi[A_i^a]|^2 > 0\, .
\ea
In this discussion we have glossed over the issue of the Gribov ambiguity which breaks the precise identification \eqref{BRSTVol}, nevertheless we can clearly regard \eqref{NAinner} as the correct non-perturbative definition of the inner product, assuming the existence of a suitable regularisation scheme to render it well-defined.

\subsubsection{Relation with temporal gauge wavefunctions}

In non-BRST approaches to gauge theories, it is common to simply choose a gauge from the outset. One particularly useful choice is the temporal (Weyl) gauge \cite{Goldstone:1978he,Greensite:2011pj,Jackiw:1995be}
\be
A_0^a(x) =0 \, .
\ee
This gauge is not unique since we can clearly perform a time independent spatial gauge transformation which acts infinitesimally as
\be
\delta A_0^a(x) = 0 \, , \quad \delta A_i^a(x) = D_i \lambda^a({\bf x}) \, .
\ee
The generators of these spatial gauge transformations are precisely the Gauss-law constraints
\be
C^a = D_i[A_i] \Pi_i^a=0 \, , 
\ee
and so a legitimate way to proceed is to follow Dirac and demand that the physical states are annihilated by these constraints
\be
\langle A| C^b |\Psi \rangle = -i D[A]_i \( \frac{\delta \Psi }{\delta A_i^b({\bf x})}\)=0\, .
\ee
This is of course just the statement that the wavefunction is invariant under small gauge transformations. 
Since temporal gauge preserves a residual spatial gauge symmetry then to define the inner product of states we must mod out by the gauge orbit, which is to say we are led to 
\ba 
 \langle \psi_1 | \psi_2 \rangle&=&\int \frac{\funcd[A_i^a]}{{\cal{V}}ol({\cal G})} \Psi_1[A_i^a]^* \Psi_2[A_i^a]\, ,
\ea
without going through the BRST route. Thus we may identify the gauge invariant part of the Lorenz/covariant gauge BRST wavefunctional $\Psi[A_i]$ with the wavefunctional $\Psi[A_i]$ in temporal gauge. This identification holds dynamically, as we see in the next section.

\subsection{Dynamical Evolution of Physical States}

In the BRST formalism the structure of the Hamiltonian is \cite{Henneaux:1985kr}
\be
\hat H = \hat H_0 + \hat s \hat {\cal F} = \hat H_0 + i \{ \hat Q, \hat {\cal F}  \} \, ,
\ee
where $\hat H_0$ is the gauge invariant part of the Hamiltonian and $\hat {\cal F} $ (which is anti-Hermitian in our conventions) is the Hamiltonian gauge fixing fermion (corresponding to the gauge-fixing term $-\partial_{\mu} B^a A_a^{\mu} + \frac{1}{2} B_a^2 + \partial_{\mu}\bar c^a D^{\mu} c^a$ in Eq.~(\ref{SNL_sec2})). We have already seen at a fixed time it is possible to parameterise BRST invariant wavefunctionals in the manner 
\be \label{waveansatz}
\psi(A_i^a ,B^a, c^a,\bar c^a,  \dots ) = e^{\hat s  G(A_i^a ,B^a, c^a,\bar c^a,  \dots )} \Psi(A_i^a \dots ) \, ,
\ee
with $\Psi(A_i^a \dots )$ gauge invariant and with $G(A_i^a ,B^a, c^a,\bar c^a,  \dots )$ the gauge fixing fermion for the wavefunction which is a functional only of the coordinate fields. Note that $ {\cal F}(A_i^a ,B^a, c^a,\bar c^a, \dots) $ and $G(A_i^a ,B^a, c^a,\bar c^a,  \dots )$ are independent.

This precise structure is not preserved by dynamical evolution, but what is preserved is a mild generalisation of the form
\be \label{waveansatz2}
|\psi \rangle = e^{\hat s \hat G} | \Psi \rangle \, ,
\ee
where $\hat G$ is now a general operator, which means that it can also be a function of conjugate momenta. We may thus regard $\hat G(t_i)$ at the initial time as the $G$ arising in \eqref{waveansatz} which is purely a function of the field coordinates.
Here $|\Psi \rangle$ encodes the gauge invariant part of the wavefunctional
\be
\langle A_i^a ,B^a, c^a,\bar c^a,  \dots  | \Psi \rangle = \Psi(A_i^a \dots ) \, ,
\ee
which automatically satisfies
\be
i\langle A_i^a ,B^a, c^a,\bar c^a,  \dots  | \hat Q | \Psi \rangle =\hat s \Psi(A_i^a \dots ) =0 \, .
\ee
As such $|\Psi \rangle$ is not a normalisable state in the Hilbert space, since there is nothing to suppress the integration over the ghost and NL fields. However, $|\psi \rangle $ is a normalisable state, as we have already seen when evaluated with ansatz \eqref{waveansatz2}.
The finite inner product is effectively being defined by
\be
\langle \psi_1 | \psi_2 \rangle = \langle \Psi_1 | e^{-\hat s \hat G_1^{\dagger}} e^{\hat s \hat G_2}| \Psi_2 \rangle \, .
\ee
This definition of the inner product is essentially that due to Batalin and Marnelius \cite{Batalin:1994rd,Fulop:1995bp,Marnelius:2000va,Duechting:1998wix}.
The dynamical evolution is governed as usual by the \Sch equation with the full Hamiltonian
\be
i \frac{\partial |\psi \rangle  }{\partial t} = \hat H | \psi \rangle \, .
\ee
 Substituting the ansatz \eqref{waveansatz} into the \Sch equation
we infer
\be
i e^{-\hat s \hat G }\frac{\partial e^{\hat s \hat G } }{\partial t} |\Psi \rangle + i \frac{\partial |\Psi \rangle}{\partial t}=  e^{-\hat s  \hat G} \hat H_0 e^{\hat s  \hat G} |\Psi \rangle+ e^{-\hat s  \hat G} \hat s \hat {\cal F}e^{\hat s  \hat G} |\Psi \rangle \, .
\ee
Using the nilpotency of the transformation it is easy to see that 
\be
e^{-\hat s  \hat G} \hat s \hat {\cal F}e^{\hat s  \hat G} =\hat s \( e^{-\hat s  \hat G}  \hat {\cal F}e^{\hat s  \hat G} \) \, ,
\ee
since $\hat s \hat X = i \{ \hat Q, \hat X \}_{\mp}$ satisfies the (anti-)Leibniz rule.
 Using further the gauge invariance of $\hat H_0$, we have $\hat s \hat H_0 = i [ \hat Q, \hat H_0]=0$ which can be used to infer that 
\be
e^{-\hat s  \hat G} \hat H_0 e^{\hat s  \hat G}=\hat H_0 + \hat s \hat \Sigma \, ,
\ee
where $\Sigma$ is 
\be
\hat \Sigma = e^{-\hat s  G} \[\hat H_0, \hat G \( \frac{e^{\hat s  \hat G}-1 }{\hat s \hat G} \)\] \, .
\ee
Thus, the \Sch equation is satisfied by demanding that the gauge invariant part of the physical states are evolved dynamically only with the gauge invariant part of the Hamiltonian
\be
i \frac{\partial |\Psi \rangle }{\partial t}=  \hat H_0  |\Psi \rangle \, ,
\ee
provided that the gauge fixing fermion operator satisfies the operator equations
\be
i \frac{\partial e^{\hat s \hat G} }{\partial t} =  \hat s \hat {\cal F}e^{\hat s  G} +e^{\hat s \hat G} \hat s \hat \Sigma = \hat s \(\hat {\cal F} e^{\hat s  G} +e^{\hat s \hat G}  \hat \Sigma \) \, .
\ee
To see that this is consistent, we note that
\ba
e^{\hat s \hat G} - 1 &=& \sum_{n=1}^{\infty} \frac{1}{n!} (\hat s \hat G)^n 
= \sum_{n=1}^{\infty} \frac{1}{n!} \hat s \hat G (\hat s \hat G)^{n-1} 
= \hat s \( \sum_{n=1}^{\infty} \frac{1}{n!} \hat G (\hat s \hat G)^{n-1} \) \\
&=& \hat s \left( \hat G + \sum_{n=2}^{\infty} \frac{1}{n!} \hat G (\hat s \hat G)^{n-1} \right) \, .
\ea
Hence, the differential equation is satisfied with
\ba \label{Gequation}
i \frac{\partial  \hat G }{\partial t} &=&   \hat {\cal F}+ \hat \Sigma  + \sum_{n=1}^{\infty} \frac{1}{n!} (\hat {\cal F} ( \hat s \hat G)^n+( \hat s \hat G)^n \hat \Sigma)  -i \frac{\partial}{\partial t} \(\sum_{n=2}^{\infty} \frac{1}{n!} \hat G (\hat s \hat G)^{n-1} \) \, .
\ea
From this $\hat G$ can be computed perturbatively in an expansion in the gauge fixing fermions, regarding $\hat {\cal F}$ and $[\hat G(0), \hat H_0]$ as the same order, with the lowest order term
\be
\hat G(t)  =\hat G(t_{\rm i})  - i \int_{t_{\rm i}}^t \d t \( \hat {\cal F} + [\hat H_0,\hat G(t_{\rm i})] \) + \dots 
\ee
This confirms that at least perturbatively \eqref{waveansatz2} describe the consistent dynamical evolution of the states \eqref{waveansatz}. In practice, it is easier to determine the evolution via the path integral where the gauge fixing fermion ${\cal F}$ enters explicitly to evolve the initial state.

Note that $|\Psi \rangle$ evolves in time with only the gauge invariant part of the Hamiltonian. This is exactly what happens in temporal gauge. Thus, at any finite time, we may regard the BRST physical state as a BRST `dressed' version of the temporal gauge wavefunction, with $e^{\hat s \hat G(t)}$ being the dressing operator. It is the temporal gauge wavefunctional that contains the meaningful physics, with the dressing a device to account for the volume of the gauge orbit in the inner product. 

\subsection{Gauge Fixing Independence of Inner Product}

Although the gauge fixing fermion operator $\hat G$ evolves with time, its precise form must be irrelevant provided that it picks out a unique representative of the gauge orbit. If this is true at the initial time $t_{\rm i}$, then it must be true at finite time $t$. In practise, this means that as long as we compute the inner product between two physical states, or an expectation value of a BRST invariant observable $\hat O$ for which $\hat s \hat O = i \{ \hat Q, \hat O \}_{\mp} $=0, we can replace $\hat G(t)$ with $\hat G(t_{\rm i})$, which is to say, we can replace a generically field and momentum dependent operator by an operator which is only a function of the fields coordinates.

The proof of this is straightforward. Consider the naive Batalin-Marnelius inner product of two states at finite time
\be
\langle \psi_1(t) | \psi_2(t) \rangle = \langle \Psi_1(t) | e^{-\hat s \hat G_1^{\dagger}(t)} e^{\hat s \hat G_2(t)}|\Psi_2(t) \rangle \, .
\ee
We first note that
\ba
e^{\hat s \hat G_2(t)} &=& e^{\hat s \hat G_2(t_{\rm i})} + \hat s \(  \hat G_2(t) \frac{(e^{\hat s \hat G_2(t)}-1)}{\hat s \hat G_2(t)} - \hat G_2(t_{\rm i}) \frac{(e^{\hat s \hat G_2(t_{\rm i})}-1)}{\hat s \hat G_2(t_{\rm i})}  \) \\
&=& e^{\hat s \hat G_2(t_{\rm i})} + \hat s \hat C_2 \, .
\ea
Hence,
\ba
 \langle \psi_1(t) | \psi_2(t) \rangle & = & \langle \Psi_1(t) | e^{-\hat s \hat G_1^{\dagger}(t)} e^{\hat s \hat G_2(t_{\rm i})}|\Psi_2(t) \rangle + \langle \Psi_1(t) | e^{-\hat s \hat G_1^{\dagger}(t)} \hat s \hat C_2|\Psi_2(t) \rangle \\
&=& \langle \Psi_1(t) | e^{-\hat s \hat G_1^{\dagger}(t)} e^{\hat s \hat G_2(t_{\rm i})}|\Psi_2(t) \rangle + \langle \Psi_1(t) |\hat s \(  e^{-\hat s \hat G_1^{\dagger}(t)}  \hat C_2 \)|\Psi_2(t) \rangle
\\
&=& \langle \Psi_1(t) | e^{-\hat s \hat G_1^{\dagger}(t)} e^{\hat s \hat G_2(t_{\rm i})}|\Psi_2(t) \rangle \\
&& + i \langle \Psi_1(t) |\hat Q \(  e^{-\hat s \hat G_1^{\dagger}(t)}  \hat C_2 \)|\Psi_2(t) \rangle+i \langle \Psi_1(t) |\(  e^{-\hat s \hat G_1^{\dagger}(t)}  \hat C_2 \)\hat Q |\Psi_2(t) \rangle \, . \nn
\ea
Given that the states $|\Psi_{1,2}(t) \rangle$ are gauge invariant at any time we have
\be
\label{psitpsit}
\hat Q |\Psi_{1,2}(t) \rangle =0 \, ,
\ee
so that
\be
\langle \psi_1(t) | \psi_2(t) \rangle =\langle \Psi_1(t) | e^{-\hat s \hat G_1^{\dagger}(t)} e^{\hat s \hat G_2(t_{\rm i})}|\Psi_2(t) \rangle  \, .
\ee
This argument can then be repeated using 
\ba
e^{-\hat s \hat G_1^{\dagger}(t)} &=& e^{-\hat s \hat G_1^{\dagger}(t_{\rm i})} + \hat s \( \hat G_1^{\dagger}(t) \frac{(e^{-\hat s \hat G_1^{\dagger}(t)}-1)}{\hat s \hat G_1^{\dagger}(t)} -\hat G_1^{\dagger}(t_{\rm i}) \frac{(e^{-\hat s \hat G_1^{\dagger}(t_{\rm i})}-1)}{\hat s\hat G_1^{\dagger}(t_{\rm i})}  \) \\
&=& e^{-\hat s \hat G_1^{\dagger}(t_{\rm i})} + \hat s \hat C_1 \, .
\ea
leading to 
\be
\langle \psi_1(t) | \psi_2(t) \rangle =\langle \Psi_1(t) | e^{-\hat s \hat G_1^{\dagger}(t_{\rm i})} e^{\hat s \hat G_2(t_{\rm i})}|\Psi_2(t) \rangle  \, .
\ee
We can further trade $\hat G_{1,2}(t_{\rm i})$ for a field field dependent, momentum independent operator $\hat G$ so that 
\ba
\langle \psi_1(t) | \psi_2(t) \rangle &=&\langle \Psi_1(t) |  e^{2\hat s \hat G}|\Psi_2(t) \rangle \\
&=& \int \funcd[A_i,B,c,\bar c , \dots] \Psi_1^*(A_i, \dots ;t) e^{2 \hat s G(A_i,B,c,\bar c , \dots)}  \Psi_2(A_i, \dots ;t) \, ,
\ea
with $\Psi_{1,2}(A_i, \dots ;t)$ the dynamically evolved gauge invariant part of the wavefunctionals.
In this form we see that $2G(A_i,B,c,\bar c , \dots)$ is just the gauge fixing fermion for the 3 dimensional functional integral over BRST fields. Thus as in the Abelian case, its purpose is just to remove the volume of the 3 dimensional gauge orbit. Hence we recover the inner product \eqref{NAinner}.

The same argument applies to expectation values of BRST invariant operators for which $\hat s \hat O(t)=0$
\ba
\langle \psi_1(t) | \hat O(t)|\psi_2(t) \rangle & = & \langle \Psi_1(t) | e^{-\hat s \hat G_1^{\dagger}(t)} \hat O(t) e^{\hat s \hat G_2(t_{\rm i})}|\Psi_2(t) \rangle + \langle \Psi_1(t) | e^{-\hat s \hat G_1^{\dagger}(t)} \hat O(t) \hat s \hat C_2|\Psi_2(t) \rangle \\
&=& \langle \Psi_1(t) | e^{-\hat s \hat G_1^{\dagger}(t)} \hat O(t) e^{\hat s \hat G_2(t_{\rm i})}|\Psi_2(t) \rangle + \langle \Psi_1(t) |\hat s \(  e^{-\hat s \hat G_1^{\dagger}(t)}  \hat O(t) \hat C_2 \)|\Psi_2(t) \rangle
\\
&=& \langle \Psi_1(t) | e^{-\hat s \hat G_1^{\dagger}(t)} \hat O(t) e^{\hat s \hat G_2(t_{\rm i})}|\Psi_2(t) \rangle \\
&& + i \langle \Psi_1(t) |\hat Q \(  e^{-\hat s \hat G_1^{\dagger}(t)} \hat O(t)  \hat C_2 \)|\Psi_2(t) \rangle+i \langle \Psi_1(t) |\(  e^{-\hat s \hat G_1^{\dagger}(t)} \hat O(t)  \hat C_2 \)\hat Q |\Psi_2(t) \rangle \, . \nn
\notag \\
&=& \langle \Psi_1(t) | e^{-\hat s \hat G_1^{\dagger}(t)} \hat O(t) e^{\hat s \hat G_2(t_{\rm i})}|\Psi_2(t) \rangle \, . 
\ea
Repeating, we have
\be
\langle \psi_1(t) | \hat O(t)|\psi_2(t) \rangle =  \langle \Psi_1(t) | e^{\hat s \hat G} \hat O(t) e^{\hat s \hat G}|\Psi_2(t) \rangle \, . 
\ee
It should be emphasized that these steps are only meaningful because at every stage, there exists some $\hat G$ that selects a representative of the gauge orbit, ensuring that the overall inner product is well defined. This confirms that (a) we may compute the inner product at any time with any choice of $\hat G$ that fixes the gauge orbit, and (b) we may always choose a $\hat G$ which is a function of the fields only.

The correlation functions of the primary fields, $A_{\mu}, B, \bar c, \dots$ are, of course, dependent on the gauge fixing terms. However, these are far more easily dealt with through the path integral, where the gauge fixing dependence is explicit, rather than trying to solve \eqref{Gequation}.

\subsection{Density matrices, Hata-Kugo trace}

 A physical density operator can be interpreted as a mixture of pure physical states
\be
\hat \rho_{\rm phys} = \sum_I p_I |\psi_{I \rm phys} \rangle \langle \psi_{I \rm phys} | \, ,
\ee
and so should similarly satisfy $\hat Q\hat \rho_{\rm phys}=\hat \rho_{\rm phys}\hat Q=0$. For example for Yang-Mills coupled to a coloured scalar we can already give an explicit expression in \Sch basis if we assume all physical states have the same gauge fixing fermion $G$
\ba
&& \langle A^{+},c^{+},\bar c^{+},\Phi^{+},{\Phi^\dagger}^{+} \vert \hat \rho_{\rm phys} 
\vert A^{-},c^{-},\bar c^{-},\Phi^{-},\Phi^{\dagger -} \rangle \nn \\
&& \quad =  e^{\hat s G[A^{+},c^{+},\bar c^{+},\Phi^{+},{\Phi^\dagger}^{+}]}
e^{(\hat s G[A^{-},c^{-},\bar c^{-},\Phi^{-},\Phi^{\dagger -}])^{\dagger}}
\,{\bf \varrho}[A_i^{-},A_j^{+},\Phi^{-},\Phi^{\dagger -},\Phi^{+},{\Phi^\dagger}^{+}]
\, .
\ea
where ${\bf \varrho}[A_i^{-},A_j^{+},\Phi^{-},\Phi^{\dagger -},\Phi^{+},{\Phi^\dagger}^{+}]$ 
is gauge invariant under two copies of spatial gauge transformations 
\be
 \delta A_i^{\pm} =D_i[A^{\pm}] \lambda^{\pm} \, , \quad  \delta \Phi^{\pm} \rightarrow i g T_{\Phi}^a \lambda_{\pm}^a \Phi^{\pm} \, ,
\ee
and admits a decomposition
\be
{\bf \varrho}[A_i^{-},A_j^{+},\Phi^{-},\Phi^{\dagger -},\Phi^{+},{\Phi^\dagger}^{+}]
=\sum_I p_I 
\Psi_I[A_i^{-},\Phi^{-},\Phi^{\dagger -}]
\Psi_I[A_j^{+},\Phi^{+},{\Phi^\dagger}^{+}]^* \, ,
\ee
with $p_I \ge 0$.

More generally, we should allow for each pure state contribution to have different gauge fixing fermions $G_I$ and so the most general physical density matrix is
\ba
&& \langle A^{+},c^{+},\bar c^{+},\Phi^{+},{\Phi^\dagger}^{+} \vert \hat \rho_{\rm phys} 
\vert A^{-},c^{-},\bar c^{-},\Phi^{-},\Phi^{\dagger -} \rangle
 \\
&& \quad = \sum_I p_I e^{\hat s G_I[A^{+},c^{+},\bar c^{+},\Phi^{+},{\Phi^\dagger}^{+}]}
e^{(\hat s G_I[A^{-},c^{-},\bar c^{-},\Phi^{-},\Phi^{\dagger -}])^{\dagger}}
\,\Psi_I[A_i^{-},\Phi^{-},\Phi^{\dagger -}] \Psi_I[A_j^{+},\Phi^{+},{\Phi^\dagger}^{+}]^* \nn 
\, .
\ea
In practice, however, the above construction of physical states is undesirably complicated. 
We can alternatively work with a density operator $\hat \rho$ that satisfies the weaker condition that it is BRST invariant $[\hat Q,\hat \rho]=0$. The physical density operator can be defined by introducing a projection operator $\hat P$ that projects back onto the physical Hilbert space
\be
\hat \rho_{\rm phys} = \hat P \hat \rho \hat P \, ,
\ee
where $\hat P^2=\hat P$ and $\hat P \hat Q = \hat Q \hat P=0$. The latter relations ensure that $ \hat Q \hat \rho_{\rm phys}=\hat \rho_{\rm phys} \hat Q=0$.  The projection operator can be shown to take the form \cite{Kugo:1979gm,Hata:1980yr,Calzetta:2008iqa}
\be
\hat P = 1- \{\hat Q, \hat R \} \, .
\ee
Physical observables should be BRST invariant, and the physical expectation values of such an observable would be
\be
\Tr[\hat \rho_{\rm phys} \hat O] = \Tr[\hat P \hat \rho \hat P \hat O] \, .
\ee
We can write
\be
\Tr[\hat P \hat \rho \hat P \hat O]=\Tr[\hat P \hat \rho \hat O]-\Tr[\hat P \hat \rho \hat Q \hat R \hat O]- \Tr[\hat P \hat \rho  \hat R \hat Q \hat O] \, .
\ee
Given that $\hat Q$ commutes with both $\hat \rho$ and $\hat O$, and that $\hat P \hat Q=\hat Q \hat P=0$ we see using the cyclic properties of the trace that the second two terms identically vanish
\be
\Tr[\hat P \hat \rho \hat P \hat O]=\Tr[\hat P \hat \rho \hat O] \, .
\ee
As it stands, we cannot repeat that argument to remove the remaining projection operator. However, there is a clever trick due to Hata and Kugo that resolves this \cite{Hata:1980yr}. Recall that the ghost charge $\hat Q_G$ has the commutation relation
\be
[\hat Q_G, \hat Q] = i \hat Q
\ee
which given $\hat Q_G = i \hat N_G$ is the statement that the BRST charge has ghost number $+1$. It then follows that
\be
e^{\theta \hat Q_G} \hat Q e^{-\theta \hat Q_G} =e^{i \theta } \hat Q \, .
\ee
By choosing $\theta=\pi$ we infer $e^{\pi \hat Q_G} \hat Q e^{-\pi \hat Q_G} =- \hat Q $
or better stated as
\be
e^{\pi \hat Q_G} \hat Q = - \hat Q e^{\pi \hat Q_G} \, .
\ee
Now, since all states in the physical Hilbert space have zero ghost number, it follows that 
\be
\Tr[\hat P \hat \rho \hat O]=\Tr[e^{\pi \hat Q_G}\hat P \hat \rho \hat O] \, .
\ee 
Substituting in the form of the projection operator
\ba
\Tr[\hat P \hat \rho \hat O]&=&\Tr[e^{\pi \hat Q_G} \hat \rho \hat O] -\Tr[e^{\pi \hat Q_G}\{\hat Q, \hat R \}  \hat \rho \hat O] \\
&=& \Tr[e^{\pi \hat Q_G} \hat \rho \hat O] -\Tr[e^{\pi \hat Q_G}\hat Q \hat R  \rho \hat O] -\Tr[e^{\pi \hat Q_G}\hat R \hat Q \hat \rho \hat O] \\
&=& \Tr[e^{\pi \hat Q_G} \hat \rho \hat O] +\Tr[\hat Q e^{ \pi \hat Q_G}\hat R  \hat \rho \hat O] -\Tr[e^{\pi \hat Q_G}\hat R \hat Q \hat \rho \hat O] \\
&=&\Tr[e^{\pi \hat Q_G} \hat \rho \hat O] \, ,
\ea
where in the last step we used the fact that $\hat \rho $ and $\hat O$ are both BRST invariant and the trace is cyclic. In summary, we conclude that the physical expectation value of a physical (BRST invariant) observable is
\be \label{thetrick}
\Tr[\hat \rho_{\rm phys} \hat O]=\Tr[e^{\pi \hat Q_G} \hat \rho \hat O] \, .
\ee
The Hata-Kugo trick allows us to work with density operators defined in the full Hilbert space which are restricted only by the requirement that they are BRST invariant in the operator sense $[\hat \rho, \hat Q]=0$. The insertion of the factor $e^{\pi \hat Q_G}$ then automatically performs the necessary projection down to the physical Hilbert space without the need to explicitly construct the projection operator. Now since $e^{\pi \hat Q_G}=e^{i \pi \hat N_G}$ it follows 
and the operators satisfy
\be
e^{\pi \hat Q_G} \hat c({\bf x}) e^{-\pi \hat Q_G}=e^{i \pi}\hat c({\bf x})=-\hat c({\bf x}) \, , \quad e^{\pi \hat Q_G} \hat{\bar c}({\bf x}) e^{-\pi \hat Q_G}=e^{-i \pi} \hat{\bar c}({\bf x})=-\hat{\bar c}({\bf x}) \, ,
\ee
then on the ghost \Sch states
\be
e^{\pi \hat Q_G} |c, \bar c \rangle = |-c, -\bar c \rangle \, .
\ee
Thus in terms of the \Sch representation, the statement is that the trace of a physical observable can be computed as
\ba
\Tr[\hat \rho_{\rm phys} \hat O]&=& \int \funcd[A_i , A_4 , c , \bar c]  \, \langle A_i,-A_4,-c,-\bar c| \rho_{\rm phys} \hat O | A_i,A_4,c,\bar c \rangle \\
&=&\int \funcd[A_i , A_4 , c , \bar c] \,  \langle A_i,-A_4,c,\bar c| \rho \hat O | A_i,A_4,c,\bar c \rangle \, .
\ea
We see that the sign in the trace from the fermionic nature of the ghosts is precisely cancelled by the sign from the Hata-Kugo prescription.

\subsection{KMS boundary conditions for ghosts}

The Hata-Kugo prescription allows us to replace arbitrary physical density operators $\hat \rho_{\rm phys}$ with a BRST invariant ($[\hat Q,\hat \rho]=0$) but not closed $\hat Q \hat \rho \neq 0$ density operator. This proves particularly useful in the definition of thermal states where since $[H,\hat Q]=0$ we may use the naive Boltzmann density operator
\be
\hat \rho = Z^{-1}e^{-\beta \hat H}=\frac{e^{-\beta \hat H}}{\Tr[e^{\pi \hat Q_G} e^{-\beta \hat H}]} \, .
\ee
This is not the true physical density operator since it automatically excites the unphysical degrees of freedom into thermal states. If the trace were computed directly from $\hat \rho$, then we would infer that the ghost two-point function satisfies an anti-periodic KMS boundary condition at finite temperature due to the fermionic nature of the ghosts. Specifically, the thermal Wightman functions satisfy 
\ba
 \Tr[\hat \rho \hat c( x) \hat{\bar c}(y)]&=& Z^{-1} \Tr[e^{-\beta \hat H} \hat c( x) \hat{\bar c}(y)]=Z^{-1} \Tr[ \hat c(x+i \beta n) e^{-\beta \hat H} \hat{\bar c}(y)] \nn \\
&=&Z^{-1} \Tr[e^{-\beta \hat H} \hat{\bar c}(y) \hat c(x+i \beta n) ]\nn  \\
&=&  -\Tr[\hat \rho \hat c(x+i \beta n) \hat{\bar c}(y)] \, ,
\ea
with $n^{\mu}=(1,0,0,0)$, where we have used the  cyclicity of the trace and the last step follows from the definition of CTP time ordering for fermions.
The virtue of the Hata-Kugo prescription is that it switches this around
\ba
G^{c\bar c}_{-+}(x,y)&=& \Tr[e^{\pi \hat Q_G}\hat \rho \hat c( x) \hat{\bar c}(y)]= Z^{-1} \Tr[e^{\pi \hat Q_G}e^{-\beta \hat H} \hat c( x) \hat{\bar c}(y)] \nn \\
&=&Z^{-1} \Tr[ e^{\pi \hat Q_G}\hat c(x+i \beta n)e^{-\pi \hat Q_G} e^{\pi \hat Q_G}e^{-\beta \hat H} \hat{\bar c}(y)]=-Z^{-1} \Tr[e^{\pi \hat Q_G}e^{-\beta \hat H} \hat{\bar c}(y) \hat c(x+i \beta n) ] \nn  \\
&=& G^{c\bar c}_{+-}(x+i \beta n,y)
\ea
This is summarised by the KMS boundary conditions, i.e.~the statement that at finite temperature ghosts satisfy periodic boundary conditions in imaginary time. This follows directly from the trace formula above.

\subsection{BRST invariance of Hata-Kugo Trace}

The reason thermal ghosts have periodic KMS boundary conditions is BRST invariance. Under a diagonal BRST transformation, the photon mixes with the ghosts. Since the former satisfies periodic boundary conditions, the latter must also, otherwise the diagonal BRST symmetry would be broken. We will see this explicitly later via the Ward-Takahashi-Slavnov-Taylor identities.

We can understand this directly at the operator level as follows. Consider an exact BRST operator of the form $\hat s \hat G = i \{ \hat Q, \hat G \}$ with $\hat G$ Grassmann odd. The physical expectation value of this should vanish, and it is indeed easy to see that
\be
\Tr[\hat \rho_{\rm phys} \{ \hat Q, \hat G \}]=\Tr[\hat \rho_{\rm phys}  \hat Q \hat G ]+\Tr[\hat \rho_{\rm phys}  \hat G \hat Q ]=0\, .
\ee
If we used the naive cyclic trace for an indefinite Hilbert space, we would have
\ba
\Tr[\hat \rho \{ \hat Q, \hat G \}]&=&\Tr[\hat \rho \hat Q \hat G]+\Tr[\hat \rho \hat G \hat Q ] \nn \\
&=&\Tr[\hat \rho \hat Q \hat G]+\Tr[\hat Q\hat \rho \hat G  ]=2 \Tr[\hat \rho \hat Q \hat G] \neq 0 \, .
\ea
Here we have used the cyclicity of the trace followed by the BRST invariance of the state. By contrast with the Hata-Kugo trace
\ba
\Tr[e^{\pi \hat Q_G}\hat \rho \{ \hat Q, \hat G \}]&=&\Tr[e^{\pi \hat Q_G}\hat \rho \hat Q \hat G]+\Tr[e^{\pi \hat Q_G}\hat \rho \hat G \hat Q ] \nn \\
&=&\Tr[e^{\pi \hat Q_G}\hat \rho \hat Q \hat G]+\Tr[\hat Q e^{\pi \hat Q_G} \hat \rho \hat G  ]
\nn \\
&=&\Tr[e^{\pi \hat Q_G}\hat \rho \hat Q \hat G]-\Tr[e^{\pi \hat Q_G} \hat Q \hat \rho \hat G  ]
\nn \\
&=& 0 \, .
\ea
In other words
\be
 \langle \hat s \hat G \rangle_{\rho}=\Tr[e^{\pi \hat Q_G} \hat \rho \hat s \hat G]=0 \, .
\ee
Thus, even without knowing the explicit projection operator which maps to the physical Hilbert space, we can infer the need for the Hata-Kugo trace simply from the requirement that it is BRST invariant.

\subsection{BRST Invariant Density Matrices}

The Hata-Kugo prescription allows us to define a physical state by a density operator $\hat \rho$ which satisfies the weaker condition that it commutes with the BRST charge $[\hat Q,\hat \rho]=0$
or equivalently that 
\be
e^{-i \eta \hat Q} \hat \rho e^{i \eta \hat Q} =\hat \rho \, .
\ee
Translating this into the NL representation, we have
\ba
&& \langle A^+_i, B^+, c^+, \bar c^+ ,\dots |e^{-i \eta \hat Q} \hat \rho e^{i \eta \hat Q} | A^-_i, B^-, c^-, \bar c^- , \dots  \rangle \nn \\
&&\quad =\langle A^+_i+\eta \hat s A^+, B^+, c^++\eta \hat s c^+, \bar c^++ \eta  \hat s \bar c^+, \dots | \hat \rho  | A^-_i+\eta \hat s A_i^-, B^-, c^-+\eta \hat s c^-, \bar c^-+\eta \hat s \bar c^- , \dots  \rangle \nn \\
&&\quad \equiv \langle A^+_i, B^+, c^+, \bar c^+ ,\dots | \hat \rho  | A^-_i, B^-, c^-, \bar c^- , \dots  \rangle \, .
\ea
Here we have anticipated the doubled branch notation of the Schwinger-Keldysh contour. With this in mind, the condition for the density matrix to be BRST invariant is then
\be
\hat s \langle A^+_i, B^+, c^+, \bar c^+ ,\dots | \hat \rho  | A^-_i, B^-, c^-, \bar c^- , \dots  \rangle=0 \, ,
\ee
where $\hat s$ is now understood to be the diagonal BRST transformation which acts on both branches in the same manner
\ba
&& \hat s {A^a}^{\pm}_{i} =  D_{i}[A_{\pm}]c^a_{\pm}  \,, \quad \hat s c_{\pm}^a = - \frac{1}{2}g f^{abc} c_{\pm}^b c_{\pm}^c  \,, \\
&& \hat s \bar c_{\pm}^a =  B_{\pm}^a  \,, \quad \hat s B_{\pm}^a = 0 \, .
\ea
Note that this condition is highly non-trivial on the types of coupling that are allowed between the two branches. To illustrate this consider the function
\be
e^{-\alpha \int \d^3 {\bf x} F_{ij}^{+a} F^{ij}_{-a}} \,.
\ee
This is invariant under the diagonal/retarded gauge transformations for which both fields transform the same way. However, it is {\bf not} invariant under the diagonal BRST transformation due to the presence of two distinct FPdW ghosts. Explicitly
\be
\hat s F_{ij}^{+a} F^{ij}_{-a} = - g f^{abc} (c_+^b-c_-^b) F_{ij}^{+c} F^{ij}_{-a} \neq 0 \, .
\ee
An acceptable coupling between the two branches is
\be
e^{-\alpha \int \d^3 {\bf x} (F_{ij}^{+a} F^{ij}_{+a})  (F_{ij}^{-a} F^{ij}_{-a})    } \,,
\ee
since it is separately gauge invariant under the independent gauge transformations on each branch and is hence automatically BRST invariant under the diagonal BRST symmetry. This example illustrates why it would be inconsistent to construct the effective action for an Open EFT, i.e.~the Feynman-Vernon influence functional, by allowing all interactions which are invariant under the diagonal gauge transformations.

\section{Schwinger-Keldysh Path Integral}

\label{Sec3}

The derivation of the path integral is straightforward once we have established the correct BRST invariant inner product and trace formula to pick out the positive norm physical states from the otherwise indefinite Hilbert space. The advantage of the \Sch representation is that it works equally well for a free theory or for an interacting one. Similarly, the existence of a BRST symmetry and ghost symmetry with the same properties $\hat Q^2=0$, $[\hat N_G, \hat Q] =  \hat Q$ for both Abelian theory or non-Abelian gauge theories means that we can follow the same recipe in each case. The only lingering subtlety in the non-Abelian case is the Gribov ambiguity, which we discuss in section~\ref{Gribov}.

\subsection{Phase-space Path Integral}

The derivation of the CTP path integral begins with the benign statement that for a \Sch picture density operator at time $t_{\rm f}$ its trace must be normalised to unity 
\be
\Tr[\hat \rho_{\rm phys}(t_{\rm f})] = \Tr[e^{\pi \hat Q_G}\hat \rho(t_{\rm f}) ]=1 \, .
\ee
We will denote bosonic matter by $\Phi$ and fermionic matter by $\psi$.
Suppressing the gauge indices on the fields so that we can treat the Abelian and non-Abelian cases at the same time, the trace can be computed as 
\be
\int \funcd [A_{i}^{\rm f},A_4^{\rm f},c_{\rm f},\bar c_{\rm f} ,\Phi_{\rm f} ,\Phi_{\rm f}^\dagger,\psi,\bar \psi]\, \mu_f[\psi_{\rm f},\bar \psi_{\rm f}] \,  \langle A_i^{\rm f}, -A_4^{\rm f} , c_{\rm f} ,\bar c_{\rm f} ,\Phi_{\rm f} , \Phi_{\rm f}^\dagger,-\bar \psi_{\rm f}|\hat \rho(t_{\rm f}) | A_i^{\rm f}, A_4^{\rm f} , c_{\rm f} ,\bar c_{\rm f} ,\Phi_{\rm f} , \Phi_{\rm f}^\dagger,\psi_{\rm f}  \rangle = 1 \, .
\ee
The only minor difference is that physical fermions are described by coherent states $|\psi\rangle  $ whose normalisation includes a measure factor $\mu_f[\psi_{\rm f},\bar \psi_{\rm f}]=e^{-\int \d^3 {\bf x}\, \bar \psi_{\rm f} \gamma^0 \psi_{\rm f}}$.  In addition, the physical fermions require a minus sign in the trace. 

The same trace in the Nakanishi-Lautrup representation is
\be
\int \funcd [A_{i}^{\rm f},B_4^{\rm f},c_{\rm f},\bar c_{\rm f} ,\Phi_{\rm f} ,\Phi_{\rm f}^\dagger,\psi,\bar \psi] \, \mu_f[\psi_{\rm f},\bar \psi_{\rm f}]  \,  \langle A_i^{\rm f}, -B_4^{\rm f} , c_{\rm f} ,\bar c_{\rm f} ,\Phi_{\rm f} , \Phi_{\rm f}^\dagger,-\bar \psi_{\rm f}|\hat \rho(t_{\rm f}) | A_i^{\rm f}, B_4^{\rm f} , c_{\rm f} ,\bar c_{\rm f} ,\Phi_{\rm f} , \Phi_{\rm f}^\dagger,\psi_{\rm f}  \rangle = 1 \, .
\ee
The Hata-Kugo factor has been used to remove the sign in the ghost trace. For notational convenience, we will label the eigenstates not by $A_4 $ and $B_4$ but with $A_0$ and $B$, with the understanding that before the final integral is taken we must rotate $A_0 \rightarrow i A_4$ and $B \rightarrow -i B_4$ {\it after} the complex conjugation is taken as discussed in Section~\ref{sec:contourrotation}. With this in mind, we denote the trace 
\be
\int \funcd [A_{i}^{\rm f},B^{\rm f},c_{\rm f},\bar c_{\rm f} ,\Phi_{\rm f} ,\Phi_{\rm f}^\dagger,\psi,\bar \psi] \, \mu_f[\psi_{\rm f},\bar \psi_{\rm f}] \,  \langle A_i^{\rm f}, B^{\rm f} , c_{\rm f} ,\bar c_{\rm f} ,\Phi_{\rm f} , \Phi_{\rm f}^\dagger,-\bar \psi_{\rm f}|\hat \rho(t_{\rm f}) | A_i^{\rm f}, B^{\rm f} , c_{\rm f} ,\bar c_{\rm f} ,\Phi_{\rm f} , \Phi_{\rm f}^\dagger,\psi_{\rm f}  \rangle = 1 \, .
\ee
The \Sch density operator evolves in time as
\be
\hat \rho(t_{\rm f}) =e^{-i \hat H(t_{\rm f}-t_{\rm i})}\hat \rho(t_{\rm i})e^{+i \hat H(t_{\rm f}-t_{\rm i})} \, , 
\ee
with $\hat H$ the full BRST Hamiltonian which is Hermitian. As we have already outlined, BRST invariance is manifest in the Nakanishi-Lautrup representation, as we will specifically use the Hamiltonian \eqref{YMHamiltonian}, or more precisely its extension including a $\theta$ term and matter fields. 

To account for the fermion sign, we include a factor of $(-1)^{\hat N_{\psi}} $ where the operator $\hat N_{\psi} = \int \mathrm{d}^3 \mathbf{x} \, \psi^{\dagger} \psi$ counts the number of physical fermions, so $N_{\psi} = \pm 1$ if the field ${\boldsymbol f}$ is a {\it physical} fermion/anti-fermion and $N_{\psi}=0$ otherwise. Crucially $N_{\psi}=0$ for $c$ and $\bar c$. The BRST Hamiltonian, being Grassmann even, satisfies $[\hat H,\hat N_{\psi}]=0$. 

Let us now introduce the following shorthand for the spatial functional integrals including the measure for physical fermions 
\be
\int \funcd {\bf f} =\int \funcd [A_{i},B,c,\bar c ,\Phi ,\Phi^\dagger,\psi,\bar \psi] \,   \mu_f[\psi,\bar \psi] \, , 
\ee
and
\be
\langle {\bf f}|e^{-i \hat H(t_{\rm f}-t_{\rm i})}|{\bf f}' \rangle =\langle A_{i},B , c ,\bar c ,\Phi , \Phi^\dagger,\bar \psi|e^{-i \hat H(t_{\rm f}-t_{\rm i})}|A_i' ,B', c' ,\bar c' ,\Phi', \Phi^{\prime\dagger},\psi'\rangle \, .
\ee
Inserting two complete sets of states (subject to the same proviso that for physical fermions we insert a complete set of coherent states which includes a measure $\mu_f$) and anticipating the two-branch $\pm$ notation of the CTP we have
\ba
&& \int \funcd {\bf f}_{\rm f} \int \funcd {\bf f}^+_{\rm i} \int \funcd {\bf f}^-_{\rm i} \;  
\langle (-1)^{\hat N_{\psi}} {\bf f}_{\rm f}|e^{-i \hat H(t_{\rm f}-t_{\rm i})}|{\bf f}_{\rm i}^+ \rangle  
\langle {\bf f}_{\rm i}^+| \hat \rho(t_{\rm i})|{\bf f}_{\rm i}^-\rangle    
\langle {\bf f}_{\rm i}^-|e^{i \hat H(t_{\rm f}-t_{\rm i})}|{\bf f}_{\rm f} \rangle \notag \\
&&\quad = \int \funcd [ {\bf f}_{\rm f} , {\bf f}^+_{\rm i} , {\bf f}^-_{\rm i} ]\; 
\langle {\bf f}_{\rm f}|e^{-i \hat H(t_{\rm f}-t_{\rm i})}|(-1)^{\hat N_{\psi}}  {\bf f}_{\rm i}^+ \rangle  
\langle {\bf f}_{\rm i}^+| \hat \rho(t_{\rm i})|{\bf f}_{\rm i}^-\rangle    
\langle {\bf f}_{\rm i}^-|e^{i \hat H(t_{\rm f}-t_{\rm i})}|{\bf f}_{\rm f} \rangle \\
&&\quad = \int \funcd [ {\bf f}_{\rm f} , {\bf f}^+_{\rm i} , {\bf f}^-_{\rm i} ] \;  
\langle {\bf f}_{\rm f}|e^{-i \hat H(t_{\rm f}-t_{\rm i})}|{\bf f}_{\rm i}^+ \rangle  
\langle (-1)^{\hat N_{\psi}}{\bf f}_{\rm i}^+| \hat \rho(t_{\rm i})|{\bf f}_{\rm i}^-\rangle    
\langle {\bf f}_{\rm i}^-|e^{i \hat H(t_{\rm f}-t_{\rm i})}|{\bf f}_{\rm f} \rangle =1 \, ,
\ea
where we made convenient use of $[\hat H,\hat N_{\psi}]$ to move the minus sign, e.g. 
\be
\langle (-1)^{\hat N_{\psi}} {\bf f}_{\rm f}|e^{-i \hat H(t_{\rm f}-t_{\rm i})}|{\bf f}_{\rm i}^+ \rangle =\langle{\bf f}_{\rm f}|e^{-i \hat H(t_{\rm f}-t_{\rm i})}| (-1)^{\hat N_{\psi}} {\bf f}_{\rm i}^+ \rangle \, .
\ee
and with a slight abuse of notation we have written
\be
(-1)^{\hat N_{\psi}} | {\bf f}_{\rm i}^+ \rangle =| (-1)^{\hat N_{\psi}} {\bf f}_{\rm i}^+ \rangle \, . 
\ee
The fermion minus sign is now integrated into the initial state so that physical fermions at the final time are continuous $\psi^+(t_{\rm f})=\psi^-(t_{\rm f})$, whereas for a CTP with a thermal branch at the initial time they are antiperiodic.

The initial state is encoded in 
\be
  \langle (-1)^{\hat N_{\psi}}{\bf f}_{\rm i}^+ | \hat \rho(t_{\rm i})|{\bf f}_{\rm i}^-\rangle
  =\langle  A^+_{{\rm i} i} ,B_{\rm i}^+ ,c_{\rm i}^+ ,{\bar c}_{\rm i}^+ ,\Phi_{\rm i}^+ , \Phi^{+\dagger}_{\rm i}, -\bar \psi^{+}_{\rm i}| 
  \hat \rho(t_{\rm i}) 
  | A^-_{{\rm i} i} ,B_{\rm i}^- ,c_{\rm i}^- ,{\bar c}_{\rm i}^- ,\Phi_{\rm i}^- , \Phi^{-\dagger}_{\rm i}, \bar \psi^{-}_{\rm i}\rangle \, .
\ee
We have already described how to construct explicit examples of BRST invariant initial states including their ghost dependence. Following the standard procedure, by discretising time and inserting a complete set of field and momentum eigenstates at each intermediate time, we infer the phase space form of the path integral that defines a transition between two states in the indefinite Hilbert space
\be \label{PSPI}
\langle {\bf f}_{\rm f}|e^{-i \hat H(t_{\rm f}-t_{\rm i})}|{\bf f}_{\rm i} \rangle =  \int_{{\boldsymbol f}(t_{\rm i})={\bf f}_{\rm i}}^{{\boldsymbol f}(t_{\rm f})={\bf f}_{\rm f}}  \pathd [{\boldsymbol{f}}] \int \pathd [ \Pi_{\boldsymbol{f}} ] \, e^{i \int_{t_{\rm i}}^{t_{\rm f}} \d t \; \( \Pi_{\boldsymbol{f}}.\partial_t {\boldsymbol{f}}- H(\boldsymbol{f},\Pi_{\boldsymbol{f}}) \)} \, .
\ee
The conjugate momenta have been specified in \eqref{conjugatemomenta} and the Hamiltonian in \eqref{YMHamiltonian} for pure Yang-Mills and these are easily extended to include additional matter states.
For the bosonic and ghost degrees of freedom \eqref{PSPI} describes a transitional amplitude between two \Sch eigenstates and so all momenta $\Pi_{\boldsymbol f}$ are integrated over. For physical fermions, $\bar \psi$ already acts as the conjugate momenta and \eqref{PSPI} describes a transition between a coherent state $|\psi_{\rm i} \rangle$ and a final state $\langle \bar \psi_{\rm f}|$ and as such the boundary conditions on the path integral are mixed
\be
\psi(t_{\rm i})=\psi_{\rm i} \, , \quad \bar \psi(t_{\rm f})=\bar \psi_{\rm f} \, .
\ee
For the bosonic and ghost degrees of freedom we may perform the path integral over the conjugate momenta, to give the configuration space path integral. In doing so, we must account for a possible measure factor we have
\be
\langle {\bf f}_{\rm f}|e^{-i \hat H(t_{\rm f}-t_{\rm i})}|{\bf f}_{\rm i} \rangle =  \int_{{\boldsymbol{f}}(t_{\rm i})={\bf f}_{\rm i}}^{{\boldsymbol{f}}(t_{\rm f})={\bf f}_{\rm f}}  \pathd [{\boldsymbol{f}}] \, \mu[{\boldsymbol{f}}] \, e^{i S[{\boldsymbol{f}}(t)] } \, ,
\ee
with $S[{\boldsymbol{f}}(t)]$ the finite time action
\be
S[{\bf f}(t)]=\int_{t_{\rm i}}^{t_{\rm f}} \d t\, L({\boldsymbol{f}}(t),{\dot{ \boldsymbol{f}}}(t)) \, .
\ee
We thus obtain
\be
\int \funcd [ {\bf f}_{\rm f}, {\bf f}^-_{\rm i} , {\bf f}^+_{\rm i} ] \int_{{\boldsymbol{f}}^-(t_{\rm i})={\bf f}^-_{\rm i}}^{{\boldsymbol{f}}^-(t_{\rm f})={\bf f}_{\rm f}}  \pathd [{\boldsymbol{f}}^-] \int_{{\boldsymbol{f}}^+(t_{\rm i})={\bf f}^+_{\rm i}}^{{\boldsymbol{f}}^+(t_{\rm f})={\bf f}_{\rm f}} \pathd [{\boldsymbol{f}}^+] \; \mu[{\boldsymbol f}^+,{\boldsymbol f}^-]\, e^{i S[{\boldsymbol f}^+]-i S[{\boldsymbol f}^-]}\langle (-1)^{\hat N_{\psi}}{\bf f}_{\rm i}^+| \hat \rho(t_{\rm i})|{\bf f}_{\rm i}^-\rangle   =1 
\ee
with $\mu[{\boldsymbol f}^+,{\boldsymbol f}^-]=\mu[{\boldsymbol f}^+]\mu[{\boldsymbol f}^-]$ and with the understanding that the integrals over physical fermions are still coherent state integrals.

For all intents and purposes this is formally the same as the equivalent path integral in a non-gauge theory, except for two subtle differences (a) there are no minus signs for the fermionic ghosts (b) the path integrals over $A_0$ ($B$) at all times should be rotated as $A_0 \rightarrow i A_4$ ($B \rightarrow -i B_4$). As already explained, since the rotation should be performed after any complex conjugation, the correct procedure is to rotate both contours in the same way
\be
A_0^{\pm} \rightarrow i A_4^{\pm} \, , \quad B^{\pm} \rightarrow -i B_4^{\pm} \, .
\ee
This ensures that the boundary conditions on the fields at the final time $A_0^{+}(t_{\rm f})=A_0^{-}(t_{\rm f})$ are preserved, i.e.~$A_4^{+}(t_{\rm f})=A_4^{-}(t_{\rm f})$.

We may further abbreviate the full path integral by simply remembering that we should integrate over the fields at all times, including the initial and final times, in each branch with the simple boundary condition (including for physical fermions)
\be
{\boldsymbol f}^+(t_{\rm f})={\boldsymbol f}^-(t_{\rm f}) \, ,
\ee
which is summarised by\footnote{The path integral measure is obtained by integrating over the conjugate momenta. An explicit one-loop expression was derived by DeWitt \cite{DeWitt:1967ub,DeWitt:2003pm} (see also the Appendix of \cite{Kaplanek:2025moq} for its connection to Matthew's theorem). More recently, several works have advanced an all-orders definition based on a novel BRST-like symmetry \cite{crossley2017effective,Liu:2018kfw,Haehl:2015foa,Haehl:2016pec}; see also \cite{Gao:2018bxz} for a related perspective. This BRST symmetry is unrelated to the one considered in the present work; in particular, there is no requirement that physical states lie in its cohomology.}
\be
\int^{{\boldsymbol f}^+(t_{\rm f})={\boldsymbol f}^-(t_{\rm f})} \pathd [{\boldsymbol f}^+,{\boldsymbol f}^-] \; \mu[{\boldsymbol f}^+,{\boldsymbol f}^-]\, e^{i S[{\boldsymbol f}^+(t)]-i S[{\boldsymbol f}^-(t)]} \langle (-1)^{\hat N_{\psi}}{\bf f}_{\rm i}^+| \hat \rho(t_{\rm i})|{\bf f}_{\rm i}^-\rangle   =1 \, ,
\ee
with the understanding that the path integral is taken over fields for $t_{\rm f} \ge t \ge t_{\rm i}$ only, so that in particular the action is a finite time integral.
For a pure initial state, the integrand factorises, and the only coupling between the two branches is at the final time. For a mixed state, they are also coupled at the initial time.

\subsection{In-in Generating Functional}

Associated with every field ${\boldsymbol f}^{\pm}$ we introduce a source ${\boldsymbol J}^{\pm}$ with the same statistics. We then formally define the connected generating function $W[{\boldsymbol J}^{+},{\boldsymbol J}^{-}]$ by the trace operation 
\ba
e^{i W[{\boldsymbol J}^{+},{\boldsymbol J}^{-}]} &=& \Tr[{\cal T} e^{-i \int_{t_{\rm i}}^{t_{\rm f}} \d t \hat H +i \int_{t_{\rm i}}^{t_{\rm f}} \d^4 x \, {\boldsymbol J}^+ \cdot \hat {\boldsymbol f}^+} \hat \rho_{\rm phys}(t_{\rm i}) \bar {\cal T} e^{+i \int_{t_{\rm i}}^{t_{\rm f}} \d t \hat H -i \int_{t_{\rm i}}^{t_{\rm f}} \d^4 x \,  {\boldsymbol J}^- \cdot \hat {\boldsymbol f}^-} ] \\
&=& \Tr[{\cal T} e^{-i \int_{t_{\rm i}}^{t_{\rm f}} \d t \hat H +i \int_{t_{\rm i}}^{t_{\rm f}} \d^4 x \,  {\boldsymbol J}^+ \cdot\hat {\boldsymbol f}_+} e^{\pi \hat Q_G} \hat \rho(t_{\rm i}) \bar {\cal T} e^{+i \int_{t_{\rm i}}^{t_{\rm f}} \d t \hat H -i \int_{t_{\rm i}}^{t_{\rm f}} \d^4 x \,  {\boldsymbol J}^- \cdot\hat {\boldsymbol f}_-} ]  \, ,
\ea
upon use of (\ref{thetrick}) and where $W[\boldsymbol{J},\boldsymbol{J}]=0$ as a consequence of unitarity. This amounts to amending the Hamiltonian density by the source term $\boldsymbol{J}^\pm \cdot \hat{\boldsymbol{f}}_\pm$ but with a different source in each branch.
Repeating the standard path integral derivation for the transition amplitude
\be
\langle {\bf f}_{\rm f}|{\cal T} e^{-i \int_{t_{\rm i}}^{t_{\rm f}} \d t \hat H +i \int_{t_{\rm i}}^{t_{\rm f}} \d^4 x \,  {\boldsymbol J}.\hat {\boldsymbol f}}|{\bf f}_{\rm i} \rangle =  \int_{{\boldsymbol f}(t_{\rm i})={\bf f}_{\rm i}}^{{\boldsymbol f}(t_{\rm f})={\bf f}_{\rm f}} \pathd [{\boldsymbol{f}}] \int \pathd[ \Pi_{\boldsymbol f} ] \, e^{i \int_{t_{\rm i}}^{t_{\rm f}} \d t \( \Pi_{\boldsymbol f}.\partial_t {\boldsymbol f}- H({\boldsymbol f},\Pi_{\boldsymbol f}) +{\boldsymbol J}.{\boldsymbol f}\)} \, ,
\ee
and so by the usual arguments
\begin{align} 
e^{i W[{\boldsymbol J}^{+},{\boldsymbol J}^{-}]} 
&= \int^{{\boldsymbol f}^+(t_{\rm f})={\boldsymbol f}^-(t_{\rm f})} \; \pathd [{\boldsymbol f}^+,{\boldsymbol f}^-,\Pi_{{\boldsymbol f}}^{+},\Pi_{{\boldsymbol f}}^{-}] \; \langle  (-1)^{\hat N_{\psi}}{\bf f}_{\rm i}^+ | \hat \rho(t_{\rm i}) | {\bf f}_{\rm i}^- \rangle \label{GF} \\
&\quad \times \, e^{ i \int_{t_{\rm i}}^{t_{\rm f}} \! \d t \,
\big[ \Pi_{\boldsymbol f}^+ \cdot \partial_t {\boldsymbol f}^+ 
   - H({\boldsymbol f}^+,\Pi_{\boldsymbol f}^+) \big]   
   + i\int \d^4x \, {\boldsymbol J}^+ \cdot {\boldsymbol f}^+  - i \int_{t_{\rm i}}^{t_{\rm f}} \! \d t \,
\big[ \Pi_{\boldsymbol f}^- \cdot \partial_t {\boldsymbol f}^-
   - H({\boldsymbol f}^-,\Pi_{\boldsymbol f}^-) \big]  
   -i \int \d^4x \, {\boldsymbol J}^- \cdot {\boldsymbol f}^- }
 \, . \nonumber
\end{align}
The key novelty of the Schwinger-Keldysh/in-in generating function is the need to include different sources on each branch. This is not achievable with a real modification of the Hamiltonian, unlike the situation for the in-out generating function
where we can legitimately interpret $e^{i W[J]}$ as the amplitude of the probability of persistence of the vacuum in the presence of an external source. Even if we set the sources to be the same on each branch so that they can be interpreted as a modification of the Hamiltonian, we have $W[{\boldsymbol J},{\boldsymbol J}]=0$. Thus, $W[{\boldsymbol J}^{+},{\boldsymbol J}^{-}]$ serves more as a mathematical device to derive in-in correlation functions than its in-out cousin.

\subsection{Covariant Formulation of Generating Function}

Path integrals are naturally derived in the phase-space formulation, but are largely utilised in the configuration space (except for physical fermions), at the price of needing to include a measure factor, mainly because the configuration space action is manifestly Lorentz invariant. In the present context, it is, however, better to work in a mixed representation where $A_i$ are in configuration space, but the pair $A_0/B$ are treated in phase space. This means that the full set of fields is $A_{\mu},B,c,\bar c, \dots$. Maintaining both $A_0$ and $B$ allows us to work with a path integral that is manifestly Lorentz invariant and exhibits the off-shell BRST symmetry.

Starting with the generating functional \eqref{GF}, we integrate out $\Pi_i$, $\Pi_c$, $\Pi_{\bar c}$ and the conjugate momenta for any bosonic matter fields. With a slight abuse of notation, we now regard $B$ as one of the fields, even though it is strictly speaking a conjugate momenta, and thus denote
\be
\int \pathd [{\boldsymbol f}]=\int \pathd [A_{\mu},B,c,\bar c, \Phi,\Phi^\dagger,\psi,\bar \psi] \, .
\ee
Furthermore, it proves convenient to introduce source terms separately for $A_0$ and $B$ and in addition to include source terms for the composite operators $D_{\mu}c^a$ and $\frac{1}{2} f^{abc} c^b c^c$ etc. which enter the BRST transformations. We denote the latter sources by ${\boldsymbol K}$ (known as antifields in the Batalin-Vilkovisky formalism \cite{Batalin:1981jr}). 
With this in mind, the full source term on each branch in the definition of the generating function is taken to be 
\ba
&& {\boldsymbol J} \cdot {\boldsymbol f}+{\boldsymbol K} \cdot \hat s {\boldsymbol f}=J_a^{\mu}A^a_{\mu}+J_B^a B^a+\bar J_c^a c^a +\bar c^a J_{c}^a+J_{\Phi}^\dagger \Phi + \Phi^\dagger J_{\Phi}+ \bar \eta \psi +\bar \psi \eta \\ \nn
&& +K_a^{\mu} D_{\mu}c^a+K_a \(- \frac{g}{2} f^{abc} c^b c^c \) + i g \kappa_{\Phi}^{\dagger} c^a T_{\Phi}^a\Phi-i g \Phi^{\dagger} c^a T_{\Phi}^a \kappa_{\Phi} +i g \kappa_{\psi}^{\dagger} c^a T_{\psi}^a\psi-i g \bar \psi c^a T_{\psi}^a \kappa_{\psi}  \, .
\ea
With this slight abuse of notation the generating functional is now
\begin{align} \label{GF2}
e^{i W[{\boldsymbol J}^{+},{\boldsymbol J}^{-};{\boldsymbol K}^{+},{\boldsymbol K}^{-}]} 
& = \int^{{\boldsymbol f}^+(t_{\rm f})={\boldsymbol f}^-(t_{\rm f})} 
\pathd[{\boldsymbol f}_+,{\boldsymbol f}_-] \;  \mu[{\boldsymbol f}_+,{\boldsymbol f}_-] \; \langle  (-1)^{\hat N_{\psi}}{\bf f}_{\rm i}^+ | \hat \rho(t_{\rm i}) | {\bf f}_{\rm i}^- \rangle\\
& \quad \times \,  
e^{ i S_{\mathrm{NL} }[{\boldsymbol f}^+]
   +i \int_{t_{\rm i}}^{t_{\rm f}} \d^4x \, ({\boldsymbol J}^+ \cdot {\boldsymbol f}^+ + {\boldsymbol K}^+ \cdot \hat s {\boldsymbol f}^+) -i S_{\mathrm{NL} }[{\boldsymbol f}^-]
   -i \int_{t_{\rm i}}^{t_{\rm f}} \d^4x \, ({\boldsymbol J}^- \cdot {\boldsymbol f}^-+ {\boldsymbol K}^-\cdot \hat s {\boldsymbol f}^-) }
 \, . \nn
\end{align}
The only caveat is that the final time boundary conditions in this form are
\be
A_i^+(t_{\rm f})=A_i^-(t_{\rm f}) \,, \quad B^+(t_{\rm f})=B^-(t_{\rm f}) \, , \quad c^+(t_{\rm f})=c^-(t_{\rm f}) \, , \quad \bar c^+(t_{\rm f})=\bar c^-(t_{\rm f}) \, , 
\ee
the initial state is defined in the $B$ representation, and $A_0$ is not integrated over at the final time (since it acts as the momentum conjugate to $B$).
Alternatively we can use
\begin{align} \label{GF2a}
e^{i W[{\boldsymbol J}^{+},{\boldsymbol J}^{-};{\boldsymbol K}^{+},{\boldsymbol K}^{-}]} 
& = \int^{{\boldsymbol f}^+(t_{\rm f})={\boldsymbol f}^-(t_{\rm f})} 
\pathd[{\boldsymbol f}^+,{\boldsymbol f}^-] \;  \mu[{\boldsymbol f}^+,{\boldsymbol f}^-] \;  \langle  (-1)^{\hat N_{\psi}}{\bf f}_{\rm i}^+ | \hat \rho(t_{\rm i}) | {\bf f}_{\rm i}^- \rangle\\
& \quad \times \;  
e^{ i \tilde S[{\boldsymbol f}^+]
   +i \int_{t_{\rm i}}^{t_{\rm f}} \d^4x \, ({\boldsymbol J}^+ \cdot {\boldsymbol f}^+ + {\boldsymbol K}^+ \cdot \hat s {\boldsymbol f}^+) -i \tilde S[{\boldsymbol f}^-]
   -i \int_{t_{\rm i}}^{t_{\rm f}} \d^4x \, ({\boldsymbol J}^- \cdot {\boldsymbol f}^-+ {\boldsymbol K}^-\cdot \hat s {\boldsymbol f}^-) }
 \, . \nn
\end{align}
where we used (\ref{boundary}), and where the final time boundary conditions are 
\be
A_{\mu}^+(t_{\rm f})=A_{\mu}^-(t_{\rm f}) \,  , \quad c^+(t_{\rm f})=c^-(t_{\rm f}) \, , \quad \bar c^+(t_{\rm f})=\bar c^-(t_{\rm f}) \,, 
\ee
with similar conditions for matter fields. Here, the initial state is defined in the $A_0$ representation, and $B$ is not integrated over at the final time (since it acts as the momentum conjugate to $A_0$).

\subsection{BRST Ward-Takahashi-Slavnov-Taylor identities}

To derive the analogue of the Ward-Takahashi-Slavnov-Taylor identities in the BRST
formalism, we study how the in-in generating functional transforms under a change of
integration variables corresponding to an infinitesimal BRST transformation.
Working with the form \eqref{GF2}, for which the action $S_{\mathrm{NL} }$ and the final-time
matching conditions are manifestly BRST invariant, and assuming that the BRST symmetry
is non-anomalous so that the functional measure is invariant, only the source couplings
contribute to the variation.

Recall that on each branch the sources are introduced via
\begin{align}
&{\boldsymbol J}\cdot{\boldsymbol f}+{\boldsymbol K} \cdot \hat s {\boldsymbol f}
= J_a^{\mu}A^a_{\mu}+J_B^a B^a+\bar J_c^a c^a +\bar c^a J_{c}^a
+J_{\Phi}^\dagger \Phi + \Phi^\dagger J_{\Phi}+ \bar J_{\psi} \psi +\bar \psi J_{\psi}  \label{source_term_full} \\
&\quad +K_a^{\mu} D_{\mu}c^a+K_a \( -\frac{g}{2} f^{abc} c^b c^c\)
+ i g K_{\Phi}^{\dagger} c^a T_{\Phi}^a\Phi
-i g \Phi^{\dagger} c^a T_{\Phi}^a K_{\Phi}
+ i g K_{\psi}^{\dagger} c^a T_{\psi}^a\psi
-i g \bar\psi\, c^a T_{\psi}^a K_{\psi} \, , \nonumber
\end{align}
where the composite sources are treated as independent external sources, so that $\hat s K=0$.
The BRST variation of the linear source couplings is
\begin{align}
\hat s \big( {\boldsymbol J} \cdot {\boldsymbol f} +{\boldsymbol K} \cdot \hat s {\boldsymbol f}\big)
&= J_a^{\mu} (\hat s A^a_{\mu})-\bar J_c^a (\hat s c^a)
+(\hat s \bar c^a) J_{c}^a
+J_{\Phi}^\dagger (\hat s \Phi)
+ (\hat s \Phi^\dagger) J_{\Phi}
- \bar J_{\psi} (\hat s \psi )
+(\hat s \bar \psi ) J_{\psi}  \nonumber\\
&= J_a^{\mu} (D_{\mu} c^a)
-\bar J_c^a \Big(-\frac{g}{2} f^{abc} c^b c^c \Big)
+B^a J_{c}^a
+J_{\Phi}^\dagger (i g T^a_{\Phi} c^a \Phi)
+ (-i g \Phi^\dagger c^a T^a_{\Phi}) J_{\Phi} \nonumber\\
& \quad - \bar J_{\psi} (i g T^a_{\psi} c^a \psi )
+(-i g \bar \psi T^a_{\psi} c^a)  J_{\psi} \, .
\end{align}
Performing the diagonal BRST change of variables on each of the two branches,
we obtain
\begin{align}
 & \int^{{\boldsymbol f}^+(t_{\rm f})={\boldsymbol f}^-(t_{\rm f})} 
\pathd[{\boldsymbol f}^+,{\boldsymbol f}^-] \; \mu[{\boldsymbol f}^+,{\boldsymbol f}^-] \, 
\Big( \hat s ( {\boldsymbol J}^+ \cdot {\boldsymbol f}^+)
      -\hat s ( {\boldsymbol J}^- \cdot {\boldsymbol f}^-) \Big) \nonumber\\
& \times
\exp\Bigg\{
 i S_{\mathrm{NL} }[{\boldsymbol f}^+]
+i \int_{t_{\rm i}}^{t_{\rm f}} \d^4x \,
\Big(
{\boldsymbol J}^+ \cdot {\boldsymbol f}^+
+
{\boldsymbol K}^+ \cdot \hat s{\boldsymbol f}^+
\Big)
-i S_{\mathrm{NL} }[{\boldsymbol f}^-]
-i \int_{t_{\rm i}}^{t_{\rm f}} \d^4x \,
\Big(
{\boldsymbol J}^- \cdot {\boldsymbol f}^-
+
{\boldsymbol K}^- \cdot \hat s{\boldsymbol f}^-
\Big)
\Bigg\} \nonumber\\
& \times
\langle  (-1)^{\hat N_{\psi}}{\boldsymbol f}_{\rm i}^+ | \hat \rho(t_{\rm i}) | {\boldsymbol f}_{\rm i}^- \rangle
=0\, .
\end{align}
This implies the functional identity
\begin{equation}
({\hat {\mathcal{W}}}_+ + {\hat {\mathcal{W}}}_-) \,
W[{\boldsymbol J}^+,{\boldsymbol J}^-;{\boldsymbol K}^+,{\boldsymbol K}^-]=0,
\end{equation}
with
\begin{equation}
\hat {\mathcal{W}}_\pm
\equiv \int \d^4x\Bigg[
J_{\pm a}^{\mu}\frac{\delta}{\delta K_{\pm a}^{\mu}}
-\bar J_{\pm c}^{a}\frac{\delta}{\delta K_{\pm a}}
+J_{\pm c}^{a}\frac{\delta}{\delta J_{\pm B}^{a}}
+J_{\pm \Phi}^{\dagger}\frac{\delta}{\delta K_{\pm\Phi}^{\dagger}}
+J_{\pm \Phi}\frac{\delta}{\delta K_{\pm\Phi}}
-\bar J_{\pm\psi}\frac{\delta}{\delta K_{\pm\psi}^{\dagger}}
+J_{\pm\psi}\frac{\delta}{\delta K_{\pm\psi}}
\Bigg] .
\label{W_ST_identity}
\end{equation}

We now perform a Legendre transform with respect to the physical sources 
$(J,\bar J,J_{\psi},\bar J_{\psi})$ only, keeping $K$ fixed. Defining the
classical fields 
\begin{align}
A_{\pm\mu}^a(x) &\equiv \pm \frac{\delta W}{\delta J_{\pm a}^{\mu}(x)}, &
B_\pm^a(x) &\equiv \pm \frac{\delta W}{\delta J_{\pm B}^{a}(x)}, &
c_\pm^a(x) &\equiv \pm \frac{\delta W}{\delta \bar J_{\pm c}^{a}(x)}, &
\bar c_\pm^a(x) &\equiv \mp \frac{\delta W}{\delta J_{\pm c}^{a}(x)}, \\
\Phi_\pm(x) &\equiv \pm \frac{\delta W}{\delta J_{\pm \Phi}^{\dagger}(x)}, &
\Phi_\pm^\dagger(x) &\equiv \pm \frac{\delta W}{\delta J_{\pm \Phi}(x)}, &
\psi_\pm(x) &\equiv \pm \frac{\delta W}{\delta \bar J_{\pm\psi}(x)}, &
\bar\psi_\pm(x) &\equiv \mp \frac{\delta W}{\delta J_{\pm\psi}(x)} \, ,
\end{align}
the effective action is defined by
\begin{align}
& \Gamma[{\boldsymbol f}_+,{\boldsymbol f}_-;{\boldsymbol K}_+,{\boldsymbol K}_-] \label{Gamma_def} \\
&  \quad \equiv W
-\int \d^4x\Big(
J_{+a}^{\mu}A^a_{+\mu}+J_{+B}^a B_+^a+\bar J_{+c}^a c_+^a+\bar c_+^a J_{+c}^a
+J_{+\Phi}^\dagger \Phi_+ +\Phi_+^\dagger J_{+\Phi}
+\bar J_{+\psi}\psi_+ +\bar\psi_+J_{+\psi}
\Big) \nonumber\\
& \qquad
+\int \d^4x\Big(
J_{-a}^{\mu}A^a_{-\mu}+J_{-B}^a B_-^a+\bar J_{-c}^a c_-^a+\bar c_-^a J_{-c}^a
+J_{-\Phi}^\dagger \Phi_- +\Phi_-^\dagger J_{-\Phi}
+\bar J_{-\psi}\psi_- +\bar\psi_-J_{-\psi}
\Big), \nonumber
\end{align}
where the opposite signs reflect the relative signs of the source couplings
in \eqref{GF2}.
This implies
\begin{align}
\frac{\delta \Gamma}{\delta A^a_{+\mu}}=-J^{\mu}_{+a},
\qquad
\frac{\delta \Gamma}{\delta A^a_{-\mu}}=+J^{\mu}_{-a},
\end{align}
and similarly for the remaining fields (up to additional signs for fermions), while
\begin{equation}
\frac{\delta \Gamma}{\delta K_{\pm a}^{\mu}(x)}
=\frac{\delta W}{\delta K_{\pm a}^{\mu}(x)},
\qquad
\frac{\delta \Gamma}{\delta K_{\pm}(x)}
=\frac{\delta W}{\delta K_{\pm}(x)} \, ,
\end{equation}
and similar for other composite sources.
Using these relations, the identity \eqref{W_ST_identity} becomes the
Zinn-Justin or Master equation (not to be confused with open systems master equations)
\begin{equation} \label{Zinn1}
\mathcal{S}(\Gamma)\equiv \mathcal{S}_+(\Gamma)-\mathcal{S}_-(\Gamma)=0,
\end{equation}
where
\begin{align}
\mathcal{S}_\pm(\Gamma)
 \equiv \int_{t_{\rm i}}^{t_{\rm f}} \d^4x\; \bigg[ &
\frac{\delta \Gamma}{\delta A^a_{\pm\mu}}
\frac{\delta \Gamma}{\delta K_{\pm a}^{\mu}}
+\frac{\delta \Gamma}{\delta c_\pm^a}
\frac{\delta \Gamma}{\delta K_{\pm a}} +B_\pm^a \frac{\delta \Gamma}{\delta \bar c_\pm^a}\label{Gamma_ST_identity} \\
& +\frac{\delta \Gamma}{\delta \Phi_\pm}
\frac{\delta \Gamma}{\delta K_{\pm\Phi}^{\dagger}}
+\frac{\delta \Gamma}{\delta \Phi_\pm^\dagger}
\frac{\delta \Gamma}{\delta K_{\pm\Phi}}
+\frac{\delta \Gamma}{\delta \psi_\pm}
\frac{\delta \Gamma}{\delta K_{\pm\psi}^{\dagger}}
+\frac{\delta \Gamma}{\delta \bar\psi_\pm}
\frac{\delta \Gamma}{\delta K_{\pm\psi}}
\bigg] . \nonumber
\end{align}
Structurally, these relations are identical to the usual in-out Zinn-Justin equation. Indeed, we could have derived them simply by accounting for the doubled contour, so that in a more compact notation\footnote{The relative minus sign in \eqref{Zinn1} between the two branches is implicit due to the way we introduce the composite operator sources on each branch $\pm {\boldsymbol K}_{\pm} \cdot \hat s {\boldsymbol f}_{\pm}$.}
\be
\mathcal{S}(\Gamma)=\int_{\mathrm{CTP}} \d^4 x \Bigg[
\frac{\delta \Gamma}{\delta A^a_{\mu}}
\frac{\delta \Gamma}{\delta K_{a}^{\mu}}
+\frac{\delta \Gamma}{\delta c^a}
\frac{\delta \Gamma}{\delta K_{a}}
+B^a \frac{\delta \Gamma}{\delta \bar c^a}
+\frac{\delta \Gamma}{\delta \Phi}
\frac{\delta \Gamma}{\delta K_{\Phi}^{\dagger}}
+\frac{\delta \Gamma}{\delta \Phi^\dagger}
\frac{\delta \Gamma}{\delta K_{\Phi}}
+\frac{\delta \Gamma}{\delta \psi}
\frac{\delta \Gamma}{\delta K_{\psi}^{\dagger}}
+\frac{\delta \Gamma}{\delta \bar\psi}
\frac{\delta \Gamma}{\delta K_{\psi}}
\Bigg]=0 \, .
\ee
However, there is a crucial difference. The usual in-out 1PI and connected generating functions are designed only to give vacuum expectation values of operators in the presence of an external source, with the 1PI effective action understood as an integral over all spacetime. By contrast, the above in-in generating functions apply for an arbitrary physical initial state, including mixed states, and apply to the finite time effective action which governs the motion between $t_{\rm i}$ and $t_{\rm f}$. The non-trivial statement is then that this equation receives no major modification from the initial state.

It is helpful to emphasise this by introducing both primary and composite sources which are localised at the initial time surface
\begin{align}
& \Delta {\boldsymbol J}_{\pm}\cdot{\boldsymbol f}_{\pm}+\Delta {\boldsymbol K}_{\pm}\cdot \hat s {\boldsymbol f}_{\pm} \label{source_term_initial} \\
& \quad = \delta(t-t_{\rm i}) \Big( 
j_{\pm a}^{i}A^a_{\pm i}
+j_{\pm B}^a B_\pm^a
+\bar j_{\pm c}^a c_\pm^a
+\bar c_\pm^a j_{\pm c}^a
+j_{\pm \Phi}^\dagger \Phi_\pm
+\Phi_\pm^\dagger j_{\pm \Phi}
+ \bar \nu_\pm \psi_\pm
+\bar \psi_\pm \nu_\pm  \nonumber\\
& \qquad
+k_{\pm a}^{i} D_{i}c_\pm^a
+k_{\pm a} \Big( -\frac{g}{2} f^{abc} c_\pm^b c_\pm^c \Big)
+ i g \sigma_{\pm\Phi}^{\dagger} c_\pm^a T_{\Phi}^a\Phi_\pm
-i g \Phi_\pm^{\dagger} c_\pm^a T_{\Phi}^a \sigma_{\pm\Phi} \nonumber\\
& \qquad
+ i g \sigma_{\pm\psi}^{\dagger} c_\pm^a T_{\psi}^a\psi_\pm
-i g \bar\psi_\pm\, c_\pm^a T_{\psi}^a \sigma_{\pm\psi}
\Big)\, , \nonumber
\end{align}
which can be used to probe the initial state. The primary sources localised on the initial surface are denoted $(j^a_i,j^a_c,\bar j^a_c, \nu, \bar \nu)$ and the composite $ (k^a_i, k^a, \sigma, \sigma^{\dagger})$.  Consistent with the NL representation, we do not introduce separate sources for $A_0$ on the initial surface so that the BRST transformation acts only on NL representation fields $(A_i^a, B^a,c^a, \bar c^a ,\dots)$ which makes it easier to deal with the initial boundary terms in the action. 

Associated with the new sources introduced at the initial time, we can define an extended 1PI effective action by performing an additional Legendre transformation with respect to the primary sources localised on the initial surface. The full dependence of the extended 1PI effective action is then
\be
\Gamma = \Gamma[A_{\mu}^a,B^a,c^a,\bar c^a,\Phi, \Phi^{\dagger},\psi,\bar \psi, K_{\mu}^a,K^a,K_{\Phi},K_{\Phi}^{\dagger},K_{\psi},K_{\psi}^{\dagger},k_i^a, k^a, \sigma_{\Phi},\sigma_{\Phi}^{\dagger},\sigma_{\psi},\sigma_{\psi}^{\dagger}] \, .
\ee
The Zinn-Justin equation is now extended to
\begin{equation} \label{Zinn2}
\mathcal{S}(\Gamma)\equiv \mathcal{S}_+(\Gamma)-\mathcal{S}_-(\Gamma)+\mathcal{S}_{\rm i +}(\Gamma)-\mathcal{S}_{\rm i -}(\Gamma)=0,
\end{equation}
where $\mathcal{S}_{i,\pm}(\Gamma_{\rm i})$ denotes the contribution from the initial time surface 
\begin{align}
\mathcal{S}_{\mathrm{i}\pm}(\Gamma)
\equiv \int \d^3 {\bf x} \Bigg[ &
\frac{\delta \Gamma}{\delta A^a_{\pm i}}
\frac{\delta \Gamma}{\delta k_{\pm a}^{i}}
+\frac{\delta \Gamma}{\delta c_\pm^a}
\frac{\delta \Gamma}{\delta k_{\pm a}}
+B_\pm^a \frac{\delta \Gamma}{\delta \bar c_\pm^a} \\
&+\frac{\delta \Gamma}{\delta \Phi_\pm}
\frac{\delta \Gamma}{\delta \sigma_{\pm\Phi}^{\dagger}}
+\frac{\delta \Gamma}{\delta \Phi_\pm^\dagger}
\frac{\delta \Gamma}{\delta \sigma_{\pm\Phi}}
+\frac{\delta \Gamma}{\delta \psi_\pm}
\frac{\delta \Gamma}{\delta \sigma_{\pm\psi}^{\dagger}}
+\frac{\delta \Gamma}{\delta \bar\psi_\pm}
\frac{\delta \Gamma}{\delta \sigma_{\pm\psi}}
\Bigg] . \nonumber
\end{align}
The initial contributions in $\Gamma$ are the terms in the effective action which determine the initial conditions, i.e.~they encode the initial state. At tree-level we have
\ba
\Gamma &=& S_{\mathrm{NL} }[{\bf f}_+]-S_{\mathrm{NL} }[{\bf f}_-]  + \sum_{\alpha=\pm}\alpha\int \d^4 x \; \Big(
K_{\alpha a}^{\mu} D_{\mu}c_\alpha^a
+K_{\alpha a} \Big( -\frac{g}{2} f^{abc} c_\alpha^b c_\alpha^c \Big)
+ i g K_{\alpha\Phi}^{\dagger} c_\alpha^a T_{\Phi}^a\Phi_\alpha
-i g \Phi_\alpha^{\dagger} c_\alpha^a T_{\Phi}^a K_{\alpha\Phi} \nn \\
&&+ i g K_{\alpha\psi}^{\dagger} c_\alpha^a T_{\psi}^a\psi_\alpha
-i g \bar\psi_\alpha\, c_\alpha^a T_{\psi}^a K_{\alpha\psi}
\Big) -i  \ln \Big(
\langle (-1)^{\hat N_{\psi}}{\bf f}_{\rm i}^+|
\hat \rho(t_{\rm i})|{\bf f}_{\rm i}^-\rangle
\Big)
+\sum_{\alpha=\pm}\alpha\int \d^3 {\bf x} \; \Big(
k_{\alpha a}^{i} D_{i}c_\alpha^a \nn \\
&&
+ k_{\alpha a} \Big( -\frac{g}{2} f^{abc} c_\alpha^b c_\alpha^c \Big)+ i g \sigma_{\alpha\Phi}^{\dagger} c_\alpha^a T_{\Phi}^a\Phi_\alpha
-i g \Phi_\alpha^{\dagger} c_\alpha^a T_{\Phi}^a \sigma_{\alpha\Phi} 
+ i g \sigma_{\alpha\psi}^{\dagger} c_\alpha^a T_{\psi}^a\psi_\alpha
-i g \bar\psi_\alpha\, c_\alpha^a T_{\psi}^a \sigma_{\alpha\psi}
\Big) \, .
\ea
where the initial density matrix is included as a contribution to the effective action localised on the initial time surface.
At tree-level the composite operators serve to generate the necessary classical BRST transformations of each field, including the fields on the initial time surface, and the Zinn-Justin equation reduces to the statement
\be
\hat s \( S_{\mathrm{NL} }[{\bf f}_+]-S_{\mathrm{NL} }[{\bf f}_-]-i  \ln \( \langle (-1)^{\hat N_{\psi}}{\bf f}_{\rm i}^+| \hat \rho(t_{\rm i})|{\bf f}_{\rm i}^-\rangle \)\)=0 \, .
\ee
We have already established that both terms are separately BRST invariant. The virtue of the Zinn-Justin equation is that it precisely encodes the way in which the BRST transformations are modified at the quantum level (information carried by the composite fields), so that the full 1PI effective action remains BRST invariant. 

It is important to stress that \eqref{Zinn2} is not genuinely different from \eqref{Zinn1}; we are just choosing to make explicit the contribution from the initial-time surface. For instance, if we compute the integral in \eqref{Zinn1} by discretizing time, as is standard in path-integral derivations, then the initial-time contribution will appear precisely as the additional term shown in \eqref{Zinn2}.

\subsection{Breaking of Advanced BRST}
\label{sec:BreakAdv}

If we ignore the boundary terms at the initial and final times, the Schwinger-Keldysh action admits two copies of BRST symmetries associated with the two original gauge symmetries. Thus naively there is a second `advanced' BRST symmetry which acts with opposite sign on the fields on each branch
\begin{equation}
  \begin{split}
    & \hat s_{\rm adv} A^{a\pm}_{\mu} = \pm D_{\mu}[A_{\pm}]c^a_{\pm} \ , \\
    & \hat s_{\rm adv} \bar c_{\pm}^a  = \pm B_{\pm}^a \ ,
  \end{split}
\qquad
  \begin{split}
    & \hat s_{\rm adv} c_{\pm}^a  = \mp \frac{1}{2}g f^{abc} c_{\pm}^b c_{\pm}^c\\
    &\hat s_{\rm adv} B_{\pm}^a = 0
  \end{split}
\end{equation}
If this were the case, then we could derive an analogous Zinn-Justin equation 
\begin{equation}
 \mathcal{S}_+(\Gamma)+\mathcal{S}_-(\Gamma)=0 ? 
\end{equation}
As discussed in \cite{Kaplanek:2025moq} this global symmetry is always broken by the final time boundary conditions, regardless of the initial state. For example, it is already broken in vacuum. This is easy to see by considering an example. Consider the following non-time ordered (Wightman) expectation value evaluated for a physical state
\be
\Tr[ \rho_{\rm phys} \hat s_{\rm adv} \( \bar c_-^a(x) A^b_{+\mu }(y)\)] \, .
\ee
If the advanced symmetry were unbroken, then this should vanish by virtue of being a physical expectation value of a BRST exact object.
Using the above BRST transformations, this would imply
\be \label{cond1}
-\Tr[e^{\pi \hat Q_G}\hat \rho B^a_-(x) A^b_{+\mu}(y)] -\Tr[e^{\pi \hat Q_G}\hat \rho  \bar c^a_-(x) D_{\mu}[A_+] c_+^b(y)] =0?
\ee
However, the retarded symmetry, which we know to be unbroken (at least in perturbation theory), gives
\be
\Tr[ e^{\pi \hat Q_G}\hat \rho \hat s \( \bar c_-^a(x) A^b_{\mu}(y)\)]=0 \, ,
\ee
which implies
\be \label{cond2}
\Tr[e^{\pi \hat Q_G}\hat \rho B^a_-(x) A^b_{+\mu}(y)] -\Tr[ e^{\pi \hat Q_G}\hat \rho  \bar c^a_-(x) D_{\mu}[A_+] c_+^b(y)] =0 \, .
\ee
The conditions \eqref{cond1} and \eqref{cond2} are in clear contradiction unless $\Tr[e^{\pi \hat Q_G}\hat \rho B^a_-(x) A^b_{+\mu}(y)]=0$ and $\Tr[ e^{\pi \hat Q_G}\hat \rho  \bar c^a_-(x) D_{\mu}[A_+] c_+^b(y)] =0$ which in the Feynman-'t Hooft gauge used here is not the case. Hence, both symmetries cannot be true, and an explicit check on the propagators for a free theory $g=0$ in a Gaussian state for which \eqref{cond2} is (written back in operator language)
\be \label{Ward1}
 \frac{\partial}{\partial x^{\nu}} \Tr[ e^{\pi \hat Q_G}\hat \rho \hat A^{a\nu}(x) \hat A_{b\mu}(y) ] + \frac{\partial}{\partial y^{\mu}} \Tr[ e^{\pi \hat Q_G}\hat \rho \hat{\bar c}^a(x) \hat c^b(y) ]
\ee
is in momentum space
\be
-i k_{\nu} \eta^{\nu \mu} 2 \pi \delta(k^2+m^2) ( \theta(k^0)+\mathfrak{n}(k)) \delta_{ab}+i  k_{\mu}  2 \pi \delta(k^2+m^2) ( \theta(k^0)+\mathfrak{n}_c(k))\delta_{ab}=0
\ee
confirms that \eqref{cond2} is the correct condition provided that the ghost occupation number matches $\mathfrak{n}_c(k)=\mathfrak{n}(k)$ which is the case, for example, for a thermal state and more generally any Gaussian spacetime translation invariant state satisfying $[\hat \rho, \hat Q]=0$.

The fundamental reason the advanced symmetry is broken is that it does not have an interpretation at the operator level. There is no operator $\hat Q_{\rm adv}$ that can treat fields in the time ordered product differently than the one in the anti-time ordered product. At the operator level, there is only one conserved BRST charge $\hat Q$, and this acts on the fields the same way, regardless of their time ordering, and it is this charge that is associated with the diagonal BRST symmety transformations $\hat s$.

\subsection{Gaussian States and \texorpdfstring{$i\epsilon$}{i epsilon} prescription}

When the state of the system is Gaussian, the entire effect of the state can be incorporated into the propagators. If further the Gaussian state is space-time translation invariant, then we may freely send $t_{\rm i} \rightarrow - \infty$ and $t_{\rm f} \rightarrow +\infty$, and encode the information in the state into the path integral by a suitable modification of the usual $i \epsilon$ prescription. This is discussed in more detail in \cite{Kaplanek:2025moq} where in particular it is shown that for a photon, when written in Keldysh variables
\ba
&& A^{\mu}_{\pm} = A^{\mu}_{\bf r}\pm \frac{1}{2} A^{\mu}_{\bf a} \, , \quad c_{\pm} =c_{\bf r}\pm \frac{1}{2} c_{\bf a} \,, \quad  \bar c_{\pm} =\bar c_{\bf r}\pm \frac{1}{2} \bar c_{\bf a} \, ,
\ea
then the appropriate $i \epsilon$ contributions to the action to describe a Gaussian space-time translation invariant state is of the form
\ba
\label{eq:photoniepsilon}
S_{i \epsilon} &=& - 2\epsilon \int \d^4 x \, \(  
A_{\mathrm{\bf a}\mu}(x) \partial_t   A^{\mu}_{\mathrm{\bf r}}(x)
-\bar c_{\mathrm{\bf a}}(x) \partial_t c_{\mathrm{\bf r}}(x)
+\bar c_{\mathrm{\bf r}}(x) \partial_t  c_{\mathrm{\bf a}}(x)
\)  \nn \\
&& +i \epsilon \int \d^4 x\int \d^4 y\, \big(
A_{\mathrm{\bf a} \mu}(x) K_A^{\mu\nu}(x,y) A_{\mathrm{\bf a}\nu}(y)
+ c_{\mathrm{\bf a}}(y) K_c(x,y) \bar c_{\mathrm{\bf a}}(x)
\big) \, .
\ea 
This is seen to be BRST invariant under the diagonal (retarded) BRST transformation,
i.e.\ $\hat s S_{i \epsilon}=0$ provided 
\be
 2 \partial_{x\mu} K_A^{\mu\nu}(x,y) + \partial_x^{\nu} K_c(x,y)  =0 \, ,
\ee
which is easily shown to be equivalent to the condition \eqref{Ward1} on solving
for the photon propagators.

Thus, to confirm that the advanced BRST is necessarily broken, we can consider
$\hat s_{\rm adv}S_{i \epsilon}$ and ask whether it vanishes. The advanced BRST symmetry mixes the roles of
the retarded and advanced fields\footnote{Here we are working in the on-shell
formulation where $B= - (\partial_{\mu}A^{\mu})$.}
\begin{equation}
  \begin{split}
    &\hat s_{\rm adv} {A}^{\bf r}_{\mu}  = \frac{1}{2}\partial_{\mu} c_{\bf a} \ ,  \\
    & \hat s_{\rm adv} c_{\bf r} = 0 \ ,  \\
    &\hat s_{\rm adv} \bar c_{\bf r}
 = -\frac{1}{2}\partial_{\mu}A^{\mu}_{\bf a} \ ,
  \end{split}
\qquad
  \begin{split}
    &\hat s_{\rm adv} {A}^{\bf a}_{\mu}
 = 2\partial_{\mu} c_{\bf r} \ , \\
&\hat s_{\rm adv} c_{\bf a}  = 0 \ ,  \\
& \hat s_{\rm adv} \bar c_{\bf a}
 = -2\partial_{\mu}A^{\mu}_{\bf r} \ ,
  \end{split}
\end{equation}
It is then straightforward to see that this breaks the symmetry, since now 
\ba
 \hat s_{\rm adv} S_{i \epsilon}  &=&
- 2\epsilon \int \d^4 x \, \(
  2\partial_{\mu} c_{\bf r}(x) \partial_t   A^{\mu}_{\mathrm{\bf r}}(x)
 +\frac{1}{2}A_{\mathrm{\bf a}\mu}(x) \partial_t   \partial^{\mu}c_{\bf a}(x)
 +2\partial_{\mu}A^{\mu}_{\bf r}(x) \partial_t c_{\mathrm{\bf r}}(x)
 -\frac{1}{2}\partial_{\mu}A^{\mu}_{\bf a}(x) \partial_t  c_{\mathrm{\bf a}}(x)
\)  \nn \\
&& +i \epsilon \int \d^4 x\int \d^4 y\, \big(
4 A_{\mathrm{\bf a} \mu}(x) K_A^{\mu\nu}(x,y) \partial_{\nu} c_{\bf r}(y)
+2 c_{\mathrm{\bf a}}(y) K_c(x,y) \partial_{\mu}A^{\mu}_{\bf r}(x)
\big) \, \nn  \\
&\neq & 0 \, .
\ea
The extra signs contained in the advanced transformations spoil the total
divergence structure which was used successfully for the retarded symmetry.
Since a free photon breaks the advanced BRST symmetry, it is automatically the case that
the non-Abelian theory in its perturbative regime violates the advanced BRST symmetry.

\subsection{Non-perturbative BRST and Gribov Ambiguity}
\label{Gribov}

The BRST construction of non-Abelian gauge theories, at least when approached from the Faddeev-Popov style argument, relies crucially on the
assumption that a gauge condition selects a unique representative on each gauge orbit,
at least locally in field space. As first made clear by Gribov, this assumption fails non-perturbatively \cite{Gribov:1977wm}. This can be seen for example in covariant
gauges, by the emergence of zero-modes for the Faddeev-Popov operator for large values of the field configurations. The boundary in field space at which these emerge is known as the Gribov horizon. In general, there are multiple Gribov copies that satisfy the same gauge condition. These Gribov copies imply that the
standard Faddeev-Popov gauge fixing is not globally well-defined on the space of gauge
orbits \cite{Gribov:1977wm,Vandersickel:2012tz}.
Since the BRST formalism is traditionally derived via the Faddeev-Popov approach, the so-called Gribov ambiguity suggests that there may be a problem with the BRST symmetry non-perturbatively, meaning that it may be explicitly broken \cite{Maggiore:1993wq,Burgio:2009xp,Li:2021wol}. If so, this creates a problem since the BRST symmetry is crucial to our definition of the physical Hilbert space. In particular, the definition of the inner product between two BRST wavefunctionals is defined via a 3 dimensional path integral whose existence relies on the BRST argument.

There is a great deal of literature on approaches to dealing with the Gribov problem, and it is fair to say that there is at present no completely satisfactory resolution. The simplest approach Gribov suggested is to restrict the functional integral to the first Gribov region, where the Faddeev-Popov-DeWitt operator is positive, however, this is inconsistent as Gribov copies also intersect this region. A more sophisticated approach is provided by the Gribov-Zwanziger
framework \cite{Zwanziger:1989mf}, in which the restriction to the Gribov region is implemented softly through a
non-local horizon functional, or equivalently through a local action with auxiliary
fields \cite{Zwanziger:1989mf,Vandersickel:2012tz,Schaden:2014bea}. Whilst resolving the Gribov ambiguity, the usual Gribov-Zwanziger action is in conflict with the standard BRST transformations, once again suggesting that the BRST is broken or at least modified.

The Gribov ambiguity is typically stated as a problem with the continuum theory, and we could take the perspective that lattice gauge theory, which is a meaningful UV regularisation, should resolve the problem. Interestingly, on the lattice the Gribov ambiguity is remediated as the Neuberger problem $0/0$ problem \cite{Neuberger:1986vv,Neuberger:1986xz}. On the lattice, performing an exact BRST gauge fixing for a compact gauge group produces a precise cancellation among all Gribov copies because there are as many $+$ signs for the determinant as $-$ signs. As a result, the gauge-fixed partition function--and therefore any expectation values--assume the indeterminate form $0/0$ \cite{Neuberger:1986vv,Neuberger:1986xz}. 

There is an extensive literature on methods for addressing this issue.
One option is to replace the conventional BRST symmetry with a modified symmetry tailored to the Gribov-Zwanziger action and its extensions
\cite{Capri:2016aqq,Ghiotti:2005ih,Schaden:2015uua}. Alternative strategies are based on averaging over Gribov copies \cite{Serreau:2012cg}.
For discussions of approaches to the lattice Neuberger problem, see \cite{Testa:1998az,Giusti:2001xf,Kalloniatis:2005if,vonSmekal:2012hfd}.

However, another perspective is that the problem does not lie with the BRST symmetry but rather with the usual class of gauge choices made. Indeed, one of the virtues of the BRST framework is that it is possible to choose gauge fixing fermions that do not have an interpretation via the usual Faddeev-Popov procedure because they do not correspond to a gauge for the gauge field alone. Furthermore, in a non-Abelian theory we would expect quantum corrections to the gauge fixing fermion to generate terms which do not have a Faddeev-Popov interpretation, which is one of the reasons why renormalisation is best addressed by the Zinn-Justin equation because it accounts for this possibility.
More recent work has suggested that the BRST procedure remains intact, provided that we choose the right type of gauge fixing term
\cite{Scholtz:1997jp,Rogers:1999zj}.

\section{Open EFTs for Gauge Theories}

\label{Sec4}

In the previous section, we developed the Schwinger-Keldysh path integral for a gauge theory, viewed as a closed system in a generic state, pure or mixed. In this formulation, the only coupling between the two branches of the CTP is at the final time, and for mixed states at the initial time (pure states factorise). The real power of the Schwinger-Keldysh formalism comes from the relatively easy way in which the formalism encodes open systems using the same path integral. Open systems arise from closed ones by tracing out subsets of degrees of freedom. These may be specific fields, or specific regions of space, or high momentum or low momentum modes of a given field or sets of fields. In the present context, there are several natural situations in which we may choose to trace out the environment. One is that there are charged/coloured states that are in some non-vacuum state, i.e.~a plasma, whose influence on the dynamics of the gluons is of interest. 
In this situation, we may be interested in an Open EFT for the gluons that accounts for the plasma without emphasising too much its precise details. 
We may also regard the gluons themselves, over some range of energy scales as the environment. An interesting example is the Hard Thermal Loop (HTL) effective theory, where the high energy gluon modes at finite temperature $k \sim T$ are integrated out in favour of an effective theory of low momenta gauge fields \cite{Braaten:1991gm,braaten1990deducing,Carrington:1997sq,Caron-Huot:2007cma,Rebhan:2008ky}. 
Regardless of the precise division, the SK formalism captures the physics of the open system by a single effective action which describes the interactions between the branches that arise from integrating the environment. This is what is often referred to as the Feynman-Vernon influence functional (IF), whose definition we come to now.

\subsection{Definition of Feynman-Vernon Influence Functional}

In simple situations, it may be possible to infer the IF directly by integrating out/tracing over the environment directly within the path integral. The procedure to do this is universal and applies to non-gauge theories, and so in this section we shall give a generic overview.

Suppose that the full field content can be split into variables `` system and `` environment, 
\be
{\boldsymbol f} \equiv ({\boldsymbol \varphi},{\boldsymbol \chi})  \,
\ee
where ${\boldsymbol \varphi}$ are the system fields and ${\boldsymbol \chi}$ denotes the environment fields to be
integrated out. 
We assume that the full closed system action has the form
\be
S[{\boldsymbol \varphi},{\boldsymbol \chi}] = S_{\rm sys}[{\boldsymbol \varphi}]+ S_{\rm env}[{\boldsymbol \varphi},{\boldsymbol \chi}] \, , 
\ee
with $ S_{\rm env}[{\boldsymbol \varphi},{\boldsymbol \chi}]$ encoded by the action for the environmental degrees of freedom and the coupling between the environment and the system. 
The naive definition of the IF is the in-in analogue of the naive definition of the Wilsonian effective action, obtained by performing the partial trace over the environment of the original closed system\footnote{Here we suppress the final-time matching conditions for notational simplicity.} 
\ba \label{IFdef}
&& \tilde \mu[{\boldsymbol \varphi}_+,{\boldsymbol \varphi}_-] e^{\,i S_{\rm IF}[{\boldsymbol \varphi}_+,{\boldsymbol \varphi}_-]} \rho_{\rm EFT}({\boldsymbol \varphi}^+_{\rm i},{\boldsymbol \varphi}^-_{\rm i}) \\
&& \equiv
\int \pathd[{\boldsymbol \chi}_+,{\boldsymbol \chi}_-]\,
\mu[{\boldsymbol \varphi}_+,{\boldsymbol \varphi}_-,{\boldsymbol \chi}_+,{\boldsymbol \chi}_-]\,
\exp\!\left\{ i S_{\rm env}[{\boldsymbol \chi}_+,{\boldsymbol \varphi}_+]
             - i S_{\rm env}[{\boldsymbol \chi}_-,{\boldsymbol \varphi}_-]\right\}
\,\rho({\boldsymbol \varphi}^+_{\rm i},{\boldsymbol \varphi}^-_{\rm i},{\boldsymbol \chi}_{\rm i}^+,{\boldsymbol \chi}_{\rm i}^-) \, , \nn
\ea
where the density matrix of the EFT is defined by tracing over the environment fields at the initial time\footnote{As usual if we are considering physical fermions the trace should include an additional minus sign which we have suppressed}
\be
\rho_{\rm EFT}({\boldsymbol \varphi}^+_{\rm i},{\boldsymbol \varphi}^-_{\rm i})=\int \funcd [{\boldsymbol \chi}_{\rm i}]\; \rho({\boldsymbol \varphi}^+_{\rm i},{\boldsymbol \varphi}^-_{\rm i},{\boldsymbol \chi}_{\rm i},{\boldsymbol \chi}_{\rm i}) \, .
\ee
This is the field representation of the operator definition of the reduced density matrix for the open system
\be
\hat \rho_{\rm EFT}(t_{\rm i}) = \Tr_{\rm env}[\rho_{\rm phys}(t_{\rm i}) ] \, .
\ee
Notice that the definition (\ref{IFdef}) is slightly more general than the usual definition of the influence functional by allowing initial system-environment entanglement, and isolating $e^{\,i S_{\rm IF}}$ as purely what generates the dynamics of the open system.
In practice, the formula \eqref{IFdef} can lead to ambiguities depending on the regularisation scheme used. For example, in the case of the in-out Wilsonian effective action, where we are integrating out heavy fields in favour of light fields, it is well-known that the use of dimensional regularisation can lead to a misidentification of the contributions from loops for which both heavy and light fields enter. In this context, the safe way to define the Wilsonian effective action is by a matching procedure at the level of the 1PI effective action. We shall follow a similar procedure here to give an unambiguous definition of the IF, or Open EFT action. 

In the in-in formulation, the CTP generating functional of the original closed system with only sources for the ${\boldsymbol \varphi}$ fields turned on is 
\ba
&& e^{i W[{\boldsymbol J}^+,{\boldsymbol J}^-,{\boldsymbol J}_{\boldsymbol \chi}^+=0,{\boldsymbol J}_{\boldsymbol \chi}^-=0]}
= \int \pathd [{\boldsymbol \varphi}_+,{\boldsymbol \varphi}_-,{\boldsymbol \chi}_+,{\boldsymbol \chi}_-]\,\mu[{\boldsymbol \varphi}_+,{\boldsymbol \varphi}_-,{\boldsymbol \chi}_+,{\boldsymbol \chi}_-] \nn\\
&& 
\exp\!\left\{ i S[{\boldsymbol \varphi}_+,{\boldsymbol \chi}_+] - i S[{\boldsymbol \varphi}_-,{\boldsymbol \chi}_-]
+i \int \d^4x \big( {\boldsymbol J}^+\!\cdot\!{\boldsymbol \varphi}_+ - {\boldsymbol J}^-\!\cdot\!{\boldsymbol \varphi}_- \big)\right\}\rho({\boldsymbol \varphi}^+_{\rm i},{\boldsymbol \varphi}^-_{\rm i},{\boldsymbol \chi}_{\rm i}^+,{\boldsymbol \chi}_{\rm i}^-)
\, ,
\label{CTP_without_IF}
\ea
where ${\boldsymbol J}^\pm$ are understood to be the sources for the system fields.
The initial density matrix can always be regarded as a boundary term at the initial time in the action, which in general links the two branches and thereby effectively constitutes part of the influence functional. Following the EFT matching definition of the Wilsonian effective action, we can define the Feynman-Vernon influence functional $S_{\rm IF}$ to be that action which couples the both branches in a generating functional which is defined via a path integral over the ${\boldsymbol \varphi}$ fields alone
\ba
&& e^{i W_{\rm EFT}[{\boldsymbol J}^+,{\boldsymbol J}^-]}
=\int \pathd[{\boldsymbol \varphi}_+,{\boldsymbol \varphi}_-]\,\tilde \mu[{\boldsymbol \varphi}_+,{\boldsymbol \varphi}_-] \nn \\
&& \exp\!\left\{ i S_{\rm sys}[{\boldsymbol \varphi}_+] - i S_{\rm sys}[{\boldsymbol \varphi}_-]
+i \int \d^4x \big( {\boldsymbol J}^+\!\cdot\!{\boldsymbol \varphi}_+ - {\boldsymbol J}^-\!\cdot\!{\boldsymbol \varphi}_- \big)\right\}
\, e^{\,i S_{\rm IF}[{\boldsymbol \varphi}_+,{\boldsymbol \varphi}_-]} \rho_{\rm EFT}({\boldsymbol \varphi}^+_{\rm i},{\boldsymbol \varphi}^-_{\rm i})\, ,
\label{CTP_with_IF}
\ea
The IF is then determined by a matching calculation at the level of the generating functionals
\be \label{matching}
W[{\boldsymbol J}^+,{\boldsymbol J}^-,{\boldsymbol J}_{\boldsymbol \chi}^+=0,{\boldsymbol J}_{\boldsymbol \chi}^-=0]
=
W_{\rm EFT}[ {\boldsymbol J}^+,{\boldsymbol J}^-] \, .
\ee
This is equivalently stated at the level of 1PI effective actions
\be
\Gamma[{\boldsymbol \varphi}_+,{\boldsymbol \varphi}_-,{\boldsymbol \chi}_+,{\boldsymbol \chi}_-] \Big|_{\frac{\delta \Gamma}{\delta {\boldsymbol \chi}_\pm}=0}=\Gamma_{\rm EFT}[{\boldsymbol \varphi}_+,{\boldsymbol \varphi}_-] \, .
\ee
In other words, if we know the 1PI CTP effective action of the original closed system, we can as usual infer that for the Open EFT simply by solving the 1PI equations of motion for the envirnoment fields and substituting back in.

It is important to stress that there is no restriction on the state of the original closed system $\rho({\boldsymbol \varphi}^+_{\rm i},{\boldsymbol \varphi}^-_{\rm i},{\boldsymbol \chi}_{\rm i}^+,{\boldsymbol \chi}_{\rm i}^-)$, in particular, we do not require it to be factorisable, nor do we require it to be Gaussian. This is the virtue of defining the IF via a matching calculation at the level of the generating functions. 
The above matching definition \eqref{matching} is consistent with the naive definition \eqref{IFdef}.

\subsection{Influence Functional with Composite Operators}

More generally we can include composite operator sources for the system fields. Denote ${\boldsymbol O}({\boldsymbol \varphi})$ as a composite operator for the system fields and ${\boldsymbol K}$ its associated source. The generating functional in the original closed theory is 
\ba
&& e^{i W[{\boldsymbol J}^+,{\boldsymbol J}^-,{\boldsymbol J}_{\boldsymbol \chi}^+=0,{\boldsymbol J}_{\boldsymbol \chi}^-=0;{\boldsymbol K}^+,{\boldsymbol K}^-]}
= \int \pathd[{\boldsymbol \varphi}_+,{\boldsymbol \varphi}_-,{\boldsymbol \chi}_+,{\boldsymbol \chi}_-]\,\mu[{\boldsymbol \varphi}_+,{\boldsymbol \varphi}_-,{\boldsymbol \chi}_+,{\boldsymbol \chi}_-] \\
&& 
\times e^{  i S[{\boldsymbol \varphi}_+,{\boldsymbol \chi}_+] - i S[{\boldsymbol \varphi}_-,{\boldsymbol \chi}_-]
+i \int \d^4x \big( {\boldsymbol J}^+\!\cdot\! \; {\boldsymbol \varphi}_+ +{\boldsymbol K}^+\!\cdot\!\; {\boldsymbol O}({\boldsymbol \varphi}_+)   - {\boldsymbol J}^-\!\cdot\! \; {\boldsymbol \varphi}_- -{\boldsymbol K}^-\!\cdot\! \; {\boldsymbol O}({\boldsymbol \varphi}_-)\big) } \rho({\boldsymbol \varphi}^+_{\rm i},{\boldsymbol \varphi}^-_{\rm i},{\boldsymbol \chi}_{\rm i}^+,{\boldsymbol \chi}_{\rm i}^-) \nn 
\, 
\label{CTP_without_IF_composite}
\ea
We then similarly define the generating functional in the EFT as
\ba
&& e^{i W_{\rm EFT}[{\boldsymbol J}^+,{\boldsymbol J}^-;{\boldsymbol K}^+,{\boldsymbol K}^-]}
=\int \pathd[{\boldsymbol \varphi}_+,{\boldsymbol \varphi}_-]\,
\tilde \mu[{\boldsymbol \varphi}_+,{\boldsymbol \varphi}_-]  \\
&&  \times e^{
i S_{\rm sys}[{\boldsymbol \varphi}_+] - i S_{\rm sys}[{\boldsymbol \varphi}_-]
+i \int \d^4x \;  \big( {\boldsymbol J}^+\!\cdot\! \; {\boldsymbol \varphi}_+ 
+{\boldsymbol K}^+\!\cdot\! \; {\boldsymbol O}({\boldsymbol \varphi}_+)
- {\boldsymbol J}^-\!\cdot\! \; {\boldsymbol \varphi}_-
-{\boldsymbol K}^-\!\cdot\! \; {\boldsymbol O}({\boldsymbol \varphi}_-)\big) + i S_{\rm IF}[{\boldsymbol \varphi}_+,{\boldsymbol \varphi}_-;{\boldsymbol K}^+,{\boldsymbol K}^-]}
\, \rho_{\rm EFT}({\boldsymbol \varphi}^+_{\rm i},{\boldsymbol \varphi}^-_{\rm i})\, , \nn
\label{CTP_with_IF_composite}
\ea

The influence functional is in general a function of the composite operator sources and is defined by the matching formula
\be \label{matching_composite}
W[{\boldsymbol J}^+,{\boldsymbol J}^-,{\boldsymbol J}_{\boldsymbol \chi}^+=0,{\boldsymbol J}_{\boldsymbol \chi}^-=0;{\boldsymbol K}^+,{\boldsymbol K}^-]
=
W_{\rm EFT}[ {\boldsymbol J}^+,{\boldsymbol J}^-;{\boldsymbol K}^+,{\boldsymbol K}^-] \, ,
\ee
which is equivalently stated as
\be
\Gamma[{\boldsymbol \varphi}_+,{\boldsymbol \varphi}_-,{\boldsymbol \chi}_+,{\boldsymbol \chi}_-;{\boldsymbol K}^+,{\boldsymbol K}^-] \Big|_{\tfrac{\delta \Gamma}{\delta {\boldsymbol \chi}_\pm}=0}=\Gamma_{\rm EFT}[{\boldsymbol \varphi}_+,{\boldsymbol \varphi}_-;{\boldsymbol K}^+,{\boldsymbol K}^-] \, .
\ee
Note that when solving the 1PI equations of motion, the solution will in general depend on the composite operator sources, and this will in turn generate non-trivial dependence in the influence functional. The usual definition of the IF is of course recovered when the composite sources are removed
\be
S_{\rm IF}[{\boldsymbol \varphi}_+,{\boldsymbol \varphi}_-]=S_{\rm IF}[{\boldsymbol \varphi}_+,{\boldsymbol \varphi}_-;{\boldsymbol K}^+=0,{\boldsymbol K}^-=0] \, .
\ee
It is nevertheless crucial to keep the composite operator sources for a gauge theory to deal with the BRST invariance. 

\subsection{BRST Invariance of Feynman-Vernon Influence Functional}

With the influence functional precisely defined, it is straightforward to translate that into a precise statement on its BRST invariance. The full CTP 1PI effective action, including composite sources for the BRST variations
\be
\Gamma=\Gamma\!\Big[
A_{\mu}^{a\pm},
B^{a\pm},
c^{a\pm},
\bar c^{a\pm},
\Phi^{\pm},
\Phi^{\dagger\pm},
\psi^{\pm},
\bar \psi^{\pm},
K_{\mu}^{a\pm},K^{a \pm},
K_{\Phi}^{\pm},
K_{\Phi}^{\dagger\pm},
K_{\psi}^{\pm},
K_{\psi}^{\dagger\pm}] \, ,
\ee
satisfies the CTP analogue of the Zinn-Justin equation
\be
\mathcal{S}(\Gamma)=\int_{\mathrm{CTP}} \d^4 x \Bigg[
\frac{\delta \Gamma}{\delta A^a_{\mu}}
\frac{\delta \Gamma}{\delta K_{a}^{\mu}}
+\frac{\delta \Gamma}{\delta c^a}
\frac{\delta \Gamma}{\delta K_{a}}
+B^a \frac{\delta \Gamma}{\delta \bar c^a}
+\frac{\delta \Gamma}{\delta \Phi}
\frac{\delta \Gamma}{\delta K_{\Phi}^{\dagger}}
+\frac{\delta \Gamma}{\delta \Phi^\dagger}
\frac{\delta \Gamma}{\delta K_{\Phi}}
+\frac{\delta \Gamma}{\delta \psi}
\frac{\delta \Gamma}{\delta K_{\psi}^{\dagger}}
+\frac{\delta \Gamma}{\delta \bar\psi}
\frac{\delta \Gamma}{\delta K_{\psi}}
\Bigg]=0 \, .
\ee
To define the Open EFT we need to decide which fields are regarded as the environment and which as the system. In the simplest situation we can regard all of the matter fields as part of the environment to give an Open EFT for the gluons. Making this choice, we can set all the composite operator sources for the matter fields ${ \boldsymbol K}$ to zero. We then define the Open EFT by matching the 1PI effective actions
\ba
&& \Gamma_{\rm EFT}\Big[
A_{\mu}^{a\pm},
B^{a\pm},
c^{a\pm},
\bar c^{a\pm},
K_{\mu}^{a\pm},K^{a \pm} \Big] \\
&& \quad = \Gamma\Big[
A_{\mu}^{a\pm},
B^{a\pm},
c^{a\pm},
\bar c^{a\pm},
\Phi^{\pm},
\Phi^{\dagger\pm},
\psi^{\pm},
\bar \psi^{\pm},
K_{\mu}^{a\pm},K^{a \pm},{ \boldsymbol K}=0 \Big] \Big|_{\tfrac{\delta \Gamma}{\delta \Phi^{\pm} }=\tfrac{\delta \Gamma}{\delta \Phi^\dagger{}^{\pm} } =\tfrac{\delta \Gamma}{\delta \psi^{\pm} }=\tfrac{\delta \Gamma}{\delta \bar \psi^{\pm} }=0} \, . \nn
\ea
Since the EFT is defined with all the sources for the environment fields set to zero, those contributions automatically drop out of the Zinn-Justin equation. It is then straightforward to see that the $\Gamma_{\rm EFT}$ satisfies its own CTP Zinn-Justin equation
\be
\mathcal{S}_{\rm EFT}(\Gamma_{\rm EFT})=\int_{\mathrm{CTP}} \d^4 x \Bigg[
\frac{\delta \Gamma_{\rm EFT}}{\delta A^a_{\mu}}
\frac{\delta \Gamma_{\rm EFT}}{\delta K_{a}^{\mu}}
+\frac{\delta \Gamma_{\rm EFT}}{\delta c^a}
\frac{\delta \Gamma_{\rm EFT}}{\delta K_{a}}
+B^a \frac{\delta \Gamma_{\rm EFT}}{\delta \bar c^a}
\Bigg]=0 \, .
\ee
This is the statement that at the quantum level, $\Gamma_{\rm EFT}$ respects BRST invariance. $\Gamma_{\rm EFT}$ is implicitly defining the Open EFT action
\ba \label{MasterEFT}
 S_{\rm EFT}\Big[
A_{\mu}^{a\pm},
B^{a\pm},
c^{a\pm},
\bar c^{a\pm},
K_{\mu}^{a\pm},K^{a \pm} \Big] &=& S_{\mathrm{NL} }[
A_{\mu}^{a+},
B^{a+},
c^{a+},
\bar c^{a+}]-S_{\mathrm{NL} }[
A_{\mu}^{a-},
B^{a-},
c^{a-},
\bar c^{a-}] \nn \\
&& + S_{\rm IF}\Big[
A_{\mu}^{a\pm},
B^{a\pm},
c^{a\pm},
\bar c^{a\pm},
K_{\mu}^{a\pm},K^{a \pm} \Big] \, .
\ea
and also via the necessary boundary conditions at the initial time, the effective density matrix
\be
\rho_{\rm EFT}(
A_{{\rm i},i}^{a\pm},
B_{\rm i}^{a\pm},
c_{\rm i}^{a\pm},
\bar c_{\rm i}^{a\pm})
\ee
via the matching procedure. Since the full quantum effective action for the EFT is BRST invariant in the sense \eqref{MasterEFT} then by considering it at tree-level we infer that
\ba
\hat s \( S_{\rm EFT}\Big[
A_{\mu}^{a\pm},
B^{a\pm},
c^{a\pm},
\bar c^{a\pm},
K_{\mu}^{a\pm},K^{a \pm} \Big]-i \ln \rho_{\rm EFT}(
A_{{\rm i},i}^{a\pm},
B_{\rm i}^{a\pm},
c_{\rm i}^{a\pm},
\bar c_{\rm i}^{a\pm})\)=0 \, .
\ea
This argument easily generalises to situations where the environment/system split is made differently. For example, in the opposite extreme we may choose to regard the gauge fields as the environment and the matter fields as the system, so that the EFT is defined via
\begin{small}
\ba
&& \Gamma_{\rm EFT}\Big[
\Phi^{\pm},
\Phi^{\dagger\pm},
\psi^{\pm},
\bar \psi^{\pm},
K_{\Phi}^{\pm},
K_{\Phi}^{\dagger\pm},
K_{\psi}^{\pm},
K_{\psi}^{\dagger\pm} \Big]= \\
&& \Gamma\Big[
A_{\mu}^{a\pm},
B^{a\pm},
c^{a\pm},
\bar c^{a\pm},
\Phi^{\pm},
\Phi^{\dagger\pm},
\psi^{\pm},
\bar \psi^{\pm},
K_{\mu}^{a\pm}=0,K^{a \pm}=0,
K_{\Phi}^{\pm},
K_{\Phi}^{\dagger\pm},
K_{\psi}^{\pm},
K_{\psi}^{\dagger\pm} \Big] \Big|_{\frac{\delta \Gamma}{\delta A_{\mu}^{a \pm} }=\frac{\delta \Gamma}{\delta B^{a \pm} } =\frac{\delta \Gamma}{\delta c^{a \pm} }=\frac{\delta \Gamma}{\delta \bar c^{a \pm} }=0} \, . \nn
\ea
\end{small}\ignorespaces
The resulting Zinn-Justin equation for the matter fields alone is
\be
\mathcal{S}_{\rm EFT}(\Gamma_{\rm EFT})=\int_{\mathrm{CTP}} \d^4 x \Bigg[
\frac{\delta \Gamma_{\rm EFT}}{\delta \Phi}
\frac{\delta \Gamma_{\rm EFT}}{\delta K_{\Phi}^{\dagger}}
+\frac{\delta \Gamma_{\rm EFT}}{\delta \Phi^\dagger}
\frac{\delta \Gamma_{\rm EFT}}{\delta K_{\Phi}}
+\frac{\delta \Gamma_{\rm EFT}}{\delta \psi}
\frac{\delta \Gamma_{\rm EFT}}{\delta K_{\psi}^{\dagger}}
+\frac{\delta \Gamma_{\rm EFT}}{\delta \bar\psi}
\frac{\delta \Gamma_{\rm EFT}}{\delta K_{\psi}}
\Bigg]=0 \, .
\ee
In practice, this Zinn-Justin equation is not particularly useful since composite operators conjugate to the sources $ K_{\Phi}$ and $K_{\psi}$ do not have a useful interpretation in the matter theory. Furthermore, since these sources have non-zero ghost number, setting ${\bf K}=0$ equivalently sets $\frac{\delta \Gamma_{\rm EFT}}{\delta {\bf K}}=0$ and so the Zinn-Justin equation is satisfied as an identity. In practice, this is telling us that BRST symmetry is realised trivially on the matter EFT if we integrate out the gauge fields as we would expect.

A more interesting example is where  the system/environment split is defined by a separation of length scales. A classic example of this is the Hard Thermal Loop (HTL) approximation where high-momentum modes of the gluon at finite temperature are integrated out to obtain a low energy effective theory for the light modes, but the general procedure can be understood for any split in energy scales. In this case we split every field into a hard and soft modes
\be
{\bf f}= {\bf f}_{\rm soft}+{\bf f}_{\rm hard} \, ,
\ee
with a similar split for composite operator sources. To derive the EFT we set to zero the hard part of the composite operator sources, and fix the hard part of the fields by requiring
\be
\frac{\delta \Gamma}{\delta {\bf f}_{\rm hard}^{\pm}}=0\, .
\ee
We then define the EFT via
\be
\Gamma_{\rm EFT}\[{\bf f}_{\rm soft}^{\pm},
K_{\mu}^{a\pm},K^{a \pm},
K_{\Phi}^{\pm},
K_{\Phi}^{\dagger\pm},
K_{\psi}^{\pm},
K_{\psi}^{\dagger\pm}\] = \Gamma\[{\bf f}_{\rm soft}^{\pm}+{\bf f}_{\rm hard}^{\pm} ,
K_{\mu}^{a\pm},K^{a \pm},
K_{\Phi}^{\pm},
K_{\Phi}^{\dagger\pm},
K_{\psi}^{\pm},
K_{\psi}^{\dagger\pm}\] \Big|_{\frac{\delta \Gamma}{\delta {\bf f}_{\rm hard}^{\pm}}=0}\, ,
\ee
where all the composite sources are regarded as soft. The resulting EFT 1PI effective action will then satisfy the remaining soft part of the Zinn-Justin equation
\ba
&& \mathcal{S}(\Gamma_{\rm EFT})= 
 \int_{\mathrm{CTP}} \d^4 x \Bigg[
\frac{\delta \Gamma_{\rm EFT}}{\delta A^a_{\mu}}
\frac{\delta \Gamma_{\rm EFT}}{\delta K_{a}^{\mu}}
+\frac{\delta \Gamma_{\rm EFT}}{\delta c^a}
\frac{\delta \Gamma_{\rm EFT}}{\delta K_{a}}
+B^a \frac{\delta \Gamma_{\rm EFT}}{\delta \bar c^a}
+\frac{\delta \Gamma_{\rm EFT}}{\delta \Phi}
\frac{\delta \Gamma_{\rm EFT}}{\delta K_{\Phi}^{\dagger}} \nn\\
&& 
+\frac{\delta \Gamma_{\rm EFT}}{\delta \Phi^\dagger}
\frac{\delta \Gamma_{\rm EFT}}{\delta K_{\Phi}}
+\frac{\delta \Gamma_{\rm EFT}}{\delta \psi}
\frac{\delta \Gamma_{\rm EFT}}{\delta K_{\psi}^{\dagger}}
+\frac{\delta \Gamma_{\rm EFT}}{\delta \bar\psi}
\frac{\delta \Gamma_{\rm EFT}}{\delta K_{\psi}}
\Bigg]\Big|_{\rm soft}=0 \, .
\ea
where it is understood that all fields and composite operator sources to be soft. 
In this sense, the soft EFT is guaranteed to be BRST invariant under the diagonal BRST transformation, although as usual the formalism is sufficiently general that it allows for the possibility that the BRST transformations receive corrections. We shall discuss the specific case of the HTL EFT in section~\ref{HTL}.

\subsection{Keldysh BRST Symmetry}

It is common in discussion of open systems written in Keldysh retarded/advanced variables, to refer to the retarded variables as `classical' and the advanced variables as `quantum'. Indeed, the classical equations of motion for an open system are written naturally in terms of the retarded fields, and the stochastic fluctuations (which may or may not be truly quantum) arise from the advanced fields. Consistent with this naming, for many systems it is often sufficient to expand the Schwinger-Keldysh action to low orders in the advanced fields, keeping the dependence on the retarded fields nonlinear. We refer to this as the Keldysh expansion. The optimal truncation point is at quadratic order in the advanced fields, since this includes both the classical equations with dissipation and the noise, characteristic of a stochastic system. Terms of higher order in the advanced fields may be interpreted as non-Gaussian stochastic noise. 

The Open EFT action expanded to quadratic order in the Keldysh expansion (i.e.~quadratic order in the advanced fields) is then in schematic notation
\be \label{Keldyshtruncation}
S_K[{\boldsymbol f}^{\bf r},{\boldsymbol f}^{\bf a}] = \int \d^4 x \,  \frac{\delta S_{\rm EFT}}{\delta {\boldsymbol f}^{\bf a}(x)} \Big|_{{\boldsymbol f}^{\bf a}=0} \cdot {\boldsymbol f}^{\bf a}(x)+ \frac{1}{2} \int \d^4 x \int \d^4 y \, {\boldsymbol f}^{\bf a}(x) \cdot \frac{\delta^2 S_{\rm EFT}}{\delta {\boldsymbol f}^{\bf a}(x) \delta {\boldsymbol f}^{\bf a}(y)}\Big|_{{\boldsymbol f}^{\bf a}=0}  \cdot {\boldsymbol f}^{\bf a}(y) \, .
\ee
When written in terms of retarded and advanced fields, the diagonal BRST transformation becomes
\begin{equation}
\begin{split}
&\hat s A^a_{\mathbf r\mu} = D_{\mu}[A_{\mathbf r}]  c_{\mathbf r}^a  
+ \frac{1}{4} g f^{abc}A_{\mathbf a \mu}^b c_{\mathbf a}^c \,, \\
&\hat s \bar c_{\mathbf r }^a = B_{\mathbf r}^a \,, \\
&\hat s A^a_{\mathbf a\mu} = D_{\mu}[A_{\mathbf r}] c_{\mathbf a}^a 
+ g f^{abc}A_{\mathbf a \mu}^b c_{\mathbf r}^c \,, \\
&\hat s \bar c_{\mathbf a}^a = B_{\mathbf a}^a \,,
\end{split}
\qquad
\begin{split}
&\hat s c^a_{\mathbf r} = - \frac{1}{2}g f^{abc} c^b_{\mathbf r} c^c_{\mathbf r} 
- \frac{1}{8}g f^{abc} c^b_{\mathbf a} c^c_{\mathbf a} \,, \\
&\hat s B^a_{\mathbf r} = 0 \,, \\
&\hat s c_{\mathbf a}^a = - g f^{abc} c_{\mathbf r}^b c_{\mathbf a}^c \,, \\
&\hat s B_{\mathbf a}^a = 0 \, ,
\end{split}
\end{equation}
with similar decompositions for the matter fields.
Crucially we note that the variations of the advanced fields are linear in advanced fields, whereas the variations of retarded fields are quadratic in the advanced fields. If we rescale all advanced fields in the manner ${\boldsymbol f}^{\bf a} \rightarrow \epsilon {\boldsymbol f}^{\bf a}$, keeping the retarded fields unchanged, the full diagonal BRST symmetry admits a contraction to the following `Keldysh' BRST symmetry
\begin{equation}
\begin{split}
&\hat s_K A^a_{\mathbf r \mu} = D_{\mu}[A_{\mathbf r}] c_{\mathbf r}^a \,, \\
&\hat s_K \bar c_{\mathbf r }^a = B_{\mathbf r}^a \,, \\
&\hat s_K A^a_{\mathbf a\mu} = D_{\mu}[A_{\mathbf r}] c_{\mathbf a}^a 
+ g f^{abc}A_{\mathbf a \mu}^b c_{\mathbf r}^c \,, \\
&\hat s_K \bar c_{\mathbf a}^a = B_{\mathbf a}^a \,,
\end{split}
\qquad
\begin{split}
&\hat s_K c^a_{\mathbf r} = - \frac{1}{2}g f^{abc} c^b_{\mathbf r} c^c_{\mathbf r} \,, \\
&\hat s_K B^a_{\mathbf r} = 0 \,, \\
&\hat s_K c_{\mathbf a}^a = - g f^{abc} c_{\mathbf r}^b c_{\mathbf a}^c \,, \\
&\hat s_K B_{\mathbf a}^a = 0 \, .
\end{split}
\end{equation}
We see that the retarded fields now transform independently of the advanced, and the advanced fields remain linear in the advanced, characteristic of an Abelian theory. 
The structure of the BRST transformation is a sum of two independent BRST transformations associated with different symmetries
\be
\hat s_K = \hat s_{\bf r} + \hat s_{\bf a} \, ,
\ee
which are separately nilpotent and anti-commute.
In detail, they are
\begin{equation}
\begin{split}
&\hat s_{\mathbf r} A^a_{\mathbf r \mu} = D_{\mu}[A_{\mathbf r}] c_{\mathbf r}^a \,, \\
& \hat s_{\mathbf r} \bar c_{\mathbf r}^a = B_{\mathbf r}^a \,, \\
& \hat s_{\mathbf r} A^a_{\mathbf a\mu} = g f^{abc}A_{\mathbf a \mu}^b c_{\mathbf r}^c \,, \\
& \hat s_{\mathbf r} \bar c_{\mathbf a}^a = 0 \,,
\end{split}
\qquad
\begin{split}
&\hat s_{\mathbf r} c^a_{\mathbf r} = - \frac{1}{2}g f^{abc} c^b_{\mathbf r} c^c_{\mathbf r} \,, \\
&\hat s_{\mathbf r} B^a_{\mathbf r} = 0 \,, \\
&\hat s_{\mathbf r} c_{\mathbf a}^a = - g f^{abc} c_{\mathbf r}^b c_{\mathbf a}^c \,, \\
& \hat s_{\mathbf r} B_{\mathbf a}^a = 0 \, ,
\end{split}
\end{equation}
and
\begin{equation}
\begin{split}
&\hat s_{\mathbf a} A^a_{\mathbf r\mu} = 0 \,, \\
&\hat s_{\mathbf a} \bar c_{\mathbf r}^a = 0 \,, \\
&\hat s_{\mathbf a} A^a_{\mathbf a\mu} = D_{\mu}[A_{\mathbf r}] c_{\mathbf a}^a \,, \\
&\hat s_{\mathbf a} \bar c_{\mathbf a}^a = B_{\mathbf a}^a \,,
\end{split}
\qquad
\begin{split}
&\hat s_{\mathbf a} c^a_{\mathbf r}= 0 \,, \\
&\hat s_{\mathbf a} B^a_{\mathbf r} = 0 \,, \\
&\hat s_{\mathbf a} c_{\mathbf a}^a = 0 \,, \\
&\hat s_{\mathbf a} B_{\mathbf a}^a = 0 \, .
\end{split}
\end{equation}
The BRST transformations generated by $\hat s_{\bf r}$ are easily seen to be the BRST transformations of the retarded gauge symmetry in which both gauge fields on both branches transform identically
\be
A_{\mu \pm} \rightarrow U(\theta_{\bf r}) A_{\mu \pm} U^{-1}(\theta_{\bf r}) + \frac{i}{g} U(\theta_{\bf r}) \partial_{\mu} U^{-1}(\theta_{\bf r}) \, .
\ee
The BRST transformation associated with $\hat s_{\bf a}$ are the independent BRST transformations of a covariant (in retarded fields) Abelian symmetry which acts only on the advanced gauge fields
\be
 A^a_{\bf a\mu} \rightarrow  A^a_{\bf a\mu} + D_{\mu}[A_{\bf r}] \theta_{\bf a} \, , \quad \quad A^a_{\bf r\mu} \rightarrow  A^a_{\bf r\mu} \, .
\ee
The full BRST symmetry includes terms quadratic in the advanced fields. Because of this, the truncated Keldysh expansion \eqref{Keldyshtruncation} will not be invariant, since the terms linear in advanced fields will mix with terms of cubic order which have been neglected. However, the truncated action \eqref{Keldyshtruncation} is invariant under the Keldysh BRST symmetry precisely because those variations quadratic in advanced fields are removed.

In summary $S_K[{\boldsymbol f}^{\bf r},{\boldsymbol f}^{\bf a}]$ is a BRST formulation of a gauge theory which exhibits a non-Abelian retarded gauge symmetry, and a covariant Abelian gauge symmetry that acts on the advanced fields. 
This pair of symmetries will govern the leading two terms in the Keldysh expansion for any Open EFT for a gauge system. However, when we go beyond quadratic order in the Keldysh expansion, then the Keldysh BRST symmetry will be broken, and we must use the full BRST symmetry. 

One important lesson in performing bottom-up\footnote{See the recent \cite{Abe:2026zlv}.} constructions of EFTs is that it is not sufficient to write down all operators consistent with the retarded/diagonal gauge symmetry only. Although the retarded gauge symmetry is necessary as part of the full BRST symmetry, it is not sufficient. The additional symmetry transformations of the advanced fields are crucial to maintain full BRST invariance, and this provides a significant restriction on the allowed operators in an Open EFT expansion.

\subsection{Hard Thermal Loop EFT as an Open Quantum System}

\label{HTL}

A standard tool in thermal gauge theories is the Hard Thermal Loop EFT
\cite{Braaten:1991gm,Frenkel:1991ts,braaten1990deducing}. At high
temperatures, it is useful to resum the effects of hard thermal modes,
whose loop momenta are of order the temperature, $q\sim T$. This resummation
produces an effective theory for soft modes with momenta $k\sim gT$.
A crude way to derive the HTL action is to compute the hard-loop
contribution to the quark and gluon self-energies, and then covariantise
the result. In Euclidean notation, this gives the non-local effective
action
\be \label{HTL1}
S_{\rm HTL}
=
m_{\psi}^2
\int \d^4x \int \frac{d\Omega_{\bf v}}{4\pi}\,
\bar\psi(x)
\frac{\gamma_\mu v^\mu}{-i\, v\cdot {\cal D}[A_E]}
\psi(x)
+
\frac{1}{2}m_D^2
\int \d^4x \int \frac{d\Omega_{\bf v}}{4\pi}\,
{\rm tr}\left[
F_{\mu\nu}(x)
\frac{v^\nu v^\rho}{\bigl(v\cdot D[A_E]\bigr)^2}
F_{\rho}{}^{\mu}(x)
\right] .
\ee
Here $v_\mu=(-i,\mathbf v)$, with $\mathbf v^2=1$, and
$d\Omega_{\bf v}$ denotes the solid-angle integral over the directions
of $\mathbf v$ and ${\cal D}[A_E]$ and $D[A_E]$ are the fundamental and adjoint covariant derivatives respectively. This expression is gauge invariant since the non-local
operators involving $v\cdot D[A_E]$ and $ v\cdot {\cal D}[A_E]$ transform covariantly.

A more sophisticated way to understand the origin of this EFT is to take
the full path integral and split all fields, including the gluons, into
soft and hard modes,
\be
{\bf f}= {\bf f}_{\rm soft}+{\bf f}_{\rm hard} \, .
\ee
The Wilsonian effective action for the soft fields may then be understood
as arising from the integration over the hard modes with thermal boundary
conditions (periodic or anti-periodic in Euclidean time)
\be
e^{-S_{\rm EFT}[{\bf f}_{\rm soft}]}
=
\int_{\tau \sim \tau + \beta} D[{\bf f}_{\rm hard}]\,
\exp\left\{
-S_{\rm BRST}[{\bf f}_{\rm soft}+{\bf f}_{\rm hard}]
\right\} .
\ee
The resulting Euclidean effective action has the schematic form
\be
S_{\rm EFT}
=
S_{\rm soft}
+
S_{\rm HTL}
+
\cdots ,
\ee
where the ellipsis denotes the terms subleading in the soft expansion. In
dimensional regularisation, this split is implemented not by an explicit
momentum cutoff, but by evaluating the loop integrals in the method-of-regions
sense: the loop momenta are taken to be hard, $q\sim T$, while all external
momenta carried by the soft fields satisfy $p\ll T$. The resulting hard-region
integrals are then expanded in powers of $p/q$.

While the effective action contribution \eqref{HTL1} is fine for Euclidean calculations, it is essentially meaningless for Lorentzian ones since the non-local operators need to be defined with a particular time ordering. Certain Lorentzian information can be obtained by analytic continuation \cite{Evans:1991ky,Frenkel:1991ts}, but ultimately the Euclidean theory is insufficient to fix the Lorentzian at finite temperature. The resolution is to recognise that because we are at finite temperature, the Lorentzian analogue of the HTL EFT is necessarily an open system, and as such should be described by an effective action with doubled fields. This was nicely made explicit in \cite{Caron-Huot:2007cma} where the Schwinger-Keldysh HTL effective action was shown to be (ignoring matter and ghost contributions)
\ba \label{HTL2}
S_{\rm HTL}^{\rm SK}[A_{\bf r},A_{\bf a}]
=
m_D^2
\int \frac{d\Omega_{\bf v}}{4\pi}
\int \d^4x\,
v^\mu A_{\mu {\bf a}}^a(x)
\left[
\frac{1}{v\cdot D[A_{\bf r}]_{\rm ret}}
\,v^i E_i[A_{\bf r}]
\right]^a(x)
\;+\; \nonumber \\
\frac{iT m_D^2}{2}
\int \frac{d\Omega_{\bf v}}{4\pi}
\int \d^4x\,
v^\mu A_{\mu {\bf a}}^a(x)
\left[
\left(
\frac{1}{v\cdot D[A_{\bf r}]_{\rm ret}}
-
\frac{1}{v\cdot D[A_{\bf r}]_{\rm adv}}
\right)
v^\nu A_{\nu {\bf a}}
\right]^a(x)
\, .
\ea
with $E_i[A_{\bf r}] = F_{i0}[A_{\bf r}]$ or equivalently 
\be
\begin{aligned}
S_{\rm HTL}^{\rm SK}[A_{\bf r},A_{\bf a}]
={}&
m_D^2
\int \frac{d\Omega_{\bf v}}{4\pi}
\int \d^4x
\int_0^\infty d\tau\,
v^\mu A_{\mu {\bf a}}^a(x)\,
U^{ab}(x,x-v\tau)[A_{\bf r}]\,
v^i E_i^b[A_{\bf r}](x-v\tau)
\\
&+
\frac{iT m_D^2}{2}
\int \frac{d\Omega_{\bf v}}{4\pi}
\int \d^4x
\int_{-\infty}^{\infty} d\tau\,
v^\mu A_{\mu {\bf a}}^a(x)\,
U^{ab}(x,x-v\tau)[A_{\bf r}]\,
v^\nu A_{\nu {\bf a}}^b(x-v\tau)
\, 
\end{aligned}
\ee
with $U$ denoting the Wilson line in the adjoint representation along a straight line path from $x$ to $y$
\be
U(x,y) = {\cal P}e^{i \int_{{\cal C}} \d z^{\mu} g T^a A^a(z)} \, .
\ee
We have used the explicit representation of the advanced and retarded Green's functions in terms of adjoint representation Wilson lines
\ba
&& \left(\frac{1}{v\cdot D_x[A_{\bf r}]}\right)^{ab}_{\rm ret}
\delta^{(4)}(x-y)
=
\int_{0}^{\infty} d\tau\,
U^{ab}[A_{\bf r}](x,x-v\tau)\,
\delta^{(4)}(x-v\tau-y)\\
&& \left(\frac{1}{v\cdot D_x[A_{\bf r}]}\right)^{ab}_{\rm adv}
\delta^{(4)}(x-y)
=
-\int_{-\infty}^{0} d\tau\,
U^{ab}[A_{\bf r}](x,x-v\tau)\,
\delta^{(4)}(x-v\tau-y)
\ea
By construction, the above effective action has been truncated at second order in the Keldysh expansion, i.e.~in the advanced fields. As such, it must be invariant under the Keldysh BRST symmetry. In practice, this means it must be separately invariant under the retarded gauge symmetry, in which the advanced field transforms covariantly, and the covariant linear gauge transformation of the advanced fields. 

The expression \eqref{HTL2} is manifestly invariant under the retarded gauge transformations since it is built entirely out of covariant objects. To check invariance under the covariant linear gauge transformation of the advanced fields we have
\be
\delta_{\bf a} A_{\mu{\bf a}}^a
=
\left(D_\mu[A_{\bf r}] \lambda_{\bf a}\right)^a,
\qquad
\delta_{\bf a} A_{\mu{\bf r}}^a=0 ,
\ee
the first term in \eqref{HTL2} varies as
\be
\begin{aligned}
\delta_{\bf a} S_{\rm HTL}^{(1)}
&=
m_D^2
\int \frac{d\Omega_{\bf v}}{4\pi}
\int \d^4x\,
\left(v\cdot D[A_{\bf r}]\lambda_{\bf a}\right)^a
\left[
\frac{1}{v\cdot D[A_{\bf r}]_{\rm ret}}
\,v^i E_i[A_{\bf r}]
\right]^a
\\
&=
- m_D^2
\int \frac{d\Omega_{\bf v}}{4\pi}
\int \d^4x\,
\lambda_{\bf a}^a
\left[
v\cdot D[A_{\bf r}]
\frac{1}{v\cdot D[A_{\bf r}]_{\rm ret}}
\,v^i E_i[A_{\bf r}]
\right]^a
\\
&=
- m_D^2
\int \d^4x\,
\lambda_{\bf a}^a(x)
\int \frac{d\Omega_{\bf v}}{4\pi}\,
v^i E_i^a[A_{\bf r}](x)
=0 ,
\end{aligned}
\ee
where we have used the fact that the angular average of a single
power of the velocity vanishes
\be
\int \frac{d\Omega_{\bf v}}{4\pi}\, v^i =0 .
\ee
For the term quadratic in $A_{\bf a}$, it is useful to define the kernel
\ba \label{Hdef}
H^{ab}[A_{\bf r}](x,y)
&\equiv&
\left(\frac{1}{v\cdot D_x[A_{\bf r}]}\right)^{ab}_{\rm ret}
\delta^{(4)}(x-y)
-
\left(\frac{1}{v\cdot D_x[A_{\bf r}]}\right)^{ab}_{\rm adv}
\delta^{(4)}(x-y)  \\
&& =\int_{-\infty}^{\infty} d\tau\,
U^{ab}[A_{\bf r}](x,x-v\tau)\,
\delta^{(4)}(x-v\tau-y) \, .
\ea
Since $H[A_{\bf r}](x,y)$ is the difference of the retarded and advanced inverses of the
same first-order covariant operator, the contact terms cancel when
$v\cdot D$ acts on it. Thus, as a distribution,
\be \label{rel1}
\left(v\cdot D_x[A_{\bf r}]\right)^{ac}
H^{cb}[A_{\bf r}](x,y)=0 .
\ee
Since a Wilson line in the adjoint representation obeys
\be
U(y,x)=U(x,y)^{-1}=U(x,y)^T
\ee
we have $U^{ab}(x,y)=U^{ba}(y,x)$
and hence
\be
H^{ab}[A_{\bf r}](x,y)=H^{ba}[A_{\bf r}](y,x) \, ,
\ee
which can be derived by redefining $\tau \rightarrow - \tau $ in \eqref{Hdef}.
With this notation, the term quadratic in $A_{\bf a}$ is
\be
S_{\rm HTL}^{(2)}
=
\frac{iT m_D^2}{2}
\int \frac{\d\Omega_{\bf v}}{4\pi}
\int \d^4x \int \d^4y\,
v^\mu A_{\mu{\bf a}}^a(x)\,
H^{ab}[A_{\bf r}](x,y)\,
v^\nu A_{\nu{\bf a}}^b(y) .
\ee
Under the covariant linear gauge transformation of the advanced field,
\be
\delta_{\bf a} A_{\mu{\bf a}}^a
=
\left(D_\mu[A_{\bf r}]\lambda_{\bf a}\right)^a ,
\qquad
\delta_{\bf a} A_{\mu{\bf r}}^a=0 ,
\ee
its variation is
\be
\begin{aligned}
\delta_{\bf a} S_{\rm HTL}^{(2)}
=
\frac{iT m_D^2}{2}
\int \frac{\d\Omega_{\bf v}}{4\pi}
\int \d^4 x \int \d^4 y\,
\Bigg\{
&
\left(v\cdot D_x[A_{\bf r}]\lambda_{\bf a}\right)^a(x)\,
H^{ab}[A_{\bf r}](x,y)\,
v^\nu A_{\nu{\bf a}}^b(y)
\\
&+
v^\mu A_{\mu{\bf a}}^a(x)\,
H^{ab}[A_{\bf r}](x,y)\,
\left(v\cdot D_y[A_{\bf r}]\lambda_{\bf a}\right)^b(y)
\Bigg\},
\end{aligned}
\ee
which by symmetry gives
\be
\delta_{\bf a} S_{\rm HTL}^{(2)}
=
iT m_D^2
\int \frac{\d\Omega_{\bf v}}{4\pi}
\int \d^4 x \int \d^4 y\,
\Bigg\{
\left(v\cdot D_x[A_{\bf r}]\lambda_{\bf a}\right)^a(x)\,
H^{ab}[A_{\bf r}](x,y)\,
v^\nu A_{\nu{\bf a}}^b(y) \Bigg\}
\ee
Now integrating by parts to put the covariant derivative in the first term on
$\lambda_{\bf a}(x)$ onto the $x$ endpoint of $H$ then using \eqref{rel1} gives
\be
\delta_{\bf a} S_{\rm HTL}^{(2)}=0 .
\ee
Thus both terms in \eqref{HTL2} are invariant under the covariant linear
gauge transformation of the advanced fields,
\be
\delta_{\bf a} S_{\rm HTL}^{\rm SK}=0 .
\ee
Together with invariance under the retarded gauge symmetry, this ensures
that the HTL action, evaluated to quadratic order in the advanced fields,
admits the appropriate Keldysh BRST symmetry,
\be
\hat s_K S_{\rm HTL}^{\rm SK}=0 \, .
\ee

The HTL EFT \eqref{HTL2} serves as the poster child of an open quantum EFT for a gauge theory. It incorporates both dissipative and stochastic (noise) contributions, which -- consistent with the absence of symmetry breaking -- are transmitted through non-local propagators whose gauge structure is determined by Wilson lines \cite{Kaplanek:2025moq}. Even though the theory is non-local, causality is preserved by the boundary conditions built into the kernels in the definition of the Open EFT. The truncated EFT remains invariant under the exact Keldysh BRST symmetry. Beyond quadratic order in the advanced fields, the full BRST of a truncated EFT at a given order in the Keldysh expansion would only be realised up to that order. Although the explicit effective action \eqref{HTL2} was derived for a purely thermal state, the general procedure will easily generalise to an arbitrary mixed state, given a suitable generalisation of the meaning of hard thermal, or equivalently of a hard/soft split. 

\subsection{Open EFTs with SSB}

The HTL EFT is an example of an open quantum gauge system where there is no spontaneous symmetry breaking (SSB). Indeed, the only symmetry actually broken is the global Lorentz invariance, specifically boosts, due to the temperature defining a preferred frame.
More generally, we can consider systems in a Higgs phase, where the gauge symmetry is spontaneously broken. Ironically, the treatment of gauge theories in a broken phase is much easier since there is a preferred gauge, unitary gauge, in which all physics may be defined. We shall for simplicity consider the case where all of the gauge symmetries are spontaneously broken. 

In the SK formalism, there are doubled gauge fields $A_{\pm}$. Before any consideration of quantisation or boundary conditions, there are naively two separate gauge redundancies. In standard notation, the gauge transformations can be written
\be
A_{\mu \pm} \rightarrow A_{\mu \pm}'
=
U(\theta_{\pm}) A_{\mu \pm} U^{-1}(\theta_\pm)
+
\frac{i}{g} U(\theta_\pm) \partial_{\mu} U^{-1}(\theta_\pm) \, .
\ee
When all the symmetries are broken, there is a preferred gauge, referred to as unitary gauge. Given its importance, we denote the gauge field in unitary gauge as ${\cal A}_{\pm}$. By definition, since unitary gauge is fixed, ${\cal A}_{\pm}$ is gauge invariant. However, we can write it in the manner of a gauge theory using the \stu procedure, which amounts to performing a gauge transformation and promoting the gauge parameter to an adjoint-valued field, usually denoted $\pi^a_{\pm}$. This is achieved via the {\it field redefinition}, (not a gauge transformation!)
\be
{\cal A}_{\mu \pm}
=
U(\pi_{\pm}) A_{\mu \pm} U^{-1}(\pi_\pm)
+
\frac{i}{g} U(\pi_\pm) \partial_{\mu} U^{-1}(\pi_\pm) \, .
\ee
Since ${\cal A}_{\mu \pm}$ is gauge invariant, consistency requires 
\be
U(\pi'_{\pm}) A'_{\mu \pm} U^{-1}(\pi'_\pm)
+
\frac{i}{g} U(\pi'_\pm) \partial_{\mu} U^{-1}(\pi'_\pm)
=
U(\pi_{\pm}) A_{\mu \pm} U^{-1}(\pi_\pm)
+
\frac{i}{g} U(\pi_\pm) \partial_{\mu} U^{-1}(\pi_\pm) \, ,
\ee
which is solved by requiring the \stu fields transform as
\be
U(\pi'_{\pm})
=
U(\pi_{\pm})U^{-1}(\theta_{\pm}) \, .
\ee
In a non-Abelian theory this is a somewhat complicated transformation, but in the Abelian case it is just the statement
\be
\pi'^a_{\pm}
=
\pi^a_{\pm}-\theta^a_{\pm} \, .
\ee

The simplest way to construct a SK effective action for a Higgsed gauge field is then to construct the action in unitary gauge, where there is no gauge redundancy. This means we can follow the rules for ordinary scalar theories. Thus, the effective action will be built as
\be
S_{\mathrm{CTP,EFT}}
=
\mathsf{S}[{\cal A}^a_{+}(x),{\cal A}^a_-(x),\partial_{\mu}] \, ,
\ee
with $\mathsf{S}$ being generically a spacetime integral over non-local scalar functions of the ingredients subject to the usual non-equilibrium constraints
\ba
&&\mathsf{S}[{\cal A}^a_{+},{\cal A}^a_+,\partial_{\mu}]
=
0 \, , \\
&& \mathsf{S}[{\cal A}^a_{+},{\cal A}^a_-,\partial_{\mu}]^*
= 
-\mathsf{S}[{\cal A}^a_{-},{\cal A}^a_+,\partial_{\mu}] \, ,
\\
&& \left.
\frac{\delta^2 \mathsf{S}}{\delta {\cal A}^a_{{\bf a}\mu}(x)\delta {\cal A}^b_{{\bf r}\nu}(y)}
\right|_{{\cal A}_{\bf a}=0} =  0
\quad
\text{unless }
x^0 \ge y^0 \, , \, (x-y)^2 \le 0 \, .
\ea
Here we have defined, as usual,
\be
{\cal A}^a_{\mu \pm}
=
{\cal A}^a_{\mu {\bf r}}
\pm
\frac{1}{2}{\cal A}^a_{\mu {\bf a}} \, .
\ee
Note that since ${\cal A}_{\pm}$ are gauge invariant we are not required to use covariant derivatives in building combinations. The first two non-equilibrium constraints follow from unitarity, and the last is causality. In addition, the validity of perturbation theory requires that the quadratic action in the Keldysh expansion has a positive semi-definite imaginary part. Schematically,
\be
S_{\rm CTP,EFT}^{(2)}
=
\int \d^dx \,
{\cal A}^a_{{\bf a}\mu}(x)
E^{a\mu}[{\cal A}_{\bf r}](x)
+
\frac{i}{2}
\int \d^dx \int \d^dy \,
{\cal A}^a_{{\bf a}\mu}(x)
N^{ab\,\mu\nu}(x,y;{\cal A}_{\bf r})
{\cal A}^b_{{\bf a}\nu}(y)
+\cdots ,
\ee
with
\be \label{noisepositivity}
\int \d^dx \int \d^dy \,
f^a_{\mu}(x)
N^{ab\,\mu\nu}(x,y;{\cal A}_{\bf r})
f^b_{\nu}(y)
\ge 0
\ee
for arbitrary test functions $f^a_\mu$. Equivalently,
\be
\Im S_{\rm CTP,EFT}^{(2)}
\ge
0 \, .
\ee
It would be incorrect, however, to assume a stronger positivity statement for the whole effective action, since convergence of the full effective action is determined only by the highest powers in fields, and in an EFT the expansion does not terminate. 

Although correct, the above formalism is cumbersome. It is better to rewrite things in a manner in which the retarded gauge symmetry is manifest, since this is the symmetry of the classical equations of motion. This is particularly useful if the effective action is only computed to quadratic order in the Keldysh expansion: at that order the retarded gauge symmetry remains manifest, while the advanced gauge symmetry acts linearly and is naturally associated with an Abelianised BRST structure. This is easily achieved by performing a different field redefinition for which the retarded gauge symmetry is made explicit. Specifically, in the Keldysh limit we define 
\ba
{\cal A}_{\mu {\bf r}}
&=&
\frac{1}{2}
\left(
{\cal A}_{\mu +}+{\cal A}_{\mu -}
\right)
=
U(\pi_{\bf r}) A_{\mu {\bf r}} U^{-1}(\pi_{\bf r})
+
\frac{i}{g} U(\pi_{\bf r}) \partial_{\mu} U^{-1}(\pi_{\bf r}) \, ,
\\
{\cal A}_{\mu {\bf a}}
&=&
{\cal A}_{\mu +}-{\cal A}_{\mu -}
=
U(\pi_{\bf r})
\left(
A_{\mu {\bf a}}
+
D_{\mu}[A_{\bf r}] \pi_{\bf a}
\right)
U^{-1}(\pi_{\bf r}) \, .
\ea
The retarded and advanced \stu fields are defined via these relations. In the Keldysh limit, the retarded gauge field $A_{\mu {\bf r}}$ transforms only via the retarded gauge transformation,
\be
A_{\mu {\bf r}}'
=
U(\theta_{\bf r}) A_{\mu {\bf r}} U^{-1}(\theta_{\bf r})
+
\frac{i}{g}
U(\theta_{\bf r}) \partial_{\mu} U^{-1}(\theta_{\bf r}) \, .
\ee
Gauge invariance of ${\cal A}_{\mu {\bf r}}$ then implies that the retarded \stu field must transform as
\be
U(\pi_{\bf r}')
=
U(\pi_{\bf r})U^{-1}(\theta_{\bf r}) \, .
\ee
Similarly, in the Keldysh limit the advanced gauge field transforms covariantly in the adjoint representation under the retarded gauge transformations and shifts in a covariant linear manner under the advanced gauge transformations,
\be
A_{\mu {\bf a}}'
=
U(\theta_{\bf r})
\left(
A_{\mu {\bf a}}
+
D_{\mu}[A_{\bf r}] \theta_{\bf a}
\right)
U^{-1}(\theta_{\bf r}) \, .
\ee
Gauge invariance of ${\cal A}_{\mu {\bf a}}$ requires that the advanced \stu fields transform covariantly in the adjoint representation under retarded gauge transformations and linearly under advanced gauge transformations,
\be
\pi_{\bf a}'
=
U(\theta_{\bf r})
\left(
\pi_{\bf a}
-
\theta_{\bf a}
\right)
U^{-1}(\theta_{\bf r}) \, .
\ee
Stated differently, the following extended covariant derivative of $\pi_{\bf a}$ transforms covariantly under retarded gauge transformations:
\be
D_{\mu}[A_{\bf r}']\pi_{\bf a}'
+
A_{\mu {\bf a}}'
=
U(\theta_{\bf r})
\left(
D_{\mu}[A_{\bf r}]\pi_{\bf a}
+
A_{\mu {\bf a}}
\right)
U^{-1}(\theta_{\bf r}) \, .
\ee
We can also define an analogous covariant derivative of the retarded \stu fields,
\be
\tilde D_{\mu}[A_{\bf r}] \pi_{\bf r}
\equiv
A_{\mu{\bf r}}
+
\frac{i}{g}
\left(
\partial_{\mu} U^{-1}(\pi_{\bf r})
\right)
U(\pi_{\bf r}) \, ,
\ee
so that
\be
\tilde D_{\mu}[A_{\bf r}'] \pi'_{\bf r}
=
U(\theta_{\bf r})
\left(
\tilde D_{\mu}[A_{\bf r}] \pi_{\bf r}
\right)
U^{-1}(\theta_{\bf r})  \, .
\ee
Putting this together, the following objects transform covariantly under retarded gauge transformations in the adjoint representation and are invariant under the linearised advanced gauge transformations:
\be
\left(
F_{\mu\nu}[A_{\bf r}],
\tilde D_{\mu}[A_{\bf r}] \pi_{\bf r},
A_{\mu{\bf a}}+D_{\mu}[A_{\bf r}]\pi_{\bf a}
\right)
\ee
and all retarded covariant derivatives thereof. 

It is now straightforward to give a general expression for the BRST gauge-fixed SK EFT, including gauge-fixing terms, in the Keldysh limit, i.e.~to quadratic order in advanced fields:
\ba
S_{K,EFT}
&=&
\int \d^dx \,
\left(
A^a_{{\bf a}\mu}(x)
+
D_{\mu}[A_{\bf r}]\pi_{\bf a}^a(x)
\right)
E^{a\mu}(x;A_{\bf r},\tilde D[A_{\bf r}] \pi_{\bf r})
\nn \\
&&
+
\frac{i}{2}
\int \d^dx  \int \d^dy \,
\left(
A^a_{{\bf a}\mu}(x)
+
D_{\mu}[A_{\bf r}]\pi_{\bf a}^a(x)
\right)
N^{ab\,\mu\nu}
\left(
x,y;
A_{\bf r},
\tilde D[A_{\bf r}] \pi_{\bf r}
\right)
\nn \\
&&
\hspace{3cm}
\times
\left(
A^b_{{\bf a}\nu}(y)
+
D_{\nu}[A_{\bf r}]\pi_{\bf a}^b(y)
\right)
\nn \\
&&
+
\hat s_K
G_K[
A_{\bf r},A_{\bf a},
\pi_{\bf r},\pi_{\bf a},
c_{\bf r},c_{\bf a},
\bar c_{\bf r},\bar c_{\bf a},
B_{\bf r},B_{\bf a}
] \, .
\ea
Keldysh BRST invariance is guaranteed by the requirement that the classical equation of motion
$E^{a\mu}(x;A_{\bf r},\tilde D[A_{\bf r}] \pi_{\bf r})$
transforms covariantly under retarded gauge transformations at the point $x$, and that the noise kernel
$N^{ab\,\mu\nu}(x,y;A_{\bf r},\tilde D[A_{\bf r}]\pi_{\bf r})$
transforms covariantly in the adjoint at the point $x$ associated with index $a$, and covariantly in the adjoint at the point $y$ associated with index $b$. In other words, under a retarded gauge transformation,
\ba
&&
\delta_{\bf r} E^{a\mu}(x)
=
-g f^{acd} \theta_{\bf r}^c(x) E^{d\mu}(x) \, ,
\\
&&
\delta_{\bf r} N^{ab\,\mu\nu}(x,y)
=
-g f^{acd} \theta_{\bf r}^c(x) N^{db\,\mu\nu}(x,y)
-g f^{bcd} \theta_{\bf r}^c(y) N^{ad\,\mu\nu}(x,y) \, .
\ea
The definition of the noise kernel is such that it can be assumed symmetric in the sense 
\be
N^{ab\,\mu\nu}(x,y)
=
N^{ba\,\nu\mu}(y,x) \, .
\ee
Note that because we are considering the Higgsed case we do not require either $E^{a\mu}(x)$ or $N^{ab\,\mu\nu}(x,y)$ to be covariantly conserved. Positivity is required in the sense \eqref{noisepositivity}. Causality requires
\ba
\frac{\delta E^{a \mu}(x)}{\delta A_{\bf r}^{\nu b}(y)} =0 \, , \quad
\text{unless }
x^0 \ge y^0 \, , \, (x-y)^2 \le 0 \, . \\
\frac{\delta E^{a \mu}(x)}{\delta \pi_{\bf r}^b(y)} =0 \,, \quad
\text{unless }
x^0 \ge y^0 \, , \, (x-y)^2 \le 0 \, .
\ea
The complete set of Keldysh BRST transformations, including those for the \stu fields, are now
\begin{equation}
\begin{split}
&\hat s_K A^a_{\mathbf r\mu}
= D_{\mu}[A_{\mathbf r}] c_{\mathbf r}^a \,, \\
&\hat s_K \bar c_{\mathbf r}^a
= B_{\mathbf r}^a \,, \\
&\hat s_K A^a_{\mathbf a\mu}
= D_{\mu}[A_{\mathbf r}] c_{\mathbf a}^a
+ g f^{abc}A_{\mathbf a\mu}^b c_{\mathbf r}^c \,, \\
&\hat s_K \bar c_{\mathbf a}^a
= B_{\mathbf a}^a \,, \\
&\hat s_K U(\pi_{\mathbf r})
= U(\pi_{\mathbf r})
\left[
- i g T_A^a c^a_{\mathbf r}
\right] \,,
\end{split}
\qquad
\begin{split}
&\hat s_K c^a_{\mathbf r}
= -\tfrac{1}{2}g f^{abc} c^b_{\mathbf r} c^c_{\mathbf r} \,, \\
&\hat s_K B^a_{\mathbf r}
= 0 \,, \\
&\hat s_K c_{\mathbf a}^a
= - g f^{abc} c_{\mathbf r}^b c_{\mathbf a}^c \,, \\
&\hat s_K B_{\mathbf a}^a
= 0 \,, \\
&\hat s_K \pi_{\mathbf a}^a
= -c_{\mathbf a}^a
- g f^{abc} c_{\mathbf r}^b \pi_{\mathbf a}^c \, .
\end{split}
\end{equation}
We have for example
\be
\hat s_K
\left(
A_{\mu{\bf a}}+D_\mu[A_{\bf r}]\pi_{\bf a}
\right)^a
=
g f^{abc}
\left(
A_{\mu{\bf a}}+D_\mu[A_{\bf r}]\pi_{\bf a}
\right)^b
c_{\bf r}^c \, .
\ee
As usual, the gauge-fixing fermion $G_K$ is essentially arbitrary, provided that it fixes the gauge. It can depend arbitrarily on the \stu fields, even without derivative suppression.

\section{Conclusions}

In this work we have constructed the Schwinger-Keldysh path-integral framework for non-Abelian gauge theories quantised via the BRST procedure in covariant gauges, allowing for arbitrary physical initial states defined at finite time. 
To include an arbitrary initial state inside the path integral we need to work with either field eigenstates or coherent states. With this in mind, we first clarified the Schrödinger-picture formulation of the indefinite BRST Hilbert space. This differs from that for scalar field theories in two ways, one is that the naive eigenstates for the temporal component of the gauge field are not normalizable. The second is that the ghosts being Grassmann odd must be dealt with differently.
The temporal component of the gauge field is consistently handled using Pauli's imaginary-eigenvalue representation, which amounts to a redefinition of the inner product.
Unlike physical fermions, when Lorenz gauge is chosen, the Faddeev-Popov-DeWitt ghosts admit a Grassmann analogue of the Schrödinger representation because their equations of motion are second order. Taken together, these provide a field-space for a BRST invariant and normalisable inner product. 

The Nakanishi-Lautrup representation proves to be particularly natural: when $B^a$ is chosen as a configuration variable, the BRST symmetry acts directly on equal-time field variables, the algebra closes off shell, and the relevant boundary contributions are properly controlled. Furthermore, the condition for physical states and density matrices to be BRST invariant becomes trivial to satisfy in field space. The generic form of the physical states and the inner product is then seen to be that of Batalin and Marnelius, realised explicitly in \Sch eigenstates.

Specifically, physical wavefunctionals are expressed as a BRST-exact dressing factor multiplying a gauge-invariant functional of the spatial gauge and matter fields. In the fixed-time functional integral that defines the inner product, this dressing plays the role of the gauge-fixing fermion and its inclusion is necessary to give a finite norm. 
The gauge invariant part of the wavefunction, is equivalent to that obtained in temporal gauge for which the Dirac constraints are imposed directly, as employed in real-time lattice approaches. 
For generic mixed states, the Hata-Kugo prescription of inserting $e^{\pi Q_G}$ inside the trace enforces the projection onto the physical Hilbert space. 
In the Schrödinger picture, this insertion compensates the usual fermionic minus sign in the ghost trace, and in the special case of thermal equilibrium it induces the standard periodic KMS boundary conditions for ghosts.

Building on these elements, we constructed the finite-time Schwinger-Keldysh path integral for general physical initial states. In the Nakanishi-Lautrup formulation, the bulk action, the matching conditions at the final time, and the initial density matrix all fit consistently within a single diagonal BRST symmetry. The corresponding CTP generating functional then obeys the associated Ward-Takahashi-Slavnov-Taylor identities, or equivalently the in-in Zinn-Justin equation, with potential boundary-source terms that encode the initial state.

An important takeaway is that doubling the Schwinger-Keldysh contour does not lead to a doubled BRST symmetry. The trace condition at the final time explicitly violates the naive advanced BRST symmetry, already at the level of the free theory. The only transformation that admits an operator formulation via the conserved BRST charge is the diagonal, or retarded, BRST transformation. 
Crucially, the diagonal/retarded BRST symmetry is not the same as the BRST symmetry of the retarded gauge transformations, since it also includes information on the advanced gauge symmetry. Thus, when constructing an open EFT for a gauge theory, demanding only retarded gauge invariance is insufficient/incorrect. 
The correct organising framework is diagonal BRST invariance, which includes both the ghost sector and the Nakanishi-Lautrup fields.

As in the in-out or Euclidean path integral, the Schwinger-Keldysh path integral for gauge theories potentially suffers non-perturbatively from the Gribov ambiguity. The precise resolution depends on the chosen treatment of Gribov copies, if indeed a treatment is necessary, and there is no clear consensus on this. One of the most interesting proposals is that the Gribov problem disappears if the correct form of the gauge fixing term is chosen \cite{Scholtz:1997jp,Rogers:1999zj}.

Finally, we implemented this framework in the context of open effective field theories. By defining the Feynman-Vernon influence functional through matching to the CTP generating functional, or equivalently to the CTP 1PI effective action, we guarantee that integrating out matter fields or high-momentum modes maintains the diagonal BRST symmetry. When the open EFT is truncated to quadratic order in the advanced fields, the full BRST symmetry contracts to an exact Keldysh BRST symmetry: standard BRST transformations associated with the retarded gauge symmetry, together with a linear BRST transformation acting on the advanced fields. As a concrete example, we discuss the Schwinger-Keldysh formulation of the Hard Thermal Loop effective action which is an example of an EFT without symmetry breaking. We then constructed, to second order in advanced fields, the generic form of the BRST gauge fixed open EFT for a gauge theory in SSB state.

\section*{Acknowledgements}

 AJT is supported by the STFC Consolidated Grant ST/X000575/1. MM is supported Kavli IPMU which was established by
the World Premier International Research Center Initiative (WPI), MEXT, Japan. MM is also grateful for the hospitality of Perimeter
Institute where part of this work was carried out. Her visit to Perimeter Institute was supported by a
grant from the Simons Foundation (1034867, Dittrich). Research at Perimeter Institute is supported
in part by the Government of Canada through the Department of Innovation, Science and Economic
Development and by the Province of Ontario through the Ministry of Colleges and Universities.

\bibliographystyle{JHEP}
\bibliography{references.bib}

\end{document}